\documentclass{aa}  

\usepackage{graphicx}
\usepackage{txfonts}
\usepackage{longtable}
\usepackage{amssymb} 
\usepackage{multirow}
\usepackage{blindtext}
\usepackage{verbatim}
\usepackage{adjustbox}
\usepackage{subfig}
\usepackage{appendix}
\usepackage[normalem]{ulem}
\usepackage{color}
\begin{document} 
\graphicspath{{images/}}
\title{Characterization of exoplanetary atmospheres with \texttt{SLOPpy}}
\author{D. Sicilia\inst{1,2}, L. Malavolta\inst{2,3}, L. Pino\inst{4}, G. Scandariato\inst{1}, V. Nascimbeni\inst{3}, G. Piotto\inst{2}, I. Pagano\inst{1}}

\institute{
\label{inst:1} INAF - Osservatorio Astrofisico di Catania, Via S. Sofia 78, 95123 Catania, Italy \and
\label{inst:2} Dipartimento di Fisica e Astronomia,  Universit\`{a} degli Studi di Padova, Vicolo dell’Osservatorio 3, 35122, Padova, Italy \and
\label{inst:3} INAF - Osservatorio Astronomico di Padova, Vicolo dell’Osservatorio 5, 35122, Padova, Italy \and
\label{inst:4} INAF - Osservatorio Astrofisico di Arcetri, Largo E. Fermi 5, 50125 Firenze, Italy}

\date{}
 
\abstract{
Transmission spectroscopy is among the most fruitful techniques to infer the main opacity sources present in the upper atmosphere of a transiting planet and to constrain the composition of the thermosphere and of the unbound exosphere. \\
Not having a public tool able to automatically extract a high-resolution transmission spectrum creates a problem of reproducibility for scientific results. As a consequence, it is very difficult to compare the results obtained by different research groups and to carry out a homogeneous characterization of the exoplanetary atmospheres. \\
In this work, we present a standard, publicly available, user-friendly tool, named \texttt{SLOPpy} (Spectral Lines Of Planets with python), to automatically extract and analyze the optical transmission spectrum of exoplanets as accurately as possible. 
Several data reduction steps are first performed by \texttt{SLOPpy} to correct the input spectra for sky emission, atmospheric dispersion, the presence of telluric features and interstellar lines, center-to-limb variation, and Rossiter-McLaughlin effect, thus making it a state-of-the-art tool.\\
The pipeline has successfully been applied to HARPS and HARPS-N data of ideal targets for atmospheric characterization. 
To first assess the code’s performance and to validate its suitability, here we present a comparison with the results obtained from the previous analyses of other works on HD 189733 b, WASP-76 b, WASP-127 b, and KELT-20 b. Comparing our results with other works that have analyzed the same datasets, we conclude that this tool gives results in agreement with the published results within 1$\sigma$ most of the time, while extracting, with \texttt{SLOPpy}, the planetary signal with a similar or higher statistical significance.}

\keywords{planets and satellites: atmospheres - techniques: spectroscopic}

\titlerunning{Characterization of exoplanetary atmospheres with \texttt{SLOPpy}}
\authorrunning{D. Sicilia et al}
\maketitle

\section{Introduction}
The last two decades of exoplanet discoveries have revealed that extrasolar systems are very common and extremely diverse in masses, radii, temperatures, and orbital parameters. Characterizing their atmospheres is necessary to better understand the formation and evolution of these systems \citep{Oberg_2011, Mordasini_2016, Cridland_2019}. Moreover, exoplanetary atmospheres are ideal laboratories for studying their chemical composition and global atmospheric dynamics, such as circulation, \citep{Snellen_2010, Kataria_2016}, the presence of thermal inversion in the pressure-temperature (P-T) profile \citep{Kreidberg_2018, Pino_2020}, escaping planetary material \citep{Spake_2018, Owen_2019}, and cloud formation \citep{Sing_2015, Helling_2019}.\\
\indent Progress has been made in detecting atmospheric signatures of exoplanets through photometric and spectroscopic methods using a variety of space-based and ground-based facilities \citep{Deming-Seager}. 
Transmission spectroscopy, pioneered by \citet{Charbonneau_2002}, is among the most fruitful techniques to infer the main opacity sources present in the atmosphere of a transiting planet and to constrain its composition from the deep layers of hot planets \citep{Sing_2016} out to the thermosphere \citep{Redfield_2008, Wyttenbach_2015} and beyond to the unbound exosphere \citep{Vidal-Madjar_2003, Ehrenreich_2015}.\\
\indent For most targets, low spectral resolution ($R \sim 10^2$) from ground-based instruments is not suitable to robustly detect elemental or molecular absorption features because 1) telluric contamination is difficult or impossible to deal with, and 2) a low-resolution transmission spectrum can only probe the deepest layers of the atmosphere since that the information about the outermost layers is encoded in the narrow core of the lines.
On the other hand, high-resolution ($R \sim 10^5$), ultra-stable spectrographs, such as HARPS (High Accuracy Radial velocity Planet Searcher) at the ESO telescope \citep{Mayor_2003}, its counterpart for the northern hemisphere HARPS-N (North) at the TNG \citep{Cosentino_2012}, and ESPRESSO at the 8-m VLT \citep{ESPRESSO}, can reach the upper levels of exoplanetary atmospheres, offering an extremely interesting opportunity in this research field.\\
\indent Thanks to high-resolution spectroscopy (HRS), broad  molecular bands, such as those from H$_2$O, CO, TiO, and CH$_4$, can be uniquely identified \citep{Snellen_2010, Brogi_2016, Allart_2017}, while single atomic lines, such as Na and K lines, are spectrally resolved and their shape and velocity components can be analyzed in great detail \citep{Brogi_2014, Keles_2019}. This also allows one to disentangle the stellar, planetary, interstellar, and telluric signals by their different radial velocity shift and to detect, at high fidelity, a specific molecule in the planetary atmosphere which would otherwise be impossible to unambiguously see from low-resolution data. 
The presence of hazes or clouds can obscure molecular features in the transmission spectra; the HRS has the potential to probe the higher altitudes above the clouds and thereby constrain the atmospheric abundances of cloudy exoplanets (e.g., \citealt{Gandhi_2020, Hood_2020}). However, the presence of aerosols can introduce some degeneracies in the retrieved values, such as the one between abundance of the species, reference pressure, and atmospheric temperature \citep{Brogi-Line}. \\
\indent The HRS also gives us access to important kinematical information about the planetary atmosphere. By analyzing the fine shape of the lines it is possible to trace a P-T profile; through the  broadening of the line profiles it is possible to detect super-rotation of the atmosphere; through the Doppler-shift of the lines we can understand if there are high altitude winds (e.g., \citealt{Wyttenbach_2020, Cauley_2021, Pai_2022}). In addition, the HRS removes the need for a reference star that is at the same time close on sky to the target, and of similar brightness – two rarely met conditions that limit the low-resolution spectrophotometry studies. In contrast, HRS data are normalized to the stellar continuum itself during the analysis process, therefore it solely requires accurate telluric correction and a careful analysis of the radial velocities of the star-planet system. However, normalising the continuum of ground-based high-resolution spectra, which is a mandatory step in data reduction, loses information about the continuum itself. Therefore, transmission spectra retrieved combining space-borne low- to medium-resolution spectroscopy, including that one of the recently-launched JWST, and ground-based HRS is highly synergic to break these degeneracies and to properly interpret complex transmission features \citep{Brogi_2017, Pino_2018, Khalafinejad_2021}, adding crucial information to their global modeling.\\
\indent Finally, there is a huge amount of archival high-resolution spectra, obtained during transits for other purposes, that are available for further analysis such as measuring the Rossiter-McLaughlin effect and its wavelength dependence \citep{Snellen_2004, Cegla_2016, Esposito_2017, Mancini_2018}. This allows to determine the sky-projected angle between the planetary orbital plane and stellar equator.\\
\indent Not having 
public tools able to automatically extract a high-resolution transmission spectrum of the planetary atmosphere causes a problem of reproducibility of scientific results,  since many details in the algorithm implementation may not be reported on a paper for the sake of readability. Using different algorithms makes the comparison of the results obtained by different working groups rather difficult.\\
\indent In this paper, we present \texttt{SLOPpy} (Spectral Lines Of Planets with python), a user-friendly, publicly available tool to homogeneously extract and analyze the transmission spectra obtained by high-resolution, ultra-stable spectrographs.
Several reduction steps are initially required to get the most reliable transmission spectrum. \texttt{SLOPpy} is the first public tool which, in addition to using current state-of-the-art techniques, also takes into account stellar effects whose treatment is very important for the purposes of analysis, such as the center-to-limb variation and the Rossiter-McLaughlin effect. 
The pipeline is modular and general enough to support new high-resolution facilities, although at the moment only HARPS and HARPS-N are supported. Furthermore, the code architecture provides easy means of modification and expansion. \\
\indent This  paper  is  organized  as  follows: section \ref{sec:the_pipeline} provides a description of the software and an overview of the stages in the pipeline;
section \ref{sec:extraction_abs_depth} illustrates two different approaches implemented in the pipeline to compute the absorption depth from the spectral features; in section \ref{sec:application_data} we report the results obtained by applying \texttt{SLOPpy} to different targets that have already been analyzed by other works: HD 189733 b \citep{Wyttenbach_2015, CB_2017}, WASP-76 b \citep{Seidel_2019}, WASP-127 b \citep{Zak, Seidel_127} and KELT-20 b \citep{CB_2019}; finally, a summary and future perspectives on the pipeline can be found in section \ref{sec:conclusions}.
\section{Software description}\label{sec:the_pipeline}
\begin{figure*}
    \centering
    \includegraphics[width=\textwidth]{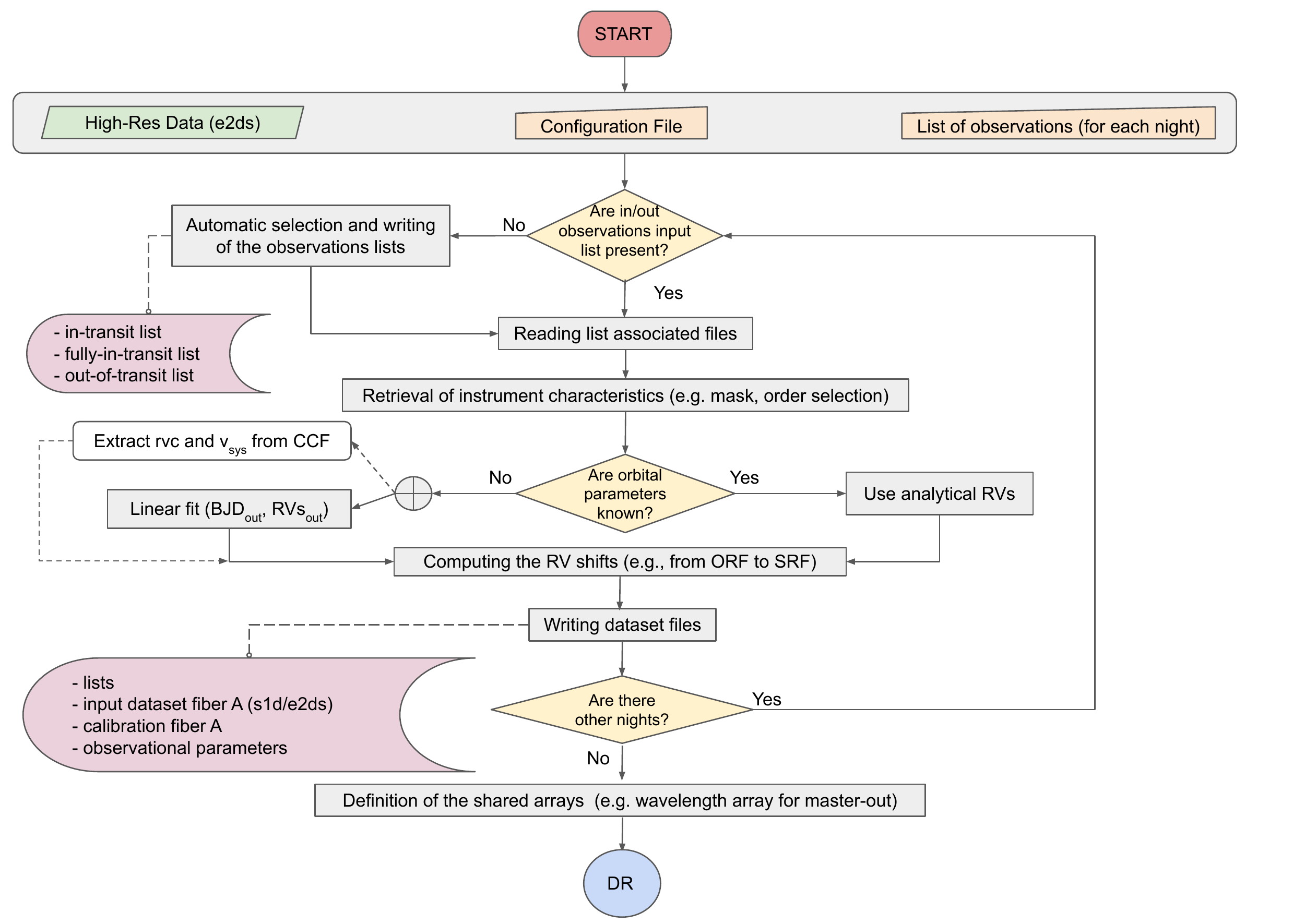}
    \caption{Processes flow scheme of $\tt{SLOPpy}$: \textit{datasets preparation}. The top area represents the user inputs (distinguishing in orange the data that must be entered manually). The pink boxes indicate the stored data. }
    \label{fig:flowchart1}
\end{figure*}
\subsection{Aim and architecture}\label{subsec:aim}
The scientific aim of the pipeline is to characterize the atmospheres of exoplanets through the detection of spectral lines in their optical transmission spectrum. Basically, to extract the transmission spectrum, what the pipeline does is to compare the spectra acquired during the transit (which contain the planetary signal) with those acquired out of transit (before the ingress and after the planet's egress).
To get a transmission spectrum that is as accurate as possible, \texttt{SLOPpy} first applies several reduction steps that are required to correct the input spectra for sky emission, atmospheric dispersion, the presence of telluric features and interstellar lines, center-to-limb variation (CLV) and Rossiter-McLaughlin (RM) effect. A detailed explanation of each reduction step is reported in section \ref{sec:data_reduction}.\\ 
\indent One of the main features of the pipeline is its \lq modularity\rq: each individual step in the analysis is performed by an independent subroutine, and the user can decide whether or not to apply it by simply adding or removing the associated keyword in a configuration file. 
Thanks to the modularity of the pipeline, the user can check the effects of each individual step on the final transmission spectrum; it is also possible to see how the results change if a particular correction is not applied. In this fashion, the user can easily understand and analyze the scientific effects of each data reduction step on the input data. Being written in separate \lq computing\rq \,and \lq plotting\rq \,modules allows a batch execution on large datasets or on server machines. Other very important features of \texttt{SLOPpy} are \lq reproducibility\rq, for example, for a given configuration file, the user always get the same result, and \lq data persistence\rq, for example, the user can extract and analyze the intermediate products and the output of a given step without running the analysis again.\\
\indent We developed our code trying to be as general as possible: all the instrumental properties, such as spectral resolution or number of echelle orders, are hard-coded in one single Python dictionary for each supported instrument, while a specific subroutine is in charge of adapting the data output from the data reduction software of the instrument to the internal standard used by \texttt{SLOPpy}. Although at the moment of writing the pipeline supports just HARPS and HARPS-N, our approach ensures a broader compatibility with other high-resolution spectrographs such as PEPSI \citep{Strassmeier_2015}, CARMENES \citep{CARMENES}, and ESPRESSO \citep{ESPRESSO} or any other spectrograph at any wavelength range if a good intra-night stability can be guaranteed.\\
\indent The \texttt{SLOPpy} pipeline is entirely written in Python 3. It is publicly available on Github\footnote{\url{https://github.com/LucaMalavolta/SLOPpy}} (along with a brief manual and some example data) and we further encourage the astronomical community to use it. 

\subsection{Datasets preparation}
\indent A flowchart representing the datasets preparation performed by \texttt{SLOPpy} is shown in Figure \ref{fig:flowchart1}. In order to successfully perform the extraction of a transmission spectrum, \texttt{SLOPpy} requires high-resolution 2-dimensional spectra and two kinds of files, which must be entered manually. The first one is a single configuration file written in \texttt{YAML}\footnote{YAML Ain't Markup Language, \url{https://yaml.org/}}, a human-readable structured data format well suited for writing complex configuration files. An overview about the overall structure of the configuration file and a brief description of the philosophy behind it can be found in the section \ref{subsec:configuration_file}.
The second kind of file required by \texttt{SLOPpy} is a list of observations to be analyzed, including both in-transit and out-of-transit observations, with each night requiring its own list. The first time \texttt{SLOPpy} is launched, for each night, three more files, listing respectively the \lq in\rq \,and \lq fully in\rq \,transit spectra and the \lq out-of-transit\rq \,spectra, will be automatically created according to the transit duration between the first and fourth contact, the transit duration between the second and third contact (when it is known), the time of exposure and the time of mid-transit of the planet provided in the configuration file. \\
\indent After reading the list of associated files and retrieving the instrument characteristics, the pipeline computes and saves in $\tt{pickle}$ files the radial velocity (RV) shifts needed to move the spectra from one reference system to another.
For example, to move the spectra from the observer reference frame (ORF) to the stellar reference frame (SRF), we need to consider the systemic velocity ($v_{sys}$), together with the RV variation due to the presence of the planet and the Barycentric Earth Radial Velocity (BERV) at each exposure. While the latter is usually provided by the observatory and/or standard data reduction software in the header of the FITS file, the other contributions can be derived directly by fitting the radial velocity on the observed spectra. However, the Rossiter-McLaughlin anomaly will not interfere when shifting the in-transit spectra in the SRF. Thus, in the absence of known orbital parameters, we recommend to use only out-of-transit spectra to linearly fit the stellar RVs.
\\
\indent Finally, after repeating the previous steps for all available nights, the pipeline defines and saves the shared arrays (e.g., the wavelength array) for all the nights. These are necessary to build the coadded spectra and the master-out (see section \ref{sec:master_out}).

\begin{figure*}[ht]
    \centering
    \includegraphics[width=\textwidth]{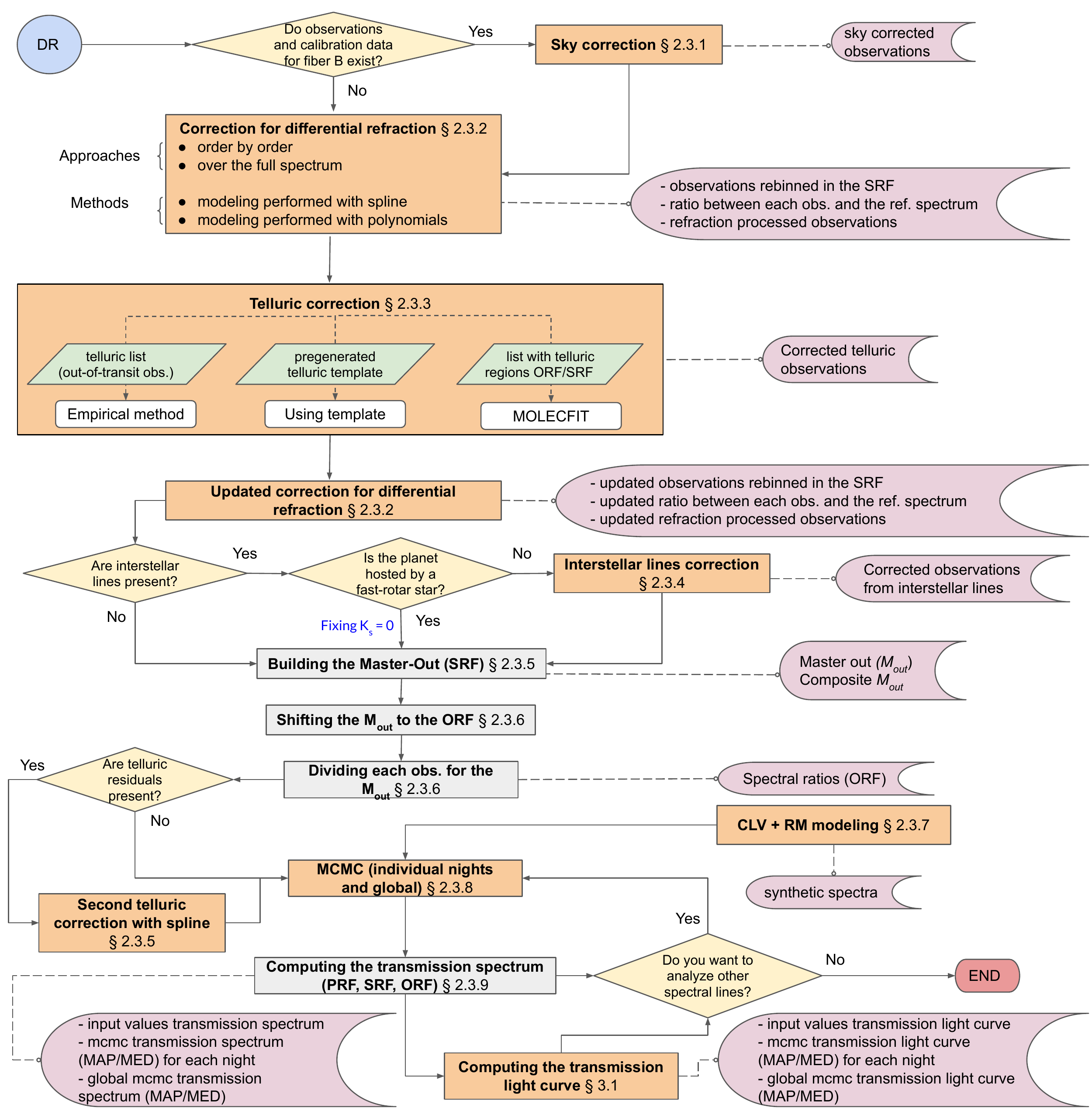}
    \caption{Processes flow scheme of $\tt{SLOPpy}$: \textit{data reduction}. The orange boxes indicate the modules that the user can switch on/off.}
    \label{fig:flowchart2}
\end{figure*}
\begin{figure*}
\centering
\includegraphics[width=\textwidth, keepaspectratio]{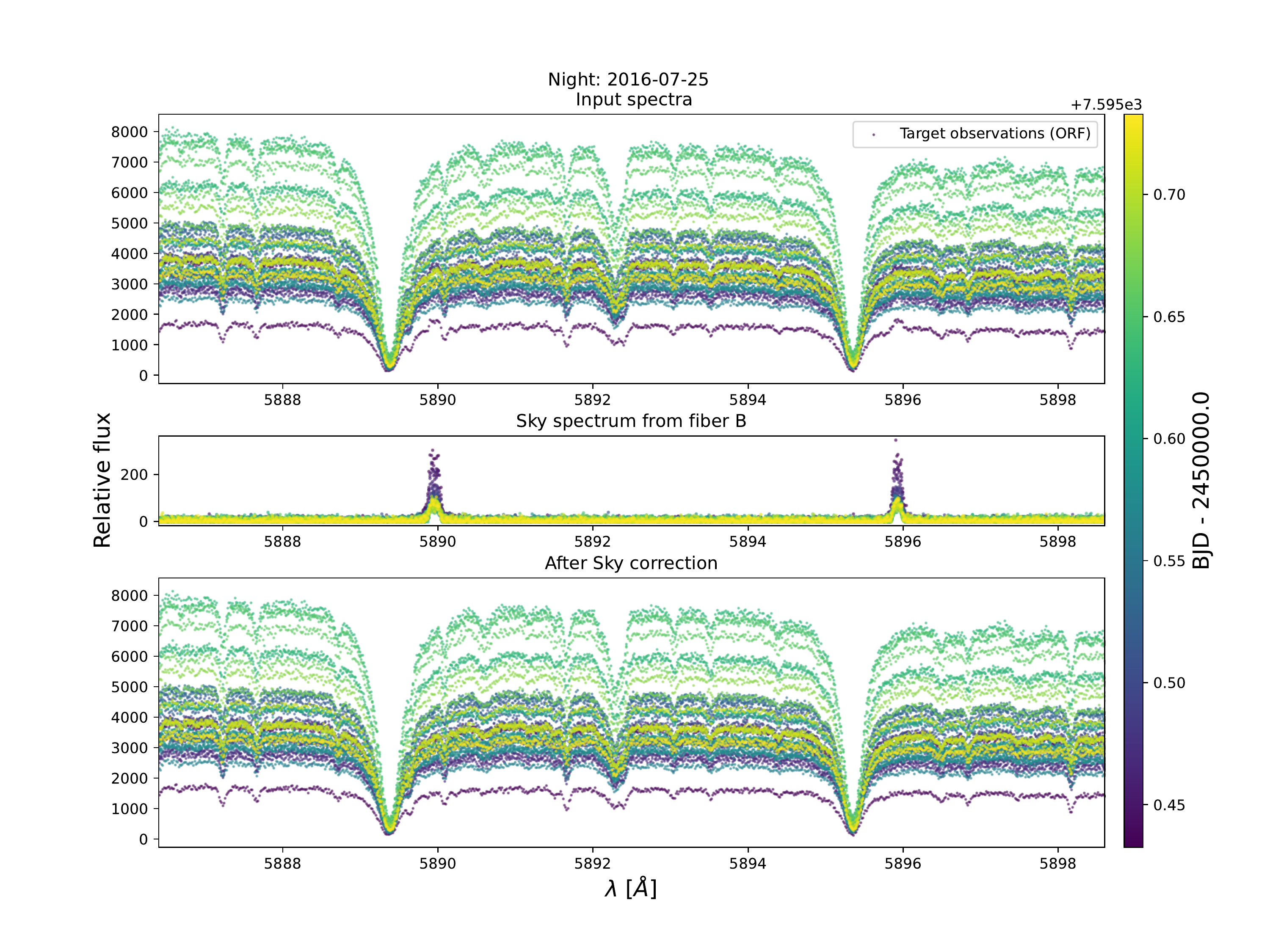}
\caption{\label{fig:sky_correction}Example of sky correction applied with \texttt{SLOPpy} to a HD 209458 b spectrum in the region of the \ion{Na}{i} doublet. \textit{Top panel:} input spectra from fiber A; \textit{middle panel:} sky emission spectra from fiber B; \textit{bottom panel:} corrected spectra. All spectra are color-coded according to the time.}
\end{figure*}
\subsection{The configuration file}\label{subsec:configuration_file}
The configuration file is thought as a place where all the possible parameters that can ultimately affect the final transmission spectrum are recorded and stored, so that the user can immediately access the details of the data reduction and analysis without delving into the code. In this context, we avoided whenever possible the use of hard-coded parameters, preferring keywords that can be explicitly specified in the configuration parameters - with a fallback value declared in a easily accessible {\it default} dictionary.\\
\indent The configuration file is divided mainly in four parts:
\begin{itemize}
    \item {\bf pipeline and plots}: here the user can list which analysis modules need to be executed and which plots to be shown, respectively. Each data reduction step is accompanied by a plotting module with the same name, so that checking the outcome of each reduction step is straightforward. It is important to remember that analyses and plots are performed independently, that is to say, the user can modify the list of plots anytime without the need of performing the time-consuming analysis again, even for intermediate steps;
    \\
    \item {\bf nights and instruments}: assuming that during a night only one transit of a given target can be observed, in this section the characteristics of each dataset gathered during a night are detailed, such as the lists of all spectra, in-transit spectra, full-transit spectra and out-of-transit spectra, the time of mid-transit ($T_c$), and the instrument used to gather the observations. For the sake of readibility, instrument properties, such as resolution and wavelength range, are detailed in a section on its own, so that it is not necessary to repeat over and over the same information if several dataset have been obtained with the same instrument; 
    \\
    \item {\bf reduction steps}: the following sections of the configuration file are dedicated to specific steps of the reduction process, and contain important parameters that can ultimately affect the final transmission spectrum. As an example, differential refraction correction (see section~\ref{sec:differential_refraction}) can be computed order-by-order or over the full spectrum at once, and it can be performed either with a polynomial or with a spline, iteratively and with a sigma-clipping removal of the outliers. All the relevant parameters can be specified by the user; additionally some sections can be duplicated under a specific dataset or instrument if a different treatment is required for a given dataset. The stellar and planetary parameters required for the change of reference systems and for the computation of the CLV are also listed in dedicated sections in this part of the configuration file;
    \item {\bf spectral lines}: in this section the spectral lines to analyze are listed. For each line the user can indicate: the spectral range over which to calculate the transmission spectrum, that must be wide enough to contain both the stellar lines and the continuous; the left and right reference bands and the central passbands for the calculation of the relative absorption depth (see section \ref{sec:extraction_abs_depth}); the fit parameters for the Markov chain Monte Carlo (MCMC) analysis to fit the final transmission spectrum (see section \ref{sec:mcmc}); the polynomial degree for the normalization of spectra and models.
\end{itemize}

\subsection{Data reduction steps}\label{sec:data_reduction}
The standard data reduction is already performed by the observatory Data Reduction Software (DRS, version 3.5-3.8 for ESO and 3.7 for TNG). The DRS produces both two-dimensional spectra ({\tt e2ds}) and one-dimensional spectra ({\tt s1d}). The first ones are float two-dimensional arrays where each line contains the extracted flux of one spectral order in photo-electrons unit. The second ones are float arrays containing the rebinned and merged spectral orders in relative flux corrected from the instrumental response.
While {\tt e2ds} spectra are referred to the ORF, {\tt s1d} spectra are corrected for the BERV, that is, {\tt s1d} spectra are referred to the barycenter of the Solar System (BRF). Rather than using the {\tt s1d} spectra, which already went through a first pass of rebinning step and a change of reference system, that is, from the ORF to the BRF, and since each rebinning is unavoidably introducing correlated noise, we decided to work on the {\tt e2ds} spectra instead. \\
\indent After the datasets preparation shown in Figure \ref{fig:flowchart1}, to extract the transmission spectrum $\tt{SLOPpy}$ performs a series of reduction steps which are described in the following sections and schematized in Figure \ref{fig:flowchart2}. To not introduce unnecessary correlated noise, we tried to avoid changing the reference system of the observed spectra whenever possible, preferring an increase of the computational cost to a decrease in the overall accuracy of the analysis (see, e.g., sections \ref{sec:transmission_spectrum_preparation} and \ref{sec:CLV_RML_effects}).

\begin{figure*}[t]
    \centering
    \includegraphics[height= 9 cm, keepaspectratio]{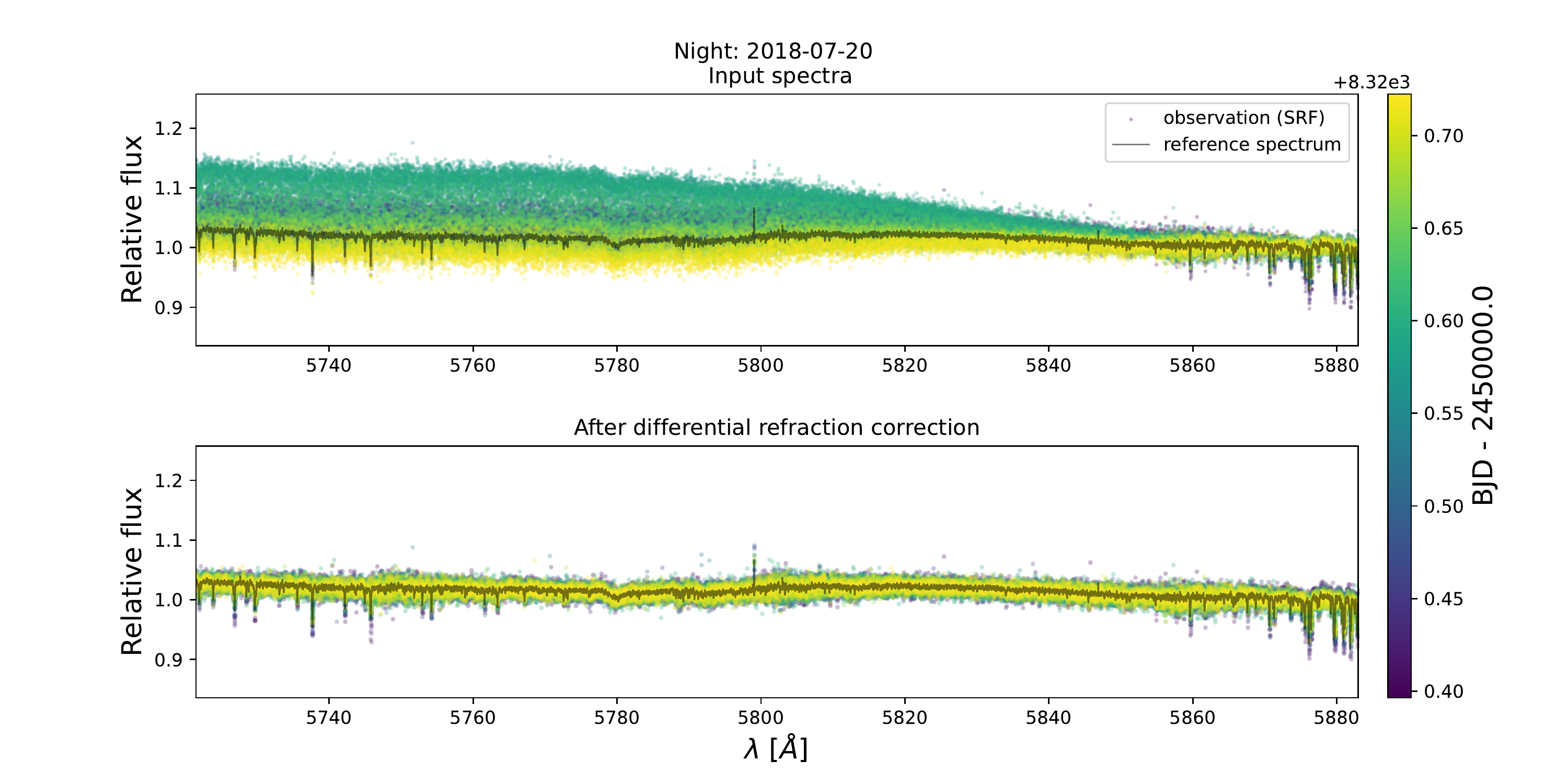}
    \caption{Example of correction of the differential refraction applied with \texttt{SLOPpy} to the input spectra of KELT-9 b retrieved with HARPS-N during the 2018-07-20 night.}
    \label{fig:differential_refraction_kelt-9}
\end{figure*}

\subsubsection{Sky correction}
\label{sec:sky_correction}
Sometimes observations are affected by sky emission features. These features evolve during the night and during the season in a different way than the telluric absorption lines. The main contributions to sky brightness are airglow and zodiacal light; in iarticular, airglow, coming from the Earth's high atmosphere (between 90 and 200 km above sea level) produces the dominant lines in the visible, namely the semi-prohibited line of neutral oxygen at 5577 Å, its doublet at 6300 and 6363 Å and the sodium doublet at 5890 and 5896 Å. These emission lines persist throughout the night and sometimes vary unpredictably.\\
\indent With HARPS/HARPS-N, sky spectrum can be retrieved simultaneously with the science observations thanks to a dedicated fiber, named fiber B, pointing at a fixed position at around 10 arcsec from the target star, ensuring exactly the same atmospheric conditions in both spectra\footnote{In this case it is not possible to use the simultaneous Fabry-Perot lamp to measure the instrumental drift, however this is not a problem since the intra-night instrumental stability of the new generation of spectrographs is sufficient for our purposes.}. The pipeline corrects for these features by subtracting the sky spectrum from the stellar spectrum, after taking into account the different transmissivity of the fibers. To do this, sky spectra are first multiplied by the ratio between the lamp flux of the two fibers.
An example is shown in Figure \ref{fig:sky_correction}. \\
\indent From the analysis of archival spectra we were not able to determine any correlation of the sky spectrum with seasons, time of observation, position in the sky or weather conditions. It is therefore strongly recommended to always gather transmission spectroscopy data with simultaneous observations of the sky.

\subsubsection{Differential refraction}\label{sec:differential_refraction}
Ground-based observations are affected by atmospheric dispersion caused by the variation of the index of refraction of air with wavelength. 
Due to differential refraction, the position of the star in the focal plane of the telescope differs depending on wavelength, with the effect getting worse at increasing airmass. If not corrected, it can affect the telluric correction and subsequently the transmission spectrum. \\
\indent To compute the correction factor as a function of wavelength, the pipeline first divides each observations with a reference spectrum obtained by coaddition of all the out-of-transit observations, in the SRF to take into account the shift of the stellar lines during the night. Then, the code models this ratio with either a low-order polynomial or a spline, depending on the choice of the user. The correction is then applied by dividing each observation by this model, after being Doppler-shifted to the ORF. The user can choose to use the same reference spectrum for all the datasets or correct each night independently, and the model can be updated after the correction for telluric absorption (section \ref{sec:telluric_correction}).
This approach works well even in a few cases where the Atmospheric Dispersion Corrector (ADC) of the telescope, which should correct for this effect, fails to update its position during the observation of a transit \citep{Borsa_2019, Pino_2020}, leading to
a considerable loss of flux on each exposures both in the blue and in the red part of the continuum (see Fig. \ref{fig:differential_refraction_kelt-9}).

\subsubsection{Telluric correction}\label{sec:telluric_correction}
One of the major difficulties of ground-based observations is to deal with the telluric imprints from the Earth's atmosphere. In the optical domain, water vapor and molecular oxygen are the main contributors of telluric absorption. Its removal is a difficult task because the intensity and position of absorption lines change with time, depending on the altitude of star over the horizon and on the weather condition of the night (airmass, water column, seeing, etc.).
\begin{figure*}[t]
    \includegraphics[height= 9 cm, keepaspectratio]{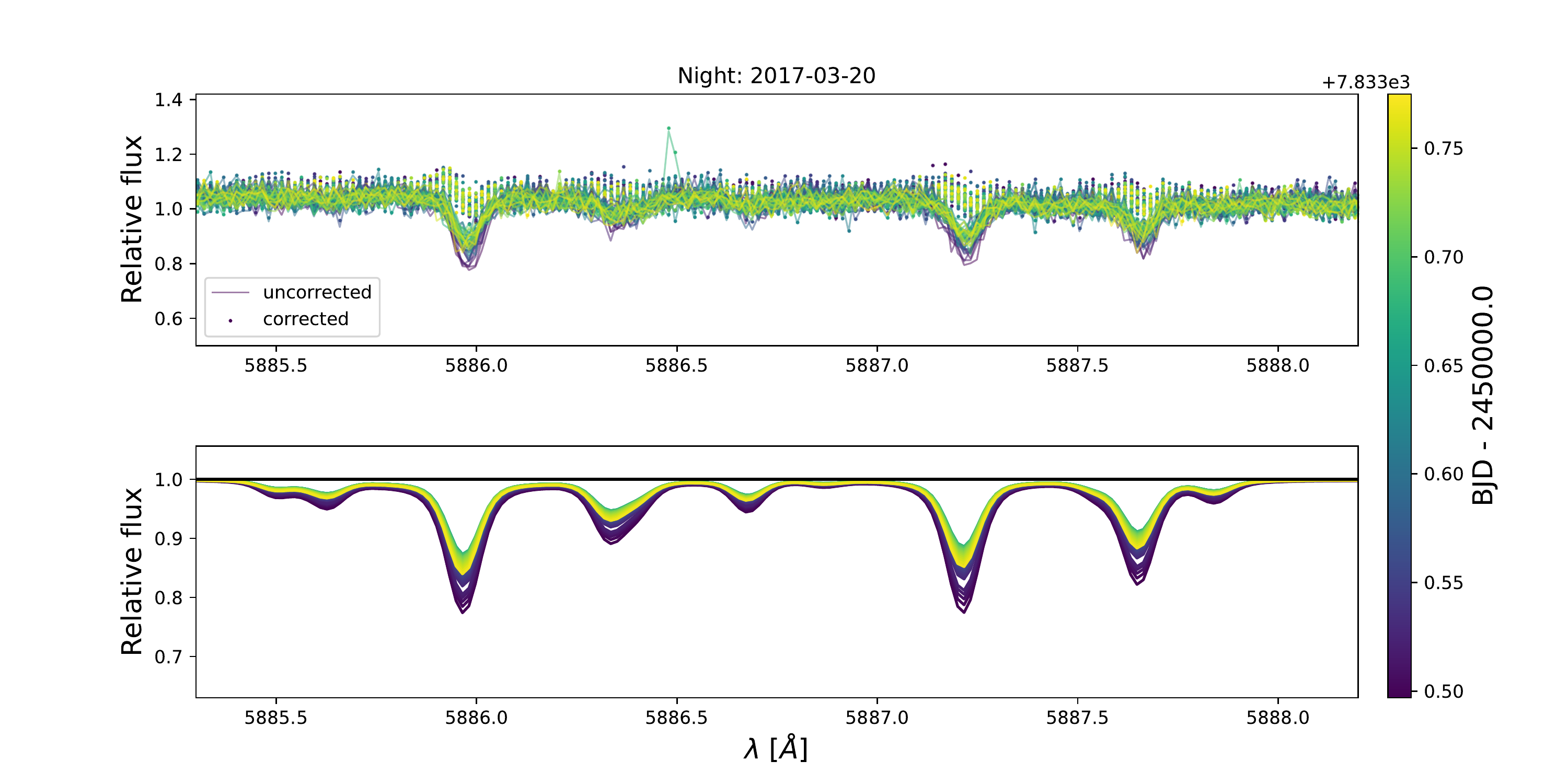}
    \caption{\label{fig:telluric_correction_molecfit} Example of telluric correction using \texttt{Molecfit} applied to a WASP-127 b spectrum. \textit{Top panel}: comparison between uncorrected and corrected spectra. \textit{Bottom panel}: telluric spectrum. All spectra are color-coded according to the time.}
\end{figure*}
Classical techniques used for telluric correction require either modeling the atmosphere using dedicated modeling tools, or taking a reference spectrum to create a telluric template.\\
\indent In \texttt{SLOPpy} different approaches have been tested and implemented: (1) empirical calulation of the telluric absorpiton based on the airmass and BERV, (2) use of a pregenerated template of the Earth’s transmission spectrum and (3) telluric modeling by the atmospheric transmission code \texttt{Molecfit} \citep{Smette_2015}. A more detailed explanation of the first two methods can be found in the Appendix \ref{Appendix_telluric_correction}. In this section, we only discuss the third method, which is certainly the most robust one and produces consistent results when applied to data from different nights with changing atmospheric conditions \citep{Langeveld_2021}.\\
\indent \texttt{Molecfit}, provided by ESO, computes a very high-resolution ($R \sim 4,000,000$) telluric spectrum using the HITRAN database and a line-by-line radiative transfer model (LBLRTM), correcting telluric lines to the noise level. \citet{Allart_2017} used it for the first time on HARPS data to search for water vapor in the transmission spectrum of HD 189733 b, using a cross-correlation technique that combines the signal of 600–900 individual lines. Then, \texttt{Molecfit} was used in high-resolution transmission spectroscopy both in the NIR for the search of helium \citep{Salz_2018, Nortmann_2018} and in the visible for the search of the sodium feature \citep{Hoeijmakers_2019, Seidel_2019}. 
\citet{Scandariato_2021} used \texttt{Molecfit} inside \texttt{SLOPpy} in the whole spectral range covered by HARPS and HARPS-N, excluding the region with wavelenghts longer than $\sim$6700 \AA \, which is heavily contaminated by saturated O$_2$ lines.\\
\indent \texttt{Molecfit} computes the reference telluric spectrum in two steps. In the first step, the LBLRTM, which is given in the ORF, is fitted to several user-defined regions of the observed spectrum by adjusting the continuum, the wavelength scale and the instrumental resolution. In addition, the molecular features are independently rescaled to take into account small differences between the atmospheric model and the actual weather. In the second step, the output parameters of the first stage are used to build a telluric absorption spectrum for the entire wavelength interval of the observations. \\
\indent Optimization of the telluric correction with this tool requires a careful selection of spectral regions with only strong telluric lines for a single molecule (H$_2$O or O$_2$), a flat continuum, and no stellar features within them, since \texttt{Molecfit} does not fit the stellar spectrum. This task is not trivial in the visible region of the spectrum, since we are in the opposite situation with respect to the ideal one for \texttt{Molecfit}. \\
\indent To perform the \texttt{Molecfit} fit, \citet{Allart_2017} provided different spectral intervals specifically for each nights of observations. In order to apply this technique to a broader range of targets, we selected two different classes, or lists, of spectral regions. The first one includes all the wavelength intervals with strong telluric features, where the wavelengths are listed in the ORF. The second list includes all the spectral regions that are devoid of stellar lines, selected in the SRF and using HD 189733 spectra as a template. The second list is updated for each night by shifting it from the SRF to the ORF, according to the systemic RV of the star and the average BERV of the night, and assuming that a variation of a few tenths of \AA ngstroms\footnote{One HARPS/HARPS-N pixel corresponds to $\sim 0.016 $ \AA, $\sim 0.08 $ km/s at 5900 \AA.} in the interval range do not significantly affect the \texttt{Molecfit} fit. The final list of wavelength intervals that \texttt{SLOPpy} passes to \texttt{Molecfit} is then given by the intersection of the two lists, resulting in a list of spectral regions with strong telluric features and almost no stellar features.\\
\indent For fainter targets, whose observations are characterized by a lower signal-to-noise ratio (S/N), the fit could become unreliable. To prevent this problem, \texttt{SLOPpy} performs an automatic coadding of consecutive exposures; the \texttt{Molecfit} analysis is then performed on the coadded spectra, assuming that weather conditions does not vary significantly within the window time of the coadding, while the second stage of the analysis (i.e., the computation of the reference telluric spectrum on the whole wavelength range) is applied on each observation individually.\\
\indent Differently from any empirical approach based on the variation of the spectral night with airmass, \texttt{Molecfit} can be applied safely also to the observations taken during the transit, since the regions used to characterize the telluric model are supposedly free from the planetary signals. As shown from Figure \ref{fig:telluric_correction_molecfit}, \texttt{Molecfit} is a very powerful tool which is able to correct telluric features to the noise level,
with a very careful choice of the fitted spectral regions being the only requirement, although it is extremely slow compared to the other algorithms.\\
\indent If, after telluric correction, some residuals are still present, the user of the pipeline can decide to execute an additional correction. Following a similar procedure to that of \citet{Snellen_2010}, when dividing each spectrum by the master-out (see section \ref{sec:master_out}), \texttt{SLOPpy} can remove the remaining telluric residuals by normalising each pixel value by its variance over time using a linear spline.

\subsubsection{Interstellar lines}\label{sec:interstellar_lines}
Depending on the distance of the star and its position in the sky, interstellar absorption lines may be present in the spectra. Since the interstellar lines are stationary with respect to the barycenter of the exoplanetary system, while the RV of star is changing due to the presence of the planet, for slow rotators ($v \sin{i} < 10 $ km$\, $s$^{-1}$) we verified that not correcting for their presence would interfere with the final transmission spectrum, especially in the sodium doublet region.
For each dataset, \texttt{SLOPpy} build a model of the interstellar absorption lines by fitting them with a spline, after stacking the spectra in the BRF and normalising for the local continuum. Each observation is then corrected by this empirical model. The results obtained from this step are still not optimal and we are trying to improve the procedure. In any case, in the data analyzed so far, it has not been necessary to use this module.\\
\indent In some cases, the interstellar lines correction can be neglected. If the host star is a fast rotator ($v \sin{i} > 15 - 20 $ km$\, $s$^{-1}$, \citealt{Stauffer_1997}), as typically happens for early-type stars, even a wavelength shift on the order of one pixel (corresponding to a RV shift of $\sim 0.8 $ km$\,$s$^{-1}$ in the case of HARPS/HARPS-N, i.e., the RV variation due to a planet of $\sim 8~M_J$) does not change the shape of the spectrum noticeably. Indeed, the spectral lines are so broadened and shallow that the variation in flux as a function of wavelength (in the rest frame) is negligible compared to the photon noise. In this case, the out-of-transit spectra can be coadded without taking into account the reflex motion due to the planet. With this assumption, interstellar lines are automatically deleted when dividing each spectrum by the master out (see section \ref{sec:master_out}). An example of this approach is provided by \citet{CB_2018} for the analysis of MASCARA-2 b/KELT-20 b, and here we confirm its validity.

\subsubsection{Building the master-out}\label{sec:master_out}
All the observations (both in-transit and out-of-transit) contain the stellar light and the telluric absorption, while in-transit observations additionally contain the exoplanet atmosphere absorption imprinted into the stellar flux. In order to isolate it and remove the stellar contribution, the pipeline computes a \lq master-out\rq \,spectrum ($M_{OUT}$), which is given by the integration of the exposures obtained before and after the transit.\\
\indent Due to the RV variation of the star during the transit of the planet, the stellar lines of all the spectra are shifted in wavelength. So, before building the $M_{OUT}$, the pipeline shifts the spectra to the SRF to align all the stellar lines at the same position.

\subsubsection{Transmission spectrum preparation}\label{sec:transmission_spectrum_preparation}
After applying the aforementioned corrections, each in-transit observation ($F_{i,in}$) is divided by the $M_{OUT}$, after being rescaled to unity ($\tilde{M_{OUT}}$) with respect to a reference wavelength range provided by the user, in order to obtain the transmission spectrum of that exposure:
\begin{equation}\label{eq:single transmission spectrum}
\tilde{\mathfrak{R_i}}\mid_{ORF}=\frac{F_{i,in}}{\tilde{M_{OUT}}\mid_{ORF}}
\end{equation}
referred to as the spectral ratio. The computation is performed in the ORF, meaning by shifting back the (high S/N) $M_{OUT}$ from the SRF, to avoid any unnecessary rebinning step on the much noisier individual in-transit observations.\\
\indent The pipeline extracts the transmission spectrum in the spectral range around the atomic species of interest, as specified in the configuration file. 
Indeed, the transmission spectra are computed across the whole spectral range of the instrument, easily allowing in the future for other approaches to search for atomic or molecular species, as the cross-correlation function \citep{Pino_ccf, Hoeijmakers_2019}, while restricting to specific wavelength ranges allows for computationally expensive analysis to be performed within a reasonable time. The resonant sodium doublet (at 5889.95 \AA \, and 5895.92 \AA), thanks to its large absorption cross section, is one of the most investigated atmospheric signature (e.g., \citealt{Snellen_2008}, \citealt{Wyttenbach_2015}, \citealt{CB_2020}).\\

\begin{figure*}
    \centering
    \includegraphics[width=0.5\textwidth, keepaspectratio]{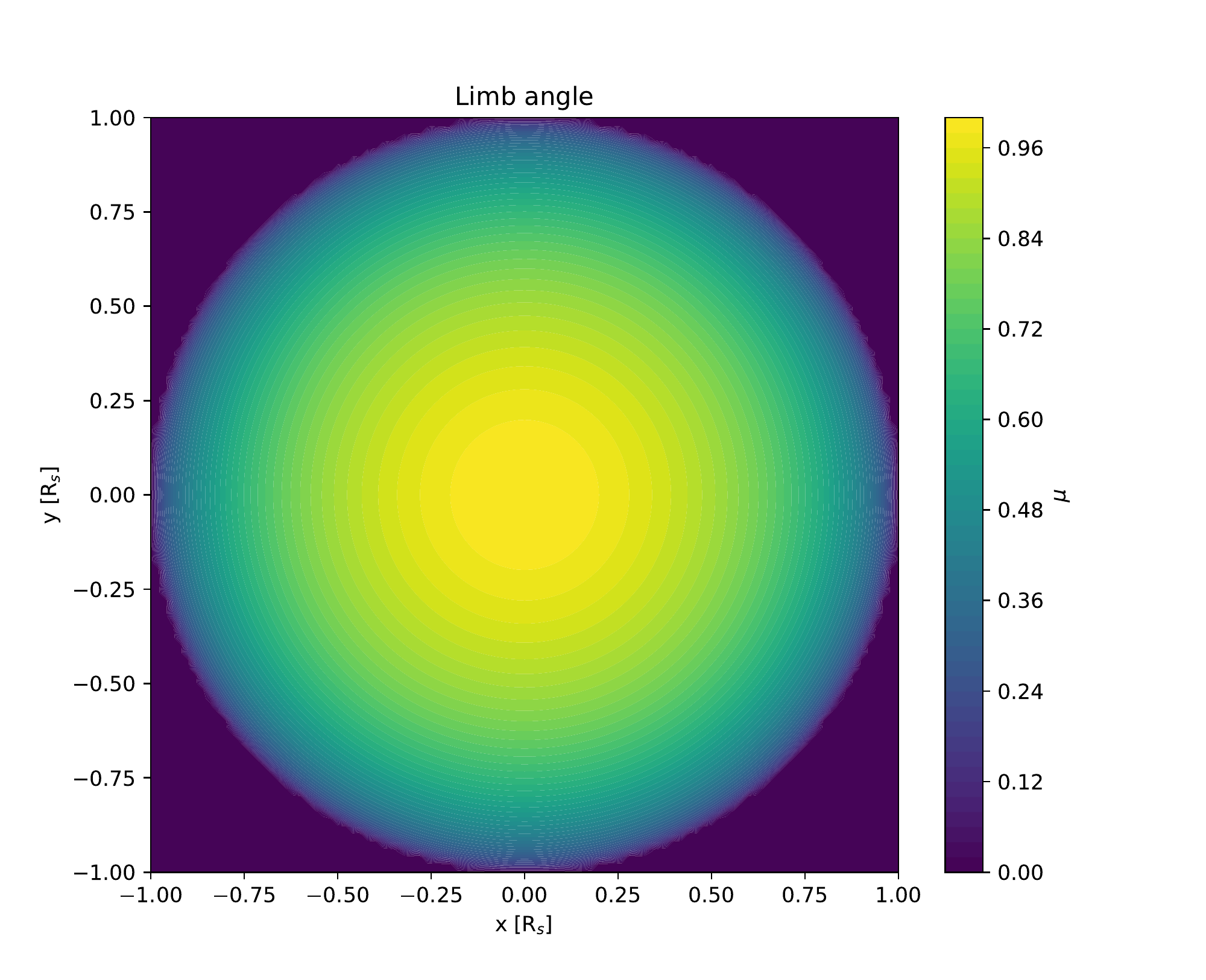}\includegraphics[width=0.5\textwidth, keepaspectratio]{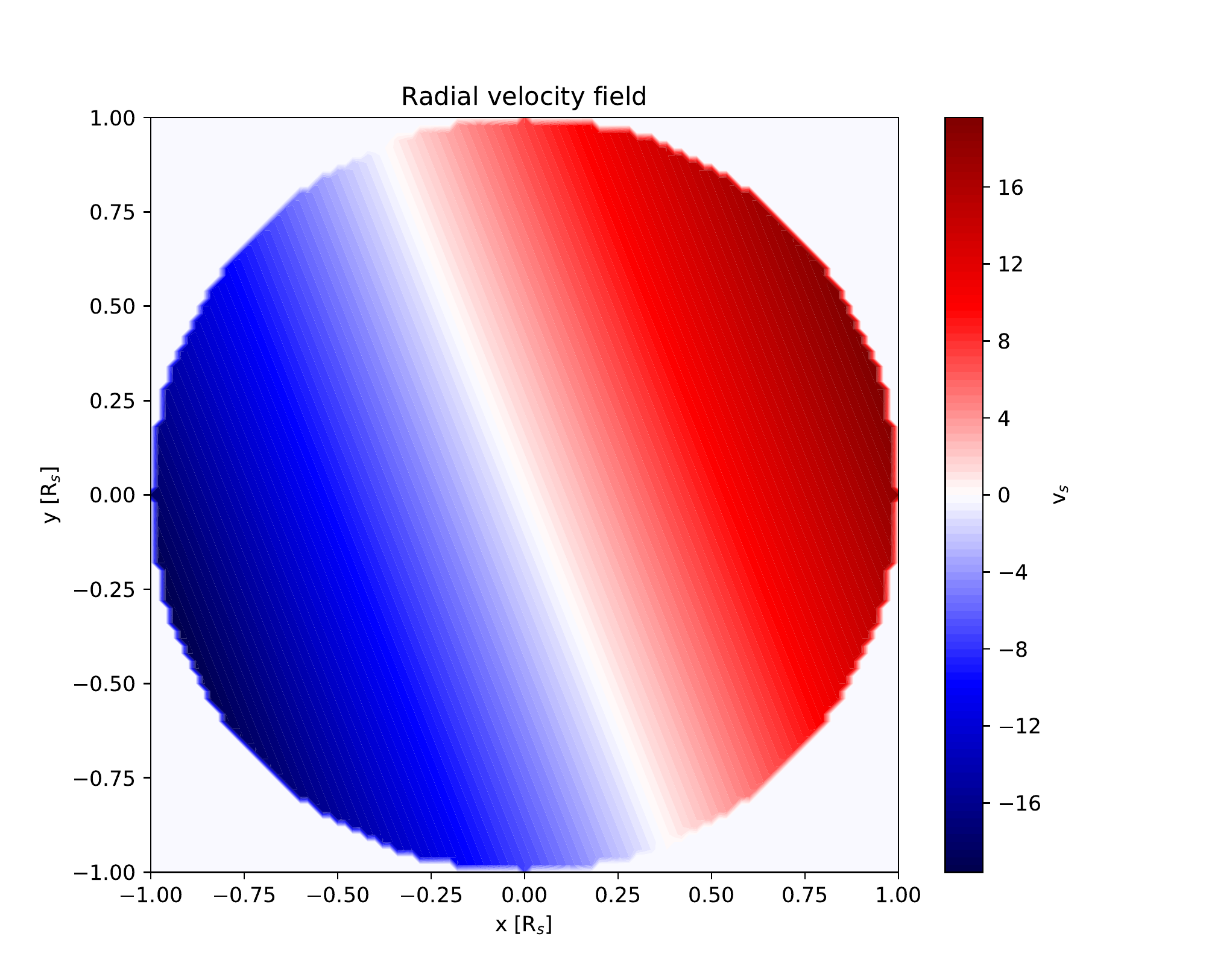}
    \caption{Different $\mu$-angles (\textit{left panel}) and radial velocities (\textit{right panel}) on the stellar disk for a exoplanetary system with $\lambda$ = (-22.1 $\pm$ 6.0)$^\circ$ e $v\sin{i}$ = (19.6 $\pm$ 0.5) km/s.}
    \label{fig:clv_modeling}
\end{figure*}

\subsubsection{Center-to-limb variation and Rossiter-McLaughlin effect}\label{sec:CLV_RML_effects}
As the planet passes in front of the stellar disk, it blocks part of the stellar lights. Depending on the properties of the spectrum in the specific location of the stellar surface covered by the planet, the integrated stellar spectrum may differ from that obtained from the out-of-transit observations, causing a deformation in the exoplanetary transmission spectrum. The two main effects altering the transmission spectra are the center-to-limb variation (CLV) and the Rossiter-McLaughlin (RM) effect.\\
\indent The RM effect is a consequence of the rotation of the star. The stellar light blocked by the planet may be red-shifted or blue-shifted, depending on which side of star the planet is covering, thus causing a deformation of each spectral line (e.g., \citealt{Louden-Wheatley}).\\
\indent The CLV effect, on the other hand, is the variation in the profile of the normalized stellar lines from the center to limb across the stellar disk, due to the fact that we observe outer, colder photospheric layers when moving toward the limb, in analogy with the photometric limb darkening. For stronger absorption lines, the CLV can vary quite significantly, becoming crucial in the detection of exoplanetary atmospheric species \citep{Yan_2017}. For slowly rotating stars the magnitude of the deformation on the absorption lines roughly scales with $(R_p/R_s)^2$. Beyond the planet-to-star radius ratio, there are a variety of parameters that affect the CLV features, including stellar parameters and planet orbital parameters. In particular, \cite{Yan_2017} shows that CLV effect for the \ion{Na}{i} D lines is stronger for stars with lower $T_{eff}$ and with an impact parameter close to 
$b$ = 0.84. Moreover, the modeled CLV feature also depends on the stellar atmosphere model; normally a non-LTE and three-dimensional model can produce more  realistic results \citep{Leenaarts_2012, Borsa_2021}.\\ 
\indent In the analysis of the transmission spectrum, it is important to take into account both effects simultaneously, since the spectrum emerging from each element of the stellar surface covered by the planet will be both shifted in wavelength (for the RM effect) and formed at a different photospheric depths (for the CLV effect). This results in the loss of a specifically limb-darkened fraction of the stellar light and with a specific RV. \\
\indent To correct these two effects, following  \citet{Yan_2017}, we start by simulating synthetic stellar spectra at 21 different $\mu$ values (ranging from zero to one with a step of 0.05), where $\mu = \cos\theta$, with $\theta$ the angle between the normal to the stellar surface and the line of sight (\lq limb angle\rq). At the stellar edge, we assume $\mu$ = 0.001 instead of $\mu$ = 0 to avoid numerical problems \citep{Czesla_2015}. Stellar spectra are obtained with Spectroscopy Made Easy (SME, \citealt{Piskunov-Valenti}) using the line list from the VALD database (\cite{Ryabchikova_2015}) and MARCS \citep{Gustafsoon_2008} or Kurucz ATLAS9 \citep{Kurucz_ATLAS} models. These spectra are computed without including the rotational broadening of the star, since they are meant to represent the emerging spectra from a given position of the star. The disk-integrated, broadened spectrum is also computed by SME. Subsequently, we divide the stellar disk into elements with a size of $0.01R_s \times 0.01R_s$; each of these elements has a $\mu$ value, so its spectrum is linearly interpolated from the previously calculated synthetic spectra. In order to consider the RM effect, as in \citet{Yan-Henning}, each spectrum is also Doppler-shifted to the projected velocity of the element dependent on its position on the disk, the geometry of the system, the rotational velocity of the star and the presence of differential rotation (Fig. \ref{fig:clv_modeling}). \\
\indent It is very important to note that modeled stellar spectra at different orbital phases are for one planetary radius ($R_p$), while the actual effective radius is larger than $R_p$ because the planetary absorption is wavelength - dependent. 
To account for this radius change, the user can decide to introduce a factor $r$ when fitting the data (see section \ref{sec:mcmc}), assuming that the corresponding line profile change is $r$ times the result obtained when the radius is 1 $R_p$; in this case, the pipeline computes the modeled stellar spectra for a grid of $r$ values ranging, for example, between 0.5 and 2.5 with steps of 0.1.\\
\indent Finally, we model the stellar spectrum of each observation taken during transit by integrating all the surface elements that are obscured by the planet, and subtracting the result for the disk-integrated stellar spectrum. The correction factor as a function of wavelength is obtained by dividing each simulated spectrum by the out-of-transit disk-integrated model. The transmission spectrum from each observation (see eq. \ref{eq:single transmission spectrum}) is then corrected for CLV and RM effects by dividing it for the corresponding correction factor. \\
\indent This correction is performed on the individual transmission spectra still in the original reference frame by rebinning the correction model instead, even if it requires a larger computational effort and a more complex structure of the algorithm. At the end, the only rebinning step on the observed spectra is performed when moving to the planetary reference frame (PRF) for the construction of the average transmission spectrum (see section \ref{sec:transmission_spectrum}).

\subsubsection{MCMC analysis}\label{sec:mcmc}
If the residuals present spectral absorption lines, the user can decide to model them and to estimate the detection significancy. As in \citet{Yan-Henning}, \texttt{SLOPpy} performs a Markov chain Monte Carlo (MCMC) analysis with the \texttt{emcee} tool \citep{MCMC}. The model assumes a Gaussian profile for the absorption lines, and a flat spectrum ($\tilde{\mathfrak{R_i}}$ = 1) otherwise. The CLV and RM modeling includes a factor $r$ to take into account the possible difference in the planetary radius in the wavelength range under analysis with respect to the value obtained with transit photometry (likely gathered in a different wavelength range).\\
\indent The free parameters of the model are: the RV semi-amplitude of the planet ($K_p$), required to model the atmospheric absorption lines in the PRF; the contrast ($h$) and the full width at half maximum (FWHM) of the Gaussian profile describing the absorption of the planet; the RV of the atmospheric wind ($v_{wind}$), which is the shift relative to the PRF transition; the effective planet radius scale factor ($r$). For each observation the spectral lines of the transmission model are shifted to the PRF according to the instantaneous RV of the planet, computed from $K_p$ and the orbital phase at the time of the observation.\\
\indent We note that the MCMC analysis is only applied to fully in-transit data  (i.e., excluding the ingress and egress), because the planetary absorption spectrum during ingress and egress is different from the absorption spectrum when fully in-transit where the RM effect most constrains the $r$ factor. While in principle it is possible to model this difference, this feature is not yet implemented in \texttt{SLOPpy}.\\ 
\indent To reduce the computational time, the analysis is performed in the SRF over individual transmission spectra binned with a step-size chosen by the user (e.g., \citealt{CB_2019} used a step size of 0.05 \AA). This choice is dictated mainly by the requirement of keeping both the usage of computer memory and the execution time of each step of the MCMC to an acceptable level. Also in this case, this is the only rebinning step performed over the data under analysis, thus minimizing the impact of systematic noise introduced by the rebinning process over low S/N spectra.
We explored the possibility of performing  the analysis on the unbinned spectra (i.e., in the original reference frame) thus avoiding any rebinning process at all, but the extremely longer computational time and the higher memory requirements (mostly due to the fact that the spectra are not on a common wavelength grid anymore), compared to the previous approach, prevented us from fully developing this option, although we do not exclude it as a feature.\\
\indent The uncertainties we report here represent the confidence interval that encompasses the 15th to the 84th percentile of the posterior distribution of the free parameters. The errors are propagated from the photon noise, however they are very often underestimated; for this reason we allow for a free jitter parameter in the fitting procedure. This parameter is then added in quadrature to the errors in order to take into account any additional systematics (e.g., bad seeing).\\
\indent In the case where there are several nights for the same target, the MCMC is performed first on the individual nights, and then a global fit including all the nights simultaneously is performed. The pipeline also returns the transmission spectrum with the parameters fixed by the user (in this case the planetary radius will not be a free parameter) and a so-called \lq average out\rq \,transmission spectrum built using only out-of-transit spectra. The average out transmission spectrum, which is expected to be flat, is used to check for any residuals of nonplanetary origin.\\
\indent In the configuration file the user can set the number of steps and walkers, the range where to compute the fit, the binning step and the parameters priors. Besides, there is a flag to decide whether to leave free the parameters $r$ and $v_{wind}$. If the MCMC analysis is to be performed on several spectral lines at the same time (e.g., the two lines of the sodium doublet or the magnesium triplet), the user can also decide whether the lines should share the same FWHM or the same radial velocity shift due to the wind.

\subsubsection{The final transmission spectrum}\label{sec:transmission_spectrum}
The final transmission spectrum is
obtained by summing all the spectral ratios of equation \eqref{eq:single transmission spectrum} in the PRF:
\begin{equation}\label{eq:transmission spectrum}
\tilde{\mathfrak{R}}\mid_{PRF}=\sum_{i=0}^N (\tilde{\mathfrak{R_i}}\mid_{PRF})
\end{equation}
where $N$ is the number of in-transit spectra. This avoids the shear of the
planetary atmospheric absorption lines that we would obtain by performing this step in the stellar or observer reference frame. If more datasets are available for the same target, for example, the target has been observed on different nights or with different instruments, the pipeline can combine all the datasets to compute the average transmission spectrum. \\
\indent The master-out must always be built in the SRF. On the other hand, the transmission spectrum can be built in either the PRF, as it has to be for the detection of a planetary signal, in the SRF, if there is a suspect that the signal could have a stellar origin, for instance, stellar activity, or in the ORF, if there is a suspect that the signal could have a local origin, such as, incorrect removal of telluric absorption.

\section{Extraction of the absorption depth}\label{sec:extraction_abs_depth}
\begin{figure}
    \centering
    \includegraphics[width=0.5\textwidth, keepaspectratio]{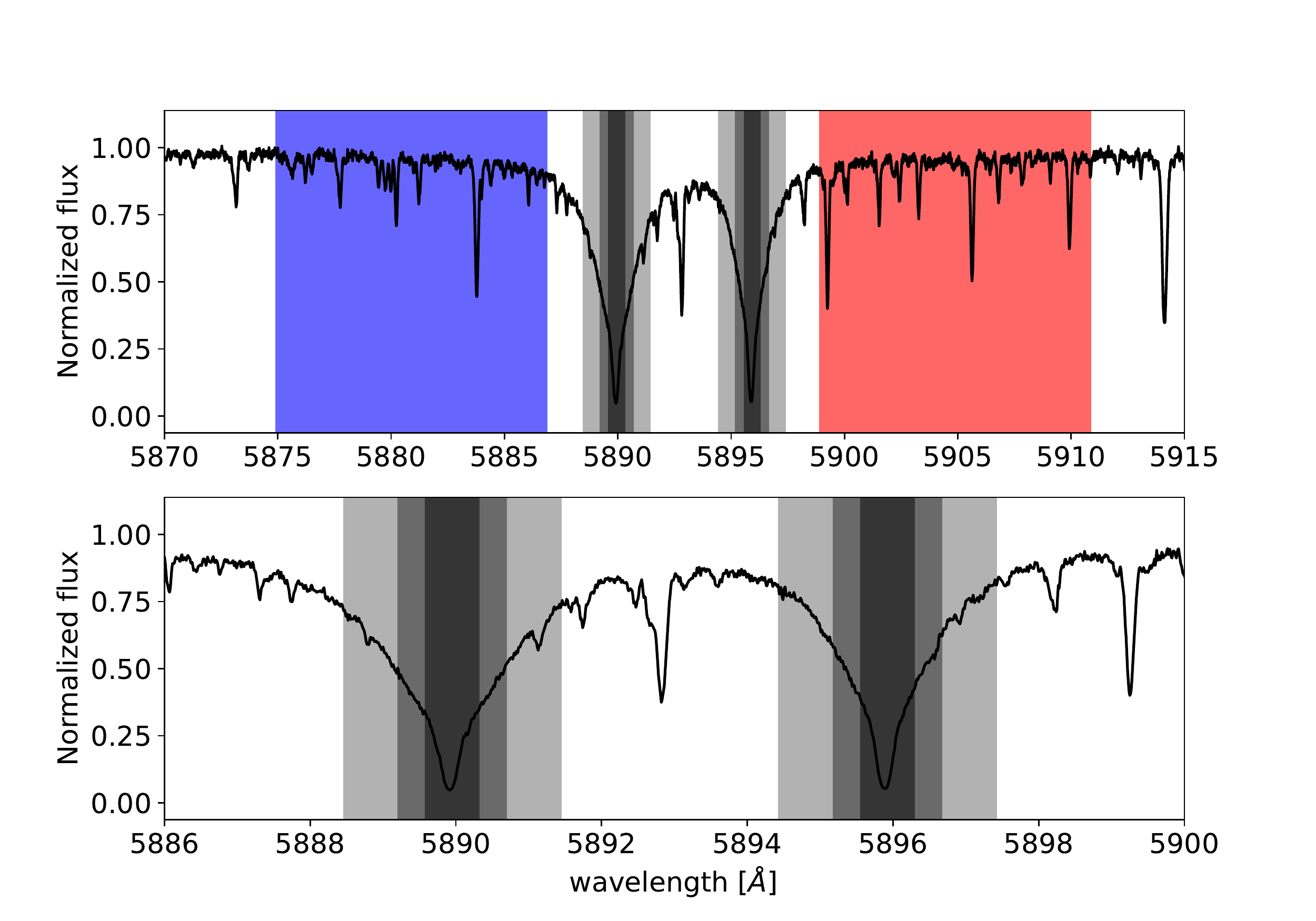}
    \caption{Observed spectrum of HD 189733 b. The top panel shows the three central bandwidths (0.75 \AA , 1.50 \AA \,  and 3.00 \AA ) used to measure the relative flux of the \ion{Na}{i} D lines. These passbands are centered at the line cores of D$_1$ and D$_2$. The reference passbands at the blue and red parts of the \ion{Na}{i} D lines are also shown. The bottom panel is the zoom at the \ion{Na}{i} D lines.}
    \label{fig:passbands}
\end{figure}

To characterize the planetary signal, if detected, we calculate the relative absorption depth ($\delta$) by integrating the flux across narrow passbands centered on the wavelength of the atmospheric species under analysis, and by comparing it with the flux integrated in a reference passband in the continuum. In the case of the sodium doublet, the central passband ($C$) containing the signal is split in two smaller passbands, one for each line of the doublet, D$_2$ and D$_1$, where we are interested in measuring the excess absorption. Because each central passband encompasses only one line, the absorption depth of the two \ion{Na}{i} doublet lines is often averaged. In order to compare our results with other works in literature (e.g., \citealt{Wyttenbach_2015, Seidel_2019}), we chose their same bandwidths (e.g., 2 $\times$ 0.75 \AA, 2 $\times$ 1.50 \AA, 2 $\times$ 3.00 \AA) for the central passbands, and their same reference passbands (e.g., $[5874.89 - 5886.89]$ \AA \, and $[5898.89 - 5910.89]$ \AA) for the continuum, taken in the blue and in the red side of the central passbands (see Fig. \ref{fig:passbands}). \\
\indent The extraction of the absorption depth can be performed by \texttt{SLOPpy} following two different approaches: analyzing the \lq transmission light curve\rq \,(\citealt{Snellen_2008}, see section \ref{subsection:excess_approach}), or analyzing the final transmission spectrum (\citealt{Redfield_2008}, see section \ref{subsection:division_approach}).

\subsection{Transmission light curve analysis}\label{subsection:excess_approach}
The presence of an absorbing species in the exoplanetary atmosphere can be seen as a relative flux decrease during the transit. This can be inferred building the transmission light curve, that is the relative flux in a specific passband as a function of time \citep{Charbonneau_2002, Snellen_2008}. \texttt{SLOPpy} derives the relative flux for each exposure ($t$) and for a given user-defined passband ($\Delta\lambda$) as: 
\begin{equation}
F_{rel}(t,\Delta\lambda)=\frac{2 \times \overline{F(C)}}{\overline{F(B)}+\overline{F(R)}}
\end{equation}
where $\overline{F(C)}$ is the weighted average flux inside the passband centered on the atomic species of interest, while $\overline{F(B)}$ and $\overline{F(R)}$ are the weighted average flux inside the two reference passbands in the blue and red side of the spectral feature. While $F(B)$ and $F(R)$ should remain unchanged during the planet’s transit, since it only contain stellar flux, $F(C)$ changes according to the additional absorption of the planet’s atmosphere.\\
\indent The  relative  absorption  depth  of the light curve for  each  passband and for each spectral line is given by:
\begin{equation}\label{eq:absorption_depth_2}
\delta(\Delta\lambda)=\frac{\overline{{F_{rel}(t_{in})}}}{\overline{F_{rel}(t_{out})}}-1, 
\end{equation}
where $\overline{F_{rel}(t_{in})}$ e $\overline{F_{rel}(t_{in})}$ are the weighted average relative flux during and outside the transit respectively. We note that this calculation is performed in the SRF. During the transit, the planetary signal will move across the spectrum according to the RV of the planet (ideally equal to zero at the central time of transit, for a circular orbit). If the central passbands are too narrow, may not capture the planetary signals in the first and final part of the transit, thus resulting in a shorter transit duration in the transmission light curve.

\subsection{Transmission spectrum analysis}\label{subsection:division_approach}
Following \citet{Redfield_2008} and \citet{Wyttenbach_2015}, in this second approach, the relative absorption depth is directly extracted from the final transmission spectrum $\tilde{\mathfrak{R}}$, given by equation \ref{eq:transmission spectrum}.
The transmission flux averaged inside the central passband (C) is compared with the transmission flux averaged inside the two adjacent control passbands, on the blue (B) and red (R) side of the central passbands.\\
\indent The relative absorption depth is given by:
\begin{equation}\label{eq:absorption_depth_1}
\delta=\frac{\sum_Cw_i\tilde{\mathfrak{R}}(\lambda_i)}{\sum_Cw_i}-\frac{1}{2}\biggl(\frac{\sum_Bw_i\tilde{\mathfrak{R}}(\lambda_i)}{\sum_Bw_i}+\frac{\sum_Rw_i\tilde{\mathfrak{R}}(\lambda_i)}{\sum_Rw_i}\biggl)
\end{equation}
where the weights $w_i$ are given by the inverse of the squared uncertainties on $\tilde{\mathfrak{R}}$.
The uncertainties are assumed to arise from photon and readout noise and are propagated from the individual spectra. \\
\indent Since the measurement of the relative absorption depth is performed in the PRF, this approach can give a good estimate of the statistical significance of the signal. Despite this, for some types of targets, for example, longer-period targets, there are other more reliable methods to confirm or not the planetary nature of the signal, such as the empirical Monte Carlo, or bootstrapping, analysis \citep{Redfield_2008}.

\begin{table}[ht]
 \caption{\label{tab:list_targets} List of targets analyzed for the pipeline validation.}
 \centering
    \begin{tabular}{c|c|c|c}
    \hline
    \hline
    Target & Date & Instrument & References \\
    \hline
    \multirow{3}{*}{HD 189733 b} & 2006-09-07 & \multirow{3}{*}{HARPS} &  \multirow{3}{*}{1, 2}\\
    & 2007-07-19 & &\\
    & 2007-08-28 & &\\
    \hline
    \multirow{3}{*}{WASP-76 b} & 2012-11-11 & \multirow{3}{*}{HARPS} &  \multirow{3}{*}{3}\\
    & 2017-10-24 & &\\
    & 2017-11-22 & &\\
    \hline
    \multirow{3}{*}{WASP-127 b} & 2017-02-27 & \multirow{3}{*}{HARPS} &  \multirow{3}{*}{4, 5}\\
    & 2017-03-20 & &\\
    & 2018-03-31 & &\\
    \hline
    \multirow{3}{*}{KELT-20 b} & 2017-08-16 & \multirow{3}{*}{HARPS-N} &  \multirow{3}{*}{6}\\
    & 2018-07-12 & &\\
    & 2018-07-19 & &\\
    \hline
    \end{tabular}
   \tablebib {
The last column indicates the references of the works with which we compare our results. (1) \citet{Wyttenbach_2015}; (2) \citet{CB_2017}; (3) \citet{Seidel_2019}; (4) \citet{Zak}; (5) \citet{Seidel_127}; (6) \citet{CB_2019}.}
\end{table} 
\section{Application to data}\label{sec:application_data}
We tested the pipeline by applying it to four benchmark targets whose datesets were acquired with HARPS or HARPS-N (see Table \ref{tab:list_targets}). All these targets were analyzed in other works searching for the resonance \ion{Na}{i} doublet, exploiting the fact that all of them have a large scale height and are hosted by a bright star, so they are favorable targets for atmospheric characterization.

\subsection{HD 189733 b}\label{subsec:HD189}
\begin{figure*}
    \centering
    \includegraphics[width=\textwidth, keepaspectratio]{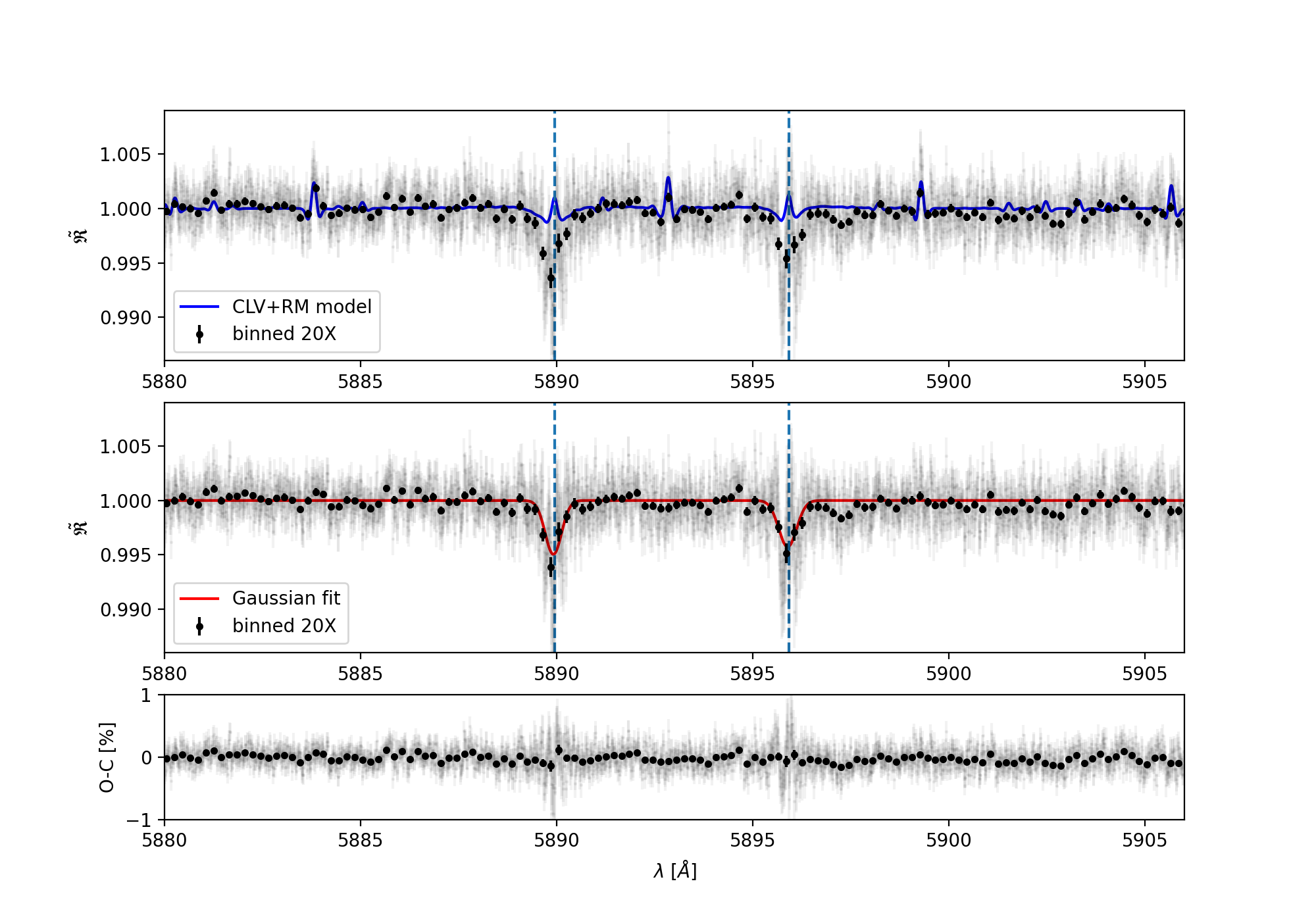}
    \caption{Final transmission spectrum of HD 189733 b centered around the sodium doublet in the planetary rest frame (light gray), also binned by 20x in black circles, without the correction of the CLV and RM effects (\textit{upper panel}) and after their correction (\textit{middle panel}). The blue line is the best-fit CLV+RM model used to correct the final transmission spectrum.
    The red line is the MCMC Gaussian fit applied to both lines of the sodium doublet. The blue dashed lines indicate the rest frame transition wavelengths of the sodium doublet. \textit{Lower panel}: residuals of the Gaussian fit.}
    \label{fig:transmission_spectrum_HD189733}
\end{figure*}
HD 189733 b is among the best-studied planets to date. It is a Hot Jupiter with approximately the same mass and radius of Jupiter, orbits a bright ($V \sim 7.7$) and active K-type star every $\sim 2.2$ days and the scale height of its atmosphere is about 200 km \citep{Desert_2011}.\\
\indent For this target, we analyzed three transits observed with HARPS. These data, retrieved from the ESO archive from programs 072.C-0488(E), 079.C-0127(A) (PI:Mayor), and 079.C-0828(A) (PI: Lecavelier des Etangs), have been analyzed in several works (e.g., \citealt{Wyttenbach_2015}, \citealt{Yan_2017}, \citealt{CB_2017}, \citealt{Borsa-Zannoni}, \citealt{Langeveld_2021}). The first and the third night were employed by \citet{Triaud_2009} to measure the RM effect. Furthermore, with these same observations, \citet{DiGloria} detected a slope in the ratios of the planet-star radius, as found by \citet{Pont_2008} and \citet{Sing_2011} using HST data, interpreting it as Rayleigh scattering.\\
\indent Of these three sequences, the first one was obtained using low-cadence (900 to 600 s) exposures, while the two others using high-cadence (300 s) exposures. The second night has no observations before transit.
All parameters used for this target are taken from \citet{Addison_2019} except the effective temperature which is taken from \citet{Stassun_2017}.\\ 
\indent In these observations, sky spectra on fiber B were gathered only for the second and third night. However, no sodium emission was observed in these sky spectra. Other authors did not apply a sky correction to any 
of these nights, so we have decided to skip as well this step for a more consistent comparison with their results. \\
\indent \cite{Yan_2017} showed that for the case of HD 189733 b, the \ion{Na}{i} absorption depth is overestimated if the CLV effect was not considered, which will result in a considerable overestimation of the \ion{Na}{i} abundance. For this reason, we decided to apply the correction of the CLV effect.\\
\indent We verified that the RM effect is clearly visible for this target as the measured RVs, as obtained from the FITS header of the spectra, deviate from the expected orbital motion of the star due to the presence of the planet. For this reason, the RM effect is taken into account together with the CLV correction (as explained in section \ref{sec:CLV_RML_effects}).\\
\indent We compare the results of our analysis to \citet{Wyttenbach_2015}, hereafter W2015, and \citet{CB_2017}, hereafter CB2017. For consistency, we only include CLV and RM correction in the comparison with CB2017.\\
\indent Figure \ref{fig:transmission_spectrum_HD189733} shows the final transmission spectrum extracted by \texttt{SLOPpy} combining all three HARPS nights without the correction of the CLV and RM effects (upper panel) and after applying this correction (middle panel). The model of the CLV and RM effects is over plotted in blue. In both spectra the two exoplanetary sodium lines peak out from the continuum but the second spectrum presents slightly deeper and thinner lines than the first, as the contribution of CLV and RM effect is greater in the wings of the doublet. \\
\begin{table*}
 \caption{\label{tab:HD189733} Summary of the best-fit parameters and 1-$\sigma$ error bars obtained with the MCMC fitting procedure for HD 189733 b.}
 \centering
 \begin{spacing}{1.4}
    \begin{tabular}{c|c|c|c|c|c|c|c}
    \hline
    \hline
    \multicolumn{8}{c}{HD 189733 b}\\
    \hline
     & \multicolumn{2}{c|}{\ion{Na}{i} D$_2$}  & \multicolumn{2}{c|}{\ion{Na}{i} D$_1$}& \multicolumn{3}{c}{\ion{Na}{i} D$_{12}$}\\
    &  $h$ & FWHM  & $h$ & FWHM & $v_{wind}$  & $r$ & $K_p$  \\
    &  [\%] & [km/s] &  [\%] & [km/s] & [km/s] & [$R_p$] & [km/s] \\
    \hline
   2006-09-07 & -0.54$_{-0.09}^{+0.10}$ & 33.1$_{-5.1}^{+6.0}$ & -1.01$_{-0.20}^{+0.21}$ & 7.10$_{-1.37}^{+1.72}$ & -8.0$_{-0.8}^{+0.9}$ & - & 89.0$_{-8.1}^{+12.3}$ \\
   2007-07-19 & -0.66$_{-0.20}^{+0.23}$ & 19.1$_{-4.7}^{+5.2}$ & -0.46$_{-0.13}^{+0.15}$ & 28.0$_{-7.0}^{+8.4}$ & +1.8$_{-2.3}^{+2.9}$ & - & 84.1$_{-5.0}^{+11.8}$ \\
   2007-08-28 & -0.55$_{-0.13}^{+0.14}$ & 24.6$_{-6.1}^{+7.8}$ & -0.58$_{-0.15}^{+0.26}$ & 26.9$_{-15.8}^{+8.7}$ & -7.3$_{-2.1}^{+3.0}$ & - & 82.0$_{-3.5}^{+8.1}$ \\
   all nights & -0.53$_{-0.07}^{+0.07}$ & 30.0$_{-3.2}^{+3.8}$ & -0.46$_{-0.07}^{+0.07}$ & 29.4$_{-3.5}^{+4.3}$ & -2.0$_{-1.1}^{+1.1}$ & - & 79.7$_{-1.8}^{+3.8}$ \\
   \hline
    \multicolumn{8}{c}{after CLV+RM correction}\\
    \hline
   2006-09-07 & -0.49$_{-0.10}^{+0.11}$ & 34.2$_{-5.9}^{+7.7}$ & -0.96$_{-0.21}^{+0.22}$ & 7.2$_{-1.4}^{+2.0s}$ & -7.9$_{-0.8}^{+1.0}$ & 0.90$_{-0.08}^{+0.07}$ & 96.7$_{-13.2}^{+14.6}$ \\
   2007-07-19 & -1.35$_{-0.77}^{+0.62}$ & 5.8$_{-1.9}^{+8.4}$ & -0.31$_{-0.12}^{+0.15}$ & 27.9$_{-8.4}^{+10.2}$ & -1.4$_{-0.8}^{+3.2}$ & 1.01$_{-0.08}^{+0.08}$ & 80.1$_{-2.9}^{+10.6}$ \\
   2007-08-28 & -0.50$_{-0.15}^{+0.16}$ & 21.4$_{-6.5}^{+8.7}$ & -0.49$_{-0.14}^{+0.23}$ & 29.5$_{-12.8}^{+11.4}$ & -7.3$_{-2.4}^{+3.0}$ & 1.09$_{-0.07}^{+0.07}$ & 88.1$_{-8.9}^{+20.9}$ \\
   all nights & -0.49$_{-0.08}^{+0.09}$ & 25.0$_{-3.5}^{+4.2}$ & -0.43$_{-0.07}^{+0.08}$ & 27.4$_{-4.2}^{+5.4}$ & -1.9$_{-1.2}^{+1.2}$ & 0.99$_{-0.04}^{+0.04}$ & 81.2$_{-3.6}^{+7.7}$ \\
    \hline
    \hline
    \end{tabular}
    \end{spacing}
\end{table*} 
\indent The contrast ($h$) and the FWHM of each sodium lines are reported in Table \ref{tab:HD189733} together with the shared values of $v_{wind}$, $r$ and $K_p$. 
Our results are compatible with W2015, while with CB2017 our results are in agreement inside the error bars only for the D$_1$ line. For the D$_2$ line, slightly larger values of the line contrast and FWHM are obtained by CB2017. We assume that this discrepancy is likely due to the different method used for telluric correction, as our results are compatible for both lines with \citet{Langeveld_2021} who analyzed the same HARPS data correcting the CLV and RM effects and using \texttt{Molecfit} to remove telluric features, like us. In any case, the D$_2$/D$_1$ found in both this work and CB2017 is compatible inside the error bars with what expected for this target (i.e., D$_2$/D$_1$ $\gtrsim$ 1.2, \citealt{Gebek-Oza}).\\
\begin{table*}[ht]
\caption{\label{tab:absorption_depths_HD189}Summary of the measured relative absorption depth in [\%] of atmospheric sodium on HD 189733 b extracted from the transmission spectrum (TS) and from the transmission light curve (TLC).}
 \centering
    \begin{tabular}{c| c c c| c c c}
    \hline
    \hline
     & \multicolumn{3}{c|}{TS} & \multicolumn{3}{c}{TLC}\\
     & 0.75 \AA & 1.50 \AA & 3.00 \AA & 0.75 \AA & 1.50 \AA & 3.00 \AA \\
     \hline
    2006-09-07 & 0.28 $\pm$ 0.04 & 0.14 $\pm$ 0.02 & 0.07 $\pm$ 0.01 & 0.32 $\pm$ 0.10 & 0.21 $\pm$ 0.06 & 0.12 $\pm$ 0.03\\
      2007-07-19 & 0.34 $\pm$ 0.05 & 0.12 $\pm$ 0.03 & 0.07 $\pm$ 0.02 & 0.40 $\pm$ 0.10 & 0.20 $\pm$ 0.07 & 0.13 $\pm$ 0.04\\
      2007-08-28 & 0.39 $\pm$ 0.04 & 0.18 $\pm$ 0.03 & 0.06 $\pm$ 0.02 & 0.40 $\pm$ 0.12 & 0.25 $\pm$ 0.07 & 0.11 $\pm$ 0.04\\
      all nights & 0.35 $\pm$ 0.02 & 0.17 $\pm$ 0.01 & 0.07 $\pm$ 0.01 & 0.38 $\pm$ 0.07 & 0.22 $\pm$ 0.04 & 0.12 $\pm$ 0.02\\
       \hline
      \multicolumn{7}{c}{after CLV+RM correction}\\
    \hline
     2006-09-07 & 0.27 $\pm$ 0.04 & 0.12 $\pm$ 0.02 & 0.07 $\pm$ 0.01 & 0.32 $\pm$ 0.09 & 0.16 $\pm$ 0.06 & 0.11 $\pm$ 0.03\\
      2007-07-19 & 0.25 $\pm$ 0.05 & 0.11 $\pm$ 0.03 & 0.07 $\pm$ 0.02 & 0.35 $\pm$ 0.10 & 0.15 $\pm$ 0.07 & 0.11 $\pm$ 0.04\\
      2007-08-28 & 0.32 $\pm$ 0.04 & 0.14 $\pm$ 0.02 & 0.05 $\pm$ 0.02 & 0.30 $\pm$ 0.12 & 0.20 $\pm$ 0.07 & 0.09 $\pm$ 0.04\\
      all nights & 0.28 $\pm$ 0.02 & 0.13 $\pm$ 0.01 & 0.06 $\pm$ 0.01 & 0.32 $\pm$ 0.07 & 0.17 $\pm$ 0.04 & 0.10 $\pm$ 0.02\\
    \hline
    \hline
    \end{tabular}
\end{table*}
\begin{figure*}
    \centering
    \includegraphics[width=\textwidth]{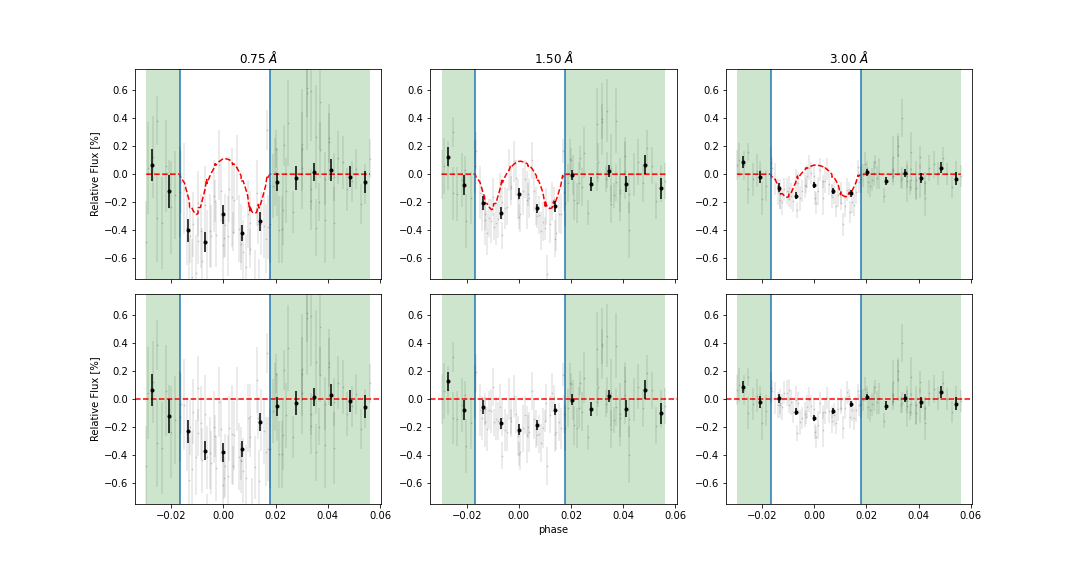}
    \caption{Transmission light curves (TLC) of HD 189733 b for all three nights combined using three different central bandwidths: 0.75 Å (left column), 1.50 Å (middle column) and 3.00 Å (right column). The gray data points show the relative absorption for each exposure; in black the data are binned by 20 spectra. The green background marks the exposures taken out-of-transit. \textit{First row}: Observed TLCs; the red dashed line shows the CLV and RM model. \textit{Second row}: TLCs after removing the contribution of the CLV and RM effects.}
    \label{fig:tlc_HD189733}
\end{figure*}
\indent Concerning the shared wind velocity extracted from the MCMC fitting procedure, we found a net blue-shift with respect to the expected wavelength position of the \ion{Na}{i} lines of $\sim$ 0.04 \AA \, in both the CLV+RM corrected and uncorrected spectra, which correspond to $\sim$ 2 km$\,$s$^{-1}$. This value is significantly smaller than the value of 8 $\pm$ 2 km$\,$s$^{-1}$ found by W2015, but consistent with the wind speed detected by \citet{Brogi_2016} (-1.7$_{-1.1}^{+1.2}$ km$\,$s$^{-1}$) using data from CRIRES (IR) and \citet{Louden-Wheatley} (-1.9$_{-0.7}^{+0.6}$ km$\,$s$^{-1}$) using the third HARPS night analyzed here. While the inclusion of the CLV+RM correction has been advocated to explain the difference between W2015 and the following works, it is not entirely clear why we do not recover such difference when switching off the correction. It is possible that the origin of the discrepancy actually resides in another step or assumption of the analysis, which we cannot verify as the results have been obtained with proprietary code. This reproducibility issue highlights again the importance of making the code publicly available.\\ 
\indent The effective radius obtained when fitting the CLV and RM model is 0.99 $\pm$ 0.04 $R_p$. On the other hand, those derived from the absorption value, assuming a continuum level of $(R_p/R_s)^2$ = 2.262 \%, are 1.10 $\pm$ 0.02 $R_p$ for D$_2$ and 1.09 $\pm$ 0.02 $R_p$ for D$_1$.\\
\indent As for the value of $K_p$, the MCMC analysis could only find an upper limit, which is about half of the theoretical value ($\sim$ 150 km s$^{-1}$). We have verified that setting a Gaussian prior of 150 $\pm$ 10 km$\,$s$^{-1}$ leads to a $K_p$ value of 121 $\pm$ 10 km$\,$s$^{-1}$ without the CLV+RM correction and 137 $\pm$ 10 km$\,$s$^{-1}$ with the correction (the MCMC correlation diagrams are shown in Figure \ref{fig:corner_plotHD189}). However, while finding a value closer to the theoretical one, the use of the prior results in the MCMC analysis fitting the D$_2$ line at a lower contrast: 0.44$_{-0.6}^{+0.7}$ \% in the transmission spectrum not corrected for the CLV and RM effects and 0.35 $\pm$ 0.07 \% after the correction; both values are not compatible with those found by W2015 and CB2017. We consider unlikely that this is due to a bug in the code, since for all other targets analyzed we get a K$_p$ value consistent with the theoretical one. Moreover, imposing a prior in the analysis of other targets does not produce discrepancies with the results of the literature. The discrepancy between the retrieved and theoretical $K_p$ value could be of astrophysical origin. For example, the profile of the line may not be symmetric, thus creating a false RV signal, or another atmospheric circulation signature, probing a different region of the atmosphere, could generate the deviation from the theoretical value (e.g., H$_\alpha$ and \ion{Fe} on KELT-9 b, \citealt{Pino_2020}).\\
\indent We extracted the relative absorption depths (ADs) using both approaches described in section \ref{sec:extraction_abs_depth} at three different central bandwidths: 0.75 \AA, 1.50 \AA \, and 3.00 \AA. To optimize the comparison, as well as CB2017, we choose the same reference passbands used by W2015: [5874.89 - 5886.89] Å for the blue and [5898.89 - 5910.89] Å for the red ward.
As it can be inferred from the results reported in Table \ref{tab:absorption_depths_HD189}, the contribution of the CLV and RM effects is significant, as if these effects are not considered, the ADs are overestimated. 
The values extracted from the transmission spectrum (TS) and from the transmission light curves (TLC) are compatible inside the error bars (which are larger in the second case). However, this is not always true when considering the wider passband; more specifically, higher ADs are extracted from the TLC where the CLV and RM effects are more evident for this target (see Fig. \ref{fig:tlc_HD189733}). CB2017 found discrepancies between the two approaches in all three passbands while W2015 reports the compatibility of the two methods, even if a discrepancy is present in the 1.5 and 3.0 \AA \, passbands when combine all three nights.\\
\indent The AD for the \ion{Na}{i} doublet is best detected with the TS approach in the smallest passband (0.75 Å), as the narrower, thinner sodium lines fill the passband better, at a level of 0.28 $\pm$ 0.02 \% ($\sim$ 11$\sigma$). In the same passband, with the TLC approach, we found a value of 0.32 $\pm$ 0.07 \%, which, even if less significant (4.8$\sigma$), it is compatible to that found with the first approach, as written above.\\
\indent Our ADs are in agreement with those obtained by W2015 and CB2017 using both approaches although, most of the times, CB2017 found higher values for the ADs with both approaches. The values of the ADs extracted from the TLC reported by CB2017 (see their table 5) refer to the combining of only two nights, namely the first and the third. We have verified that if we also combine only these two nights, we find: 0.32 $\pm$ 0.09 in 0.75 \AA, 0.19 $\pm$ 0.05 in 1.50 \AA \, and 0.10 $\pm$ 0.03 in 3.00 \AA. Thus, the only incompatibility with CB2017 is in the largest passband, where they find a higher value, but which is in agreement with the value extracted by the TS. Again, the discrepancy is likely due to the differences between the methods applied for the telluric correction.

\begin{figure*}
 \centering
    \includegraphics[width=\textwidth, keepaspectratio]{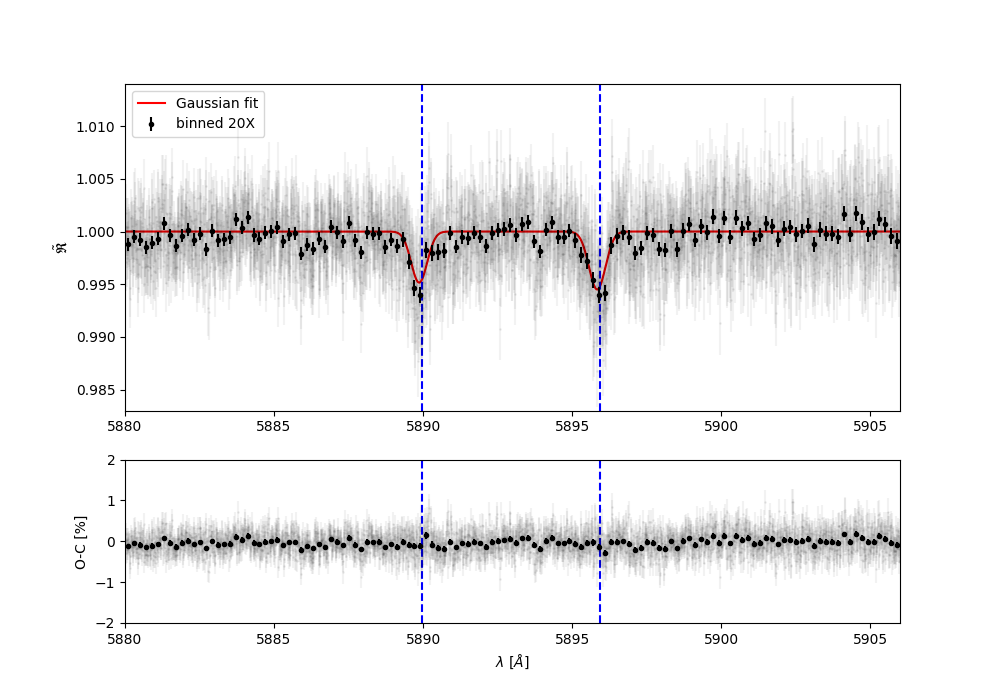}
    \caption{\label{fig:WASP-76_transmission_spectrum_HARPS} Final transmission spectrum of WASP-76 b centered around the sodium doublet in the planetary rest frame. \textit{Upper panel}: full-resolution data in light gray, and binned by 20x in black circles. The MCMC Gaussian fit is shown in red, and the rest frame transition wavelengths of the sodium doublet are indicated with blue dashed lines. \textit{Lower panel}: residuals of the Gaussian fit.}
\end{figure*}
\begin{table*}
 \caption{\label{tab:WASP-76} Summary of the best-fit parameters and 1-$\sigma$ error bars obtained with the MCMC fitting procedure for WASP-76 b.}
 \centering
 \begin{spacing}{1.4}
    \begin{tabular}{c|c|c|c|c|c|c|c}
    \hline
    \hline
    \multicolumn{8}{c}{WASP-76 b}\\
    \hline
     & \multicolumn{2}{c|}{\ion{Na}{i} D$_2$}  & \multicolumn{2}{c|}{\ion{Na}{i} D$_1$}& \multicolumn{3}{c}{\ion{Na}{i} D$_{12}$}\\
    &  $h$ & FWHM  & $h$ & FWHM & $v_{wind}$  & $r$ & $K_p$  \\
    &  [\%] & [km/s] &  [\%] & [km/s] & [km/s] & [$R_p$] & [km/s] \\
    \hline
   2012-11-11 & -0.75$_{-0.23}^{+0.35}$ & 22.7$_{-11.4}^{+15.1}$ & -0.65$_{-0.17}^{+0.18}$ & 29.5$_{-8.3}^{+11.1}$ & -3.5$_{-2.3}^{+2.3}$ & - & 148.8$_{-15.4}^{+15.1}$ \\
   2017-10-24 & -0.56$_{-0.15}^{+0.15}$ & 26.0$_{-6.6}^{+10.5}$ & -0.37$_{-0.10}^{+0.11}$ & 41.7$_{-10.4}^{+14.3}$ & -6.1$_{-3.1}^{+2.8}$ & - & 214.8$_{-20.9}^{+19.3}$ \\
   2017-11-22 & -0.31$_{-0.04}^{+0.04}$ & 93.4$_{-9.7}^{+4.8}$ & -0.61$_{-0.10}^{+0.10}$ & 34.5$_{-6.7}^{+8.8}$ & -2.4$_{-2.4}^{+2.3}$ & - & 206.5$_{-22.5}^{+19.1}$ \\
   all nights & -0.43$_{-0.09}^{+0.10}$ & 39.7$_{-11.8}^{+17.4}$ & -0.56$_{-0.06}^{+0.07}$ & 34.7$_{-4.6}^{+5.3}$ & -3.7$_{-1.7}^{+1.8}$ & - & 196.6$_{-12.3}^{+11.9}$ \\
   \hline
    \multicolumn{8}{c}{after CLV+RM correction}\\
    \hline
   2012-11-11 & -0.77$_{-0.20}^{+0.24}$ & 29.0$_{-11.1}^{+11.9}$ & -0.70$_{-0.17}^{+0.19}$ & 27.4$_{-7.5}^{+9.7}$ & -3.9$_{-2.2}^{+2.4}$ & 0.96$_{-0.05}^{+0.05}$ & 142.3$_{-14.1}^{+16.0}$ \\   
   2017-10-24 & -0.59$_{-0.13}^{+0.14}$ & 40.1$_{-11.1}^{+15.2}$ & -0.44$_{-0.10}^{+0.11}$ & 52.8$_{-11.8}^{+19.0}$ & -8.1$_{-3.3}^{+3.0}$ & 0.96$_{-0.05}^{+0.05}$ & 207.0$_{-25.4}^{+22.3}$ \\
   2017-11-22 & -0.38$_{-0.13}^{+0.16}$ & 27.4$_{-11.4}^{+27.2}$ & -0.56$_{-0.10}^{+0.10}$ & 28.1$_{-5.3}^{+6.5}$ & -2.9$_{-2.1}^{+2.3}$ & 0.97$_{-0.05}^{+0.05}$ & 205.3$_{-17.6}^{+14.5}$ \\
   all nights & -0.48$_{-0.08}^{+0.09}$ & 32.2$_{-7.1}^{+10.3}$ & -0.55$_{-0.06}^{+0.07}$ & 33.3$_{-4.1}^{+4.7}$ & -4.2$_{-1.6}^{+1.6}$ & 0.90$_{-0.04}^{+0.05}$ & 190.6$_{-11.7}^{+11.4}$ \\
    \hline
    \hline
    \end{tabular}
    \end{spacing}
\end{table*} 
\subsection{WASP-76 b}
WASP-76 b is a hot gas giant planet of approximately one Jupiter mass, but roughly twice its radius. It orbits a F7 star with a visual apparent magnitude of V = 9.5 every $\sim 1.8$ days. The scale height of its atmosphere is estimated to be about 1212 km, making it an excellent system for the analysis by transmission spectroscopy \citep{Kabath_2019}; this target is indeed one of the most studied target in the field of exoplanetary atmospheres \citep{Seidel_2019, Seidel_2020, Edwards_2020, Tabernero_2021}.\\
\indent For this target, we analyzed two transits as part of the HEARTS survey (ESO programme: 100.C-0750; PI: Ehrenreich) and combined them with another transit (ESO programme: 090.C-0540; PI: Triaud), all of them were observed with the HARPS spectrograph. 
The exposure times of the observations varied from 300 to 600 s depending on seeing conditions on the respective nights. In the third night, all data after planet transit were excluded from the analysis due to clouds, so the master-out of the specified night was created using just the spectra taken before the transit. All parameters used for this target are taken from \citet{West_2016}.
The results found from these observations have been compared with the ones obtained by \citet{Seidel_2019}, hereafter S2019.\\
\indent All transits were observed with one fiber on the target (fiber A) and one fiber on the sky (fiber B). This allowed us to apply the sky correction to all the observations.\\
\indent S2019 did not apply the correction of the CLV and RM effects as both effects should be negligible for this target. Indeed, for earlier-type stars like WASP-76, CLV effect is less pronounced \citep{Kostogryz-Berdyugina, Czesla_2015, Yan_2017} and the amplitude of RM effect is too small ($\sim$ 2 ms) to imprint a change in the line shape on the transmission spectrum beyond the noise level.
This is due to the fact that WASP-76 is a slow rotator ($v\sin{i}$ = 3.3 $\pm$ 0.6 km/s) and has its planet in a polar orbit \citep{Brown_2017}; as a conclusion, the planet always masks an area of the star with almost zero velocity during the whole transit. Nevertheless, we decided to also apply the correction for CLV and RM effect in order to quantify their contribution on this target.\\
\begin{figure*}[ht]
    \centering
    \includegraphics[width=\textwidth]{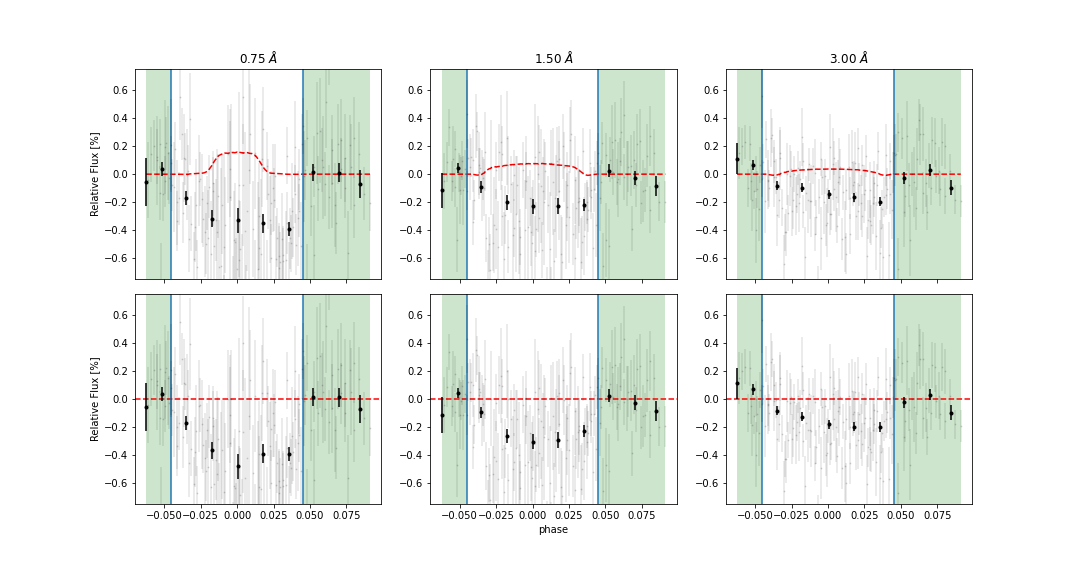}
    \caption{Same as Fig. \ref{fig:tlc_HD189733} but for WASP-76 b.}
    \label{fig:tlc_WASP76}
\end{figure*}
\indent Figure \ref{fig:WASP-76_transmission_spectrum_HARPS} shows the final transmission spectrum centered around the sodium doublet in the PRF with the CLV and RM correction.
The \ion{Na}{i} D absorption lines from the planetary atmosphere can be clearly seen. The results obtained from the MCMC analysis are listed in Table \ref{tab:WASP-76}, while the corresponding corner plot is shown in Fig. \ref{fig:corner_plotWASP76}. Taking into account the uncertainties, both the contrasts and the FWHMs are perfectly compatible with those found by S2019. When applying the CLV and RM correction, the values found are not very different from the previous ones, differing by only 0.02 - 0.05 \% for contrasts and 0.05 - 0.20 \% for FWHMs. In any case, they are compatible with each other.\\
\indent Concerning the shared wind velocity, we found a net blue-shift of $\sim$ -0.07 \AA \, which corresponds to $\sim$ -4 km s$^{-1}$. S2019 did not reported any shift.\\
\indent The planetary radius is a free parameter only when CLV and RM correction is applied. In this case, we found an effective radius of 0.90 $R_p$, while considering the absorption value and assuming a continuum level of 1.172 \%, we found 1.19 $\pm$ 0.03 $R_p$ for D$_2$ and 1.21 $\pm$ 0.02 for D$_1$. The fact that the best-fit value of $r$ is slightly smaller than 1 could be due to the intrinsic error of the stellar models (e.g., the abundances of the lines). However, the importance of the model fitting, here, is to reproduce the CLV and RM effects as accurately as possible, even if the best-fit $r$ value is not compatible with what we expected.\\
\indent The $K_p$ value found by the MCMC fitting procedure is 190.6$^{+11.4}_{-11.7}$ km s$^{-1}$. It is in agreement with the theoretical one ($\sim$ 196 km s$^{-1}$).\\
\indent The relative absorption depth for the different passbands was computed by S2019 just applying equation \ref{eq:absorption_depth_1}. However, we decided to use the TLC approach as well.
For a better comparison, we chose their same reference bands, which are the same used for HD 189733 b. The results are listed in Table \ref{tab:absorption_depths_WASP76}.\\ 
\begin{table*}[ht]
\caption{\label{tab:absorption_depths_WASP76}Summary of the measured relative absorption depth in [\%] of atmospheric sodium on WASP-76 b extracted from the transmission spectrum (TS) and from the transmission light curve (TLC).}
 \centering
    \begin{tabular}{c| c c c| c c c}
    \hline
    \hline
     & \multicolumn{3}{c|}{TS} & \multicolumn{3}{c}{TLC}\\
     & 0.75 \AA & 1.50 \AA & 3.00 \AA & 0.75 \AA & 1.50 \AA & 3.00 \AA \\
     \hline
      2012-11-11 & 0.49 $\pm$ 0.06 & 0.30 $\pm$ 0.05 & 0.13 $\pm$ 0.03 & 0.54 $\pm$ 0.11 & 0.36 $\pm$ 0.08 & 0.23 $\pm$ 0.06\\
      2017-10-24 & 0.41 $\pm$ 0.05 & 0.23 $\pm$ 0.04 & 0.11 $\pm$ 0.03 & 0.29 $\pm$ 0.09 &  0.16 $\pm$ 0.06 & 0.08 $\pm$ 0.05\\
      2017-11-22 & 0.39 $\pm$ 0.04 & 0.27 $\pm$ 0.03 & 0.17 $\pm$ 0.02 & 0.33 $\pm$ 0.05 & 0.15 $\pm$ 0.04 & 0.15 $\pm$ 0.03\\
      all nights & 0.41 $\pm$ 0.03 & 0.26 $\pm$ 0.02 & 0.15 $\pm$ 0.01 & 0.39 $\pm$ 0.05 & 0.24 $\pm$ 0.04 & 0.16 $\pm$ 0.03\\
       \hline
      \multicolumn{7}{c}{after CLV+RM correction}\\
    \hline
      2012-11-11 & 0.51 $\pm$ 0.06 & 0.32 $\pm$ 0.05 & 0.13 $\pm$ 0.03 & 0.56 $\pm$ 0.11 & 0.39 $\pm$ 0.08 & 0.26 $\pm$ 0.06\\
      2017-10-24 & 0.44 $\pm$ 0.05 & 0.26 $\pm$ 0.04 & 0.12 $\pm$ 0.03 & 0.33 $\pm$ 0.09 &  0.21 $\pm$ 0.06 & 0.11 $\pm$ 0.05\\
      2017-11-22 & 0.41 $\pm$ 0.04 & 0.28 $\pm$ 0.03 & 0.17 $\pm$ 0.02 & 0.37 $\pm$ 0.05 &  0.21 $\pm$ 0.04 & 0.17 $\pm$ 0.03\\
      all nights & 0.43 $\pm$ 0.03 & 0.27 $\pm$ 0.02 & 0.15 $\pm$ 0.01 & 0.42 $\pm$ 0.05 &  0.28 $\pm$ 0.04 & 0.18 $\pm$ 0.03\\
    \hline
    \hline
    \end{tabular}
\end{table*}
\indent As expected, for this target, the CLV and RM contribution is not significant on the measurement of the ADs. The values extracted from the TS before and after their removal differ by only 0.04 - 0.05 \%, while the ones extracted from the TLC differ by 0.07 - 0.17 \%. In any case, they are compatible within the error bars, even if slightly deeper ADs are found when correcting the CLV and RM effects. This time, compared to HD 189733 b, the ADs found with the two methods are always compatible with each other, regardless the passband; in fact, as can be seen in Figure \ref{fig:tlc_WASP76}, the CLV+RM model does not make a significant contribution, especially in the larger passband.\\
\indent Also the ADs we find are in agreement with S2019 in all three passbands here considered and lead to a higher significance.
In particular, the AD is best detected in the 0.75 Å passband with an absorption depth of 0.41 $\pm$ 0.03 \% ($\sim$ 15$\sigma$).
In the same passband S2019 found their best detection at a level of 0.371 $\pm$ 0.034 \% (10.7 $\sigma$).

\subsection{WASP-127 b}
\begin{figure*}
 \centering
       \includegraphics[width=\textwidth]{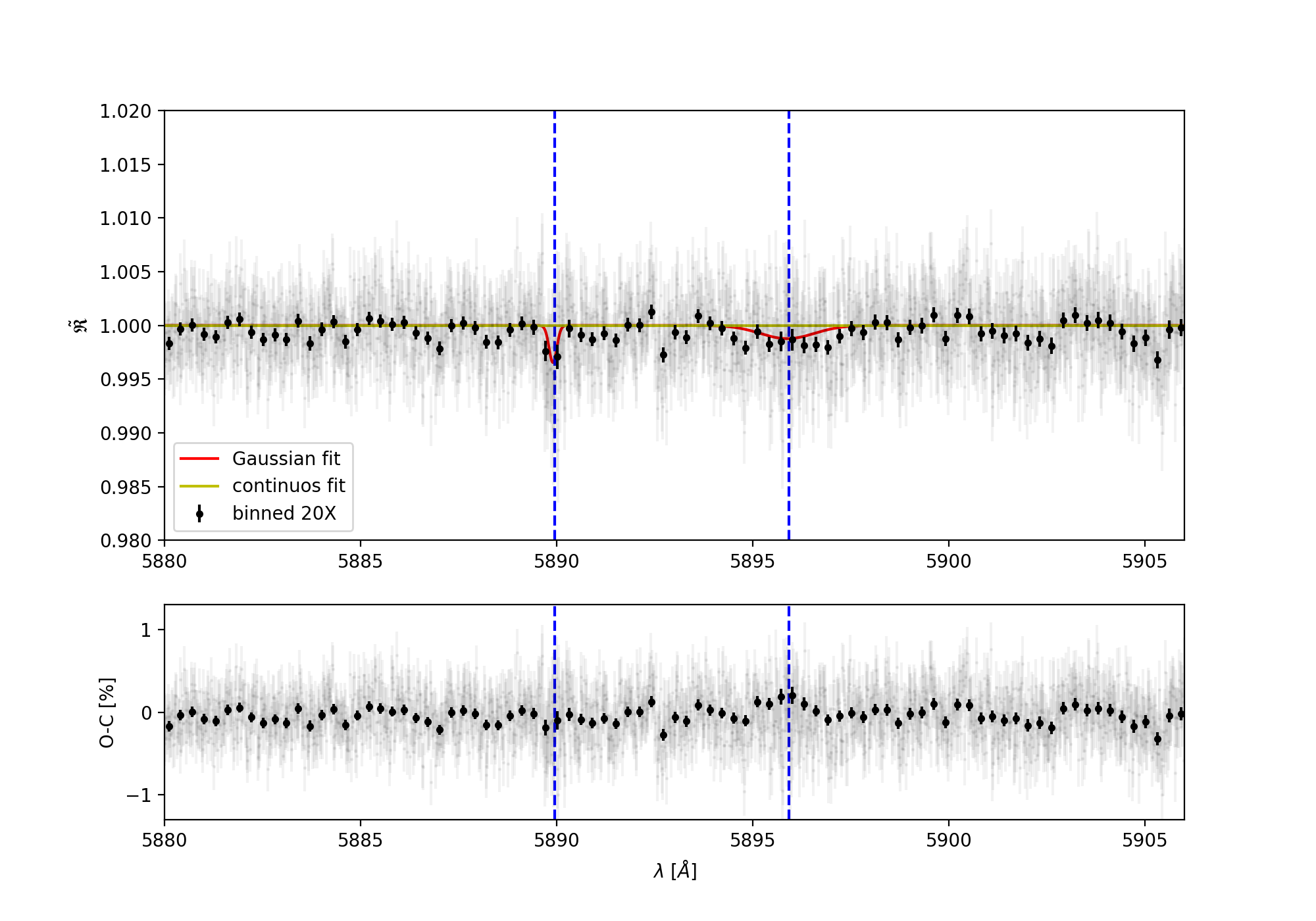}
     \caption{Same as Fig. \ref{fig:WASP-76_transmission_spectrum_HARPS}, but for WASP-127 b. The linear fit (yellow line) is strongly favored over the Gaussian fit (red line).}
     \label{fig:WASP-127_all}
\end{figure*}
\begin{table*}
 \caption{\label{tab:WASP-127} Summary of the best-fit parameters and 1-$\sigma$ error bars obtained with the MCMC fitting procedure for WASP-127 b.}
 \centering
 \begin{spacing}{1.4}
    \begin{tabular}{c|c|c|c|c|c|c}
    \hline
    \hline
    \multicolumn{7}{c}{WASP-127 b}\\
      \hline
     & \multicolumn{2}{c|}{\ion{Na}{i} D$_2$}  & \multicolumn{2}{c|}{\ion{Na}{i} D$_1$}& \multicolumn{2}{c}{\ion{Na}{i} D$_{12}$}\\
    &  $h$ & FWHM  & $h$ & FWHM & $v_{wind}$  & $K_p$  \\
    &  [\%] & [km/s] & [\%] & [km/s] & [km/s] & [km/s] \\
    \hline
   2017-02-27 & -0.41$_{-0.22}^{+0.39}$ & 26.1$_{-14.8}^{+38.0}$ & -0.29$_{-0.13}^{+0.24}$ & 40.0$_{-21.8}^{+33.3}$ & -3.5$_{-5.0}^{+6.3}$ &  137.5$_{-49.8}^{+43.4}$ \\
   2017-03-20 & -0.74$_{-0.62}^{+1.65}$ & 7.47$_{-5.28}^{+16.51}$ & -0.21$_{-0.16}^{+0.98}$ & 15.1$_{-14.2}^{+40.8}$  & -1.0$_{-11}^{+3.8}$ & 95.4$_{-35.4}^{+69.2}$\\
   2018-03-31 & -0.12$_{-0.08}^{+0.19}$ & 44.4$_{-39.6}^{+36.3}$ & -0.18$_{-0.07}^{+0.08}$ & 81.9$_{-39.1}^{+13.1}$ & +17$_{-20}^{+6.1}$ & 122.4$_{-51.1}^{+55.7}$\\
   all nights & -0.35$_{-0.25}^{+0.52}$ & 11.9$_{-7.9}^{+20.1}$ & -0.12$_{-0.04}^{+0.04}$ & 82.3$_{-24.0}^{+12.7}$ & -2.5$_{-5.7}^{+6.5}$ & 124.5$_{-53.5}^{+48.6}$\\
   \hline
   \hline
    \end{tabular}
    \end{spacing}
\end{table*}
\begin{figure}
    \centering
    \includegraphics[width=0.5\textwidth]{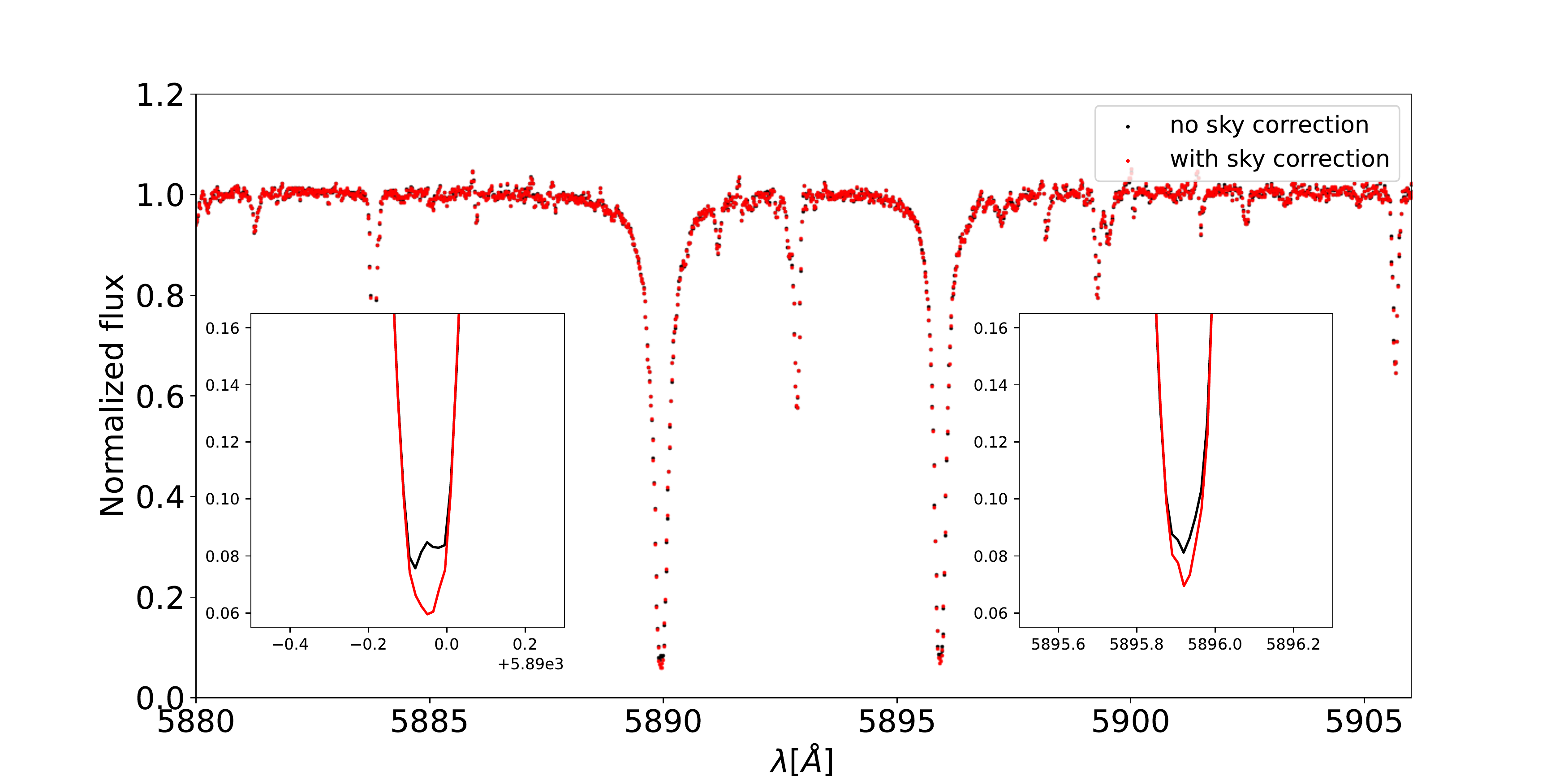}
    \caption{WASP-127 b master-out spectrum normalized to unity around the sodium doublet (night 2) before (\textit{black}) and after (\textit{red}) the sky correction. A zoom into the line center of the sodium D lines is shown in the insets.}
    \label{fig:master_out_wasp127}
\end{figure}
WASP-127 b is a heavily bloated gaseous exoplanet with one of the lowest densities discovered to date. With a sub-Saturn mass ($M_p$ = 0.18 $\pm$ 0.02 $M_J$) and super-Jupiter radius ($R_p$ = 1.37 $\pm$ 0.04 $R_J$), it orbits a bright G5 star (V=10.2) that is about to leave the main-sequence. Its scale height is 2500 $\pm$ 400 km \citep{Palle_2017}, making it another ideal target for transmission spectroscopy.\\ 
\indent For this target, we analyzed publicly available HARPS data, covering three transit events as part of the HEARTS survey (ESO programme: 098.C-0304, 100.C-0750; PI: Ehrenreich).
The first night has no observations after transit while the second and the third night have a very short post-transit phase. This is due to the visibility constraints of the target, especially its long transit duration (> 4 hours). With the first two nights, \citet{Zak} reported the detection of sodium in the atmosphere of WASP-127 b at a 4–8$\sigma$ level of significance, confirming earlier results based on low-resolution spectroscopy \citep{Palle_2017,Chen_2018}. However, \citet{Seidel_127}, hereafter S2020, shows that this sodium detection was actually due to contamination from telluric sodium emissions and the low S/N in the core of the deep stellar sodium lines.  The more recent work by \citet{Allart_2020} who analyzed two other transits of this target observed with ESPRESSO in the frame of the GTO Consortium, reported the detection of the sodium line core at a 9$\sigma$ confidence level.\\ 
\indent To rule out any false-positive detections due to stellar activity, \citet{Zak} monitored the \ion{Mg}{i} and \ion{Ca}{i} lines, finding no features (even if these measurements included spectra with large noise which could mask a signal and hide stellar activity). In addition, S2020 presented EulerCam light curves taken simultaneously to the spectroscopic data, showing no photometric variability.\\
\indent The RM effect is not measurable in WASP-127; indeed, it is a slow rotator  ($v\sin{i}$ = 0.3 $\pm$ 0.2 km/s) and its RV variation is estimated to be $\sim$ 2 m s$^{-1}$, too small to have any real impact on the depth of the \ion{Na}{i} D lines \citep{Nortmann_2018}. In this case, we decided to not apply the correction for the CLV and RM effects, as well as S2020.\\
\indent We set the configuration file with the same parameters used by S2020. We excluded from the analysis the last spectrum of the first night, which was only partially taken, and the last one of the third night because of the high airmass ($\sim 2.7$). Differently from S2020, we kept the spectra with the planetary signal overlapping the low-S/N core of the stellar lines. Besides, telluric residuals have been removed after the \lq transmission spectrum preparation\rq \,which computes the ratio between each spectrum and the out-of-transit master spectrum (see section \ref{sec:transmission_spectrum_preparation}).\\
\indent The final transmission spectrum for the three nights combined extracted with \texttt{SLOPpy} is shown in Figure \ref{fig:WASP-127_all} while the MCMC best-fit parameters are listed in Table \ref{tab:WASP-127}. In this case, the sodium features do not clearly peak out from the continuum.  
Furthermore, carrying out a model comparison between the Gaussian and a flat line using the Bayesian Information Criterion (BIC), we found a $\Delta$BIC < 0, indicating that the linear fit is favored compared to the Gaussian model. S2020 found the absorption feature only in the D$_2$ line at a level of 0.73 $\pm$ 0.45 \%, corresponding to 1.6$\sigma$. This value is compatible with what we found inside the error bars. \\
\begin{figure}
    \centering
    \includegraphics[ width=0.5\textwidth]{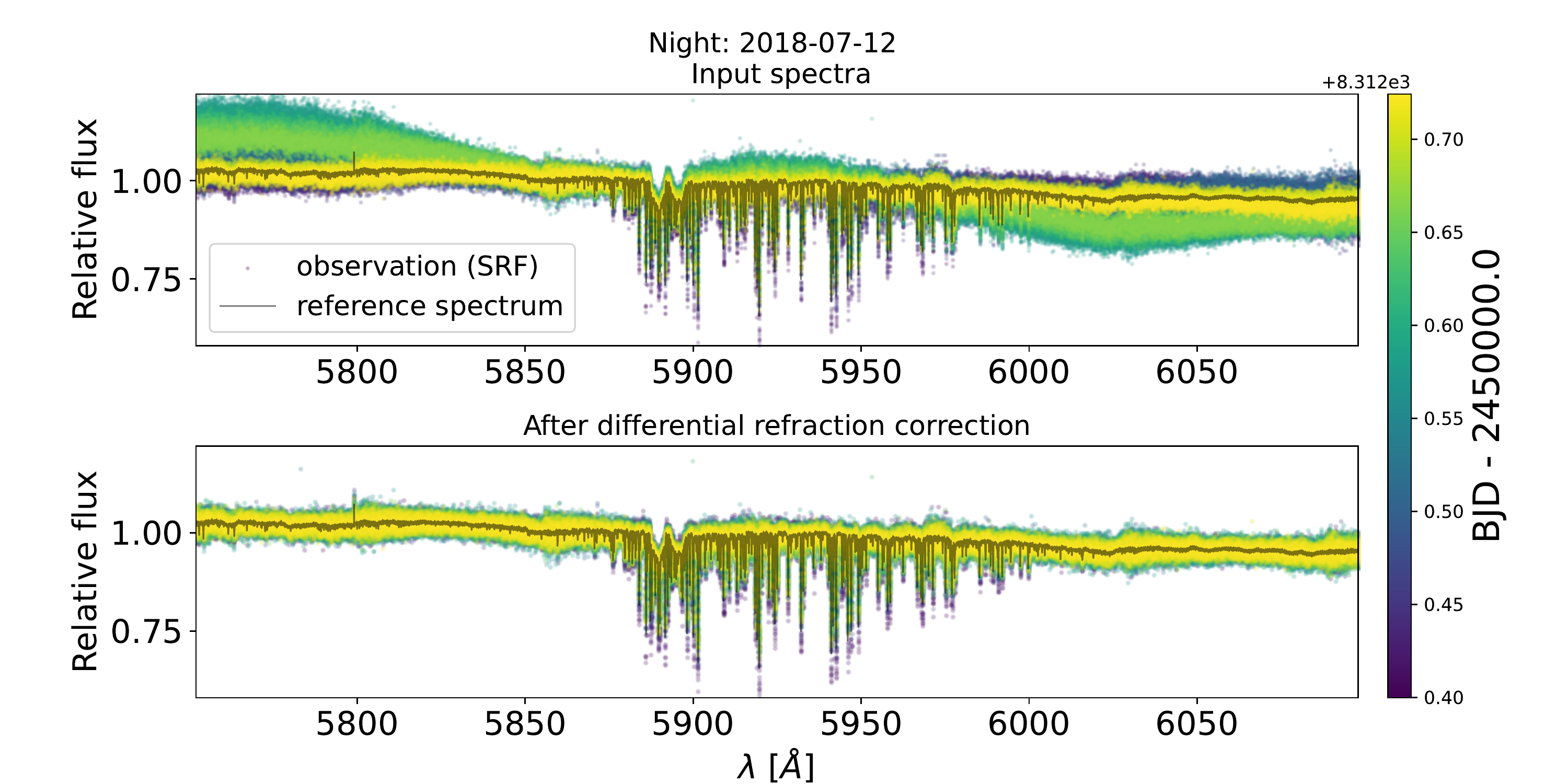}
    \caption{Differential refraction correction applied with \texttt{SLOPpy} to the second night of KELT-20 b.}
    \label{fig:differential_refraction_kelt20}
\end{figure}
\indent As noted by S2020, the second night, which is the one with the strongest detection in the  D$_2$ line, reveals an emission feature in the master-out sodium line core which stems from telluric sodium (see Fig. 4 in S2020). This emission creates an artificially deep residual when dividing by the master-out to extract the transmission spectrum. In order to correct this, while S2020 masked out the wavelength range of the emission feature in the SRF, we applied the sky correction implemented in \texttt{SLOPpy} (section \ref{sec:sky_correction}). Figure \ref{fig:master_out_wasp127} shows how the expected line shape of the stellar sodium doublet seems to be recovered when the sky correction is applied. Our approach has the additional advantage of not requiring any ad-hoc selection or removal of in-transit spectra. However, we remark that it could still be possible that some residual contamination is left.\\   
\indent In this case, we calculated the absorption depth from the TS using the smaller central passbands 0.375 \AA \, and 0.188 \AA, finding respectively 0.248 $\pm$ 0.082 \% and 0.421 $\pm$ 0.137 \% ($\sim 3\sigma$). The measurement by S2020 of 0.456 $\pm$ 0.198 \% is in a smaller passband than the 12 \AA \, originally reported in their paper (J. Seidel, priv. comm.).\\
\indent In conclusion, from this analysis, we confirm S2020's findings, which is that HARPS data cannot either confirm or confidently rule out the detection of sodium in WASP-127 b. 

\subsection{KELT-20 b}
\begin{figure*}
 \centering
    \includegraphics[width=\textwidth]{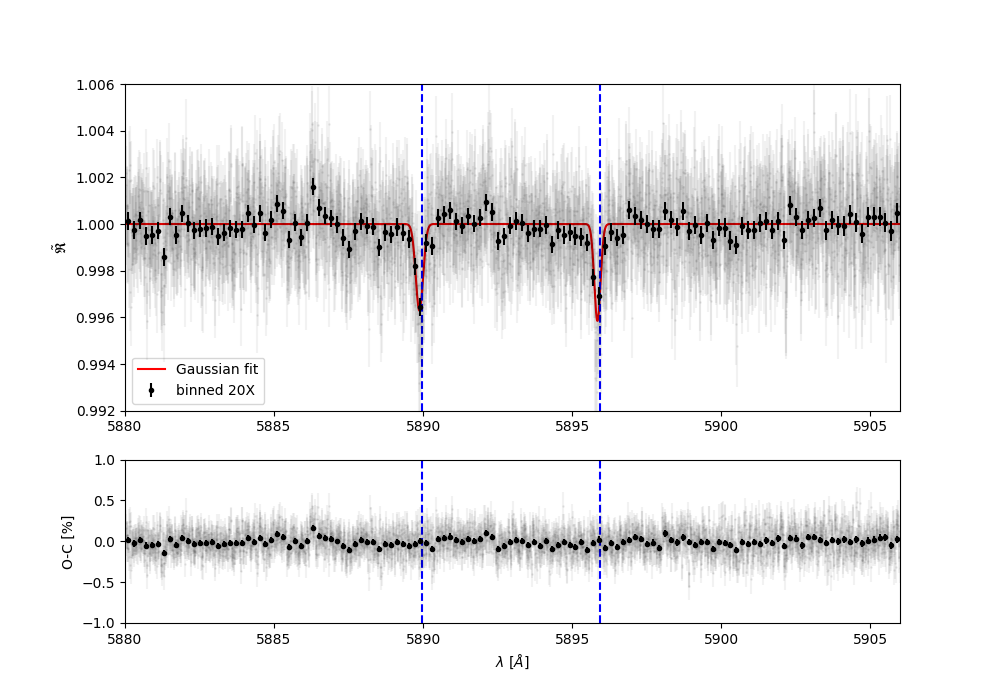}
    \caption{\label{fig:Kelt20_transmission_spectrum} Same as Fig. \ref{fig:WASP-76_transmission_spectrum_HARPS} but for KELT-20 b.}
\end{figure*}
KELT-20 b \citep{Lund_2017}, also named MASCARA-2 b \citep{Talens_2018}, is a hot Jupiter ($R_p$ = 1.83 $\pm$ 0.07 $R_J$, $M_p$ < 3.510 $M_J$) transiting a rapidly rotating ($v \sin{i} = 115.9 \pm 3.4 $ km$\, $s$^{-1}$) A-type star with an orbital period of $\sim$ 3.5 days. 
This planet receives strong irradiation from its host star ($T_{eff} \sim $ 9000 K), leading to a high equilibrium temperature of $\sim$ 2260 K which classify it as an ultra-hot Jupiter.\\
\indent For this target we analyzed three transits retrieved with HARPS-N (TNG archive, programs: CAT17A$\_$38 and CAT18A$\_$34), which have been analyzed by \citet{CB_2019} (hereafter CB2019) searching for \ion{Ca}{ii}, \ion{Fe}{ii}, \ion{Na}{i} and Balmer series of H. The first and second night were acquired with an exposure time of 200 s while in the third night 300 s were used, in order to obtain a higher S/N. As in CB2019, in the second night we discarded 8 consecutive out-of-transit spectra which presented a lower S/N due to clouds and one in-transit spectrum which also presented a similar low S/N. Furthermore, two of the three nights were affected by variations of the continuum caused by an error in the control software of the ADC of the telescope; CB2019 mainly removed this effect applying a very broad median filter, while we take advantage of the \lq differential refraction\rq \,module of \texttt{SLOPpy}, as explained in section \ref{sec:differential_refraction} and as shown in Figure \ref{fig:differential_refraction_kelt20}. \\
\begin{table*}
 \caption{\label{tab:Kelt20} Summary of the best-fit parameters and 1-$\sigma$ error bars obtained with the MCMC fitting procedure for KELT-20 b.}
 \centering
 \begin{spacing}{1.4}
    \begin{tabular}{c c|c|c|c|c|c|c}
    \hline
    \hline
    \multicolumn{8}{c}{KELT-20 b}\\
    \hline
    & & Contrast & FWHM  & $v_{wind}$  & $r$ &  $r (h)$  & $K_p$  \\
    & & [\%] & [km s$^{-1}$] & [km s$^{-1}$] & [$R_p$] & [$R_p$] & [km s$^{-1}$] \\
    \hline
     H$_\alpha$ & 2017-08-16 & -0.61$_{-0.20}^{+0.21}$ & 31.9$_{-9.9}^{+19.3}$ & +3.8$_{-4.2}^{+4.8}$ & 1.12$_{-0.15}^{+0.13}$ & 1.20$_{-0.06}^{+0.07}$ & 160.6$_{-42.3}^{+50.8}$\\
    & 2018-07-12 & -0.55$_{-0.11}^{+0.12}$ & 27.3$_{-6.3}^{+7.2}$ & -9.7$_{-2.7}^{+3.0}$ & 1.26$_{-0.07}^{+0.07}$ & 1.18$_{-0.04}^{+0.04}$ & 152.9$_{-34.0}^{+38.3}$\\
    & 2018-07-19 & -0.79$_{-0.14}^{+0.15}$ & 24.2$_{-4.0}^{+4.7}$ & -3.8$_{-2.8}^{+2.8}$ & 1.27$_{-0.10}^{+0.09}$ & 1.25$_{-0.05}^{+0.05}$ & 174.7$_{-38.3}^{+40.2}$\\
    & all nights & -0.61$_{-0.07}^{+0.08}$ & 27.9$_{-3.3}^{+3.7}$ & -5.2$_{-1.7}^{+1.8}$ & 1.24$_{-0.05}^{+0.05}$& 1.20$_{-0.03}^{+0.03}$ & 156.3$_{-27.0}^{+28.5}$\\
       \hline
    \ion{Na}{i} D$_2$ & 2017-08-16 & -0.66$_{-0.23}^{+0.25}$ & 4.53$_{-1.56}^{+2.81}$ & -3.8$_{-0.8}^{+0.7}$ & 0.92$_{-0.07}^{+0.07}$ & 1.22$_{-0.07}^{+0.08}$ & 172.8$_{-11.2}^{+17.7}$\\
    & 2018-07-12 & -0.32$_{-0.05}^{+0.06}$ & 26.9$_{-6.3}^{+6.7}$ & -1.4$_{-2.4}^{+2.6}$ & 1.05$_{-0.04}^{+0.04}$ & 1.11$_{-0.02}^{+0.02}$ & 202.8$_{-32.7}^{+28.6}$\\
    & 2018-07-19 & -0.32$_{-0.07}^{+0.08}$ & 18.5$_{-4.8}^{+5.5}$ & -5.6$_{-2.9}^{+2.6}$ & 0.99$_{-0.05}^{+0.05}$ & 1.11$_{-0.02}^{+0.03}$ & 149.0$_{-31.5}^{+36.1}$\\
    & all nights & -0.37$_{-0.05}^{+0.05}$ & 14.8$_{-2.4}^{+2.8}$ & -3.7$_{-0.9}^{+0.9}$ & 1.00$_{-0.03}^{+0.03}$ & 1.13$_{-0.02}^{+0.02}$ & 170.0$_{-13.5}^{+14.9}$\\
    \hline
    \ion{Na}{i} D$_1$ & 2017-08-16 & -0.39$_{-0.10}^{+0.12}$ & 19.3$_{-6.1}^{+8.2}$ & -2.9$_{-2.1}^{+2.6}$ & 0.98$_{-0.07}^{+0.07}$ & 1.13$_{-0.03}^{+0.04}$ & 186.2$_{-47.1}^{+33.8}$\\
    & 2018-07-12 & -0.41$_{-0.09}^{+0.10}$ & 12.6$_{-3.4}^{+5.2}$ & -5.4$_{-1.3}^{+1.2}$ & 0.95$_{-0.05}^{+0.05}$ & 1.14$_{-0.03}^{+0.04}$ & 190.6$_{-20.3}^{+17.4}$\\
    & 2018-07-19 & -0.44$_{-0.10}^{+0.10}$ & 10.4$_{-2.2}^{+3.5}$ & -1.9$_{-1.2}^{+1.1}$ & 0.89$_{-0.05}^{+0.05}$ & 1.15$_{-0.03}^{+0.04}$ & 176.3$_{-18.7}^{+17.7}$\\
    & all nights & -0.41$_{-0.06}^{+0.06}$ & 12.5$_{-2.3}^{+2.9}$ & -3.8$_{-0.7}^{+0.7}$ & 0.94$_{-0.03}^{+0.03}$ & 1.14$_{-0.02}^{+0.02}$ & 192.5$_{-13.5}^{+12.4}$\\
    \hline
    \hline
    \end{tabular}
    \tablefoot{For this target, we also listed $r(h)$, the effective radius derived from the absorption value ($h$), assuming a continuum level of $(R_p/R_s)^2$ = 1.382 \%.}
\end{spacing}
\end{table*} 
\indent Although sky spectra were gathered on fiber B, we decided to do the same as CB2019 and not apply the sky correction since no emission feature was present on any of the nights, and neglect the stellar reflex motion since that the planet is orbiting a fast rotator star. For the same reason, interstellar sodium lines, which are present in the spectra of KELT-20 b, are automatically corrected when dividing each spectrum by the master-out in the SRF, as explained in section \ref{sec:interstellar_lines}.
Finally, as telluric correction with \texttt{Molecfit} left residuals, we normalized the final transmission spectrum using a linear spline (see section \ref{sec:telluric_correction}).\\ 
\indent This target is affected by CLV and RM effects. Following as close as possible CB2019, we modeled the stellar spectra using LTE Kurucz ATLAS9 models, with solar abundance for the hydrogen lines and [Na/H] = 0.98.\\
\indent The final transmission spectrum for the sodium doublet is shown in Figure \ref{fig:Kelt20_transmission_spectrum}. In order to better compare our results (listed in Table \ref{tab:Kelt20}) with CB2019's, we decided to apply the MCMC fitting procedure to each sodium D lines separately this time, so that we have individual $K_p$, $r$ and $v_{wind}$ values. For this target, we also compared the results obtained for the H$_{\alpha}$ line, whose transmission spectrum is shown in Figure \ref{fig:Kelt20_Halpha}. The MCMC correlation diagrams are shown in Appendix \ref{Appendix_cornerplot} (Figs. \ref{fig:corner_plotKelt20_D2}, \ref{fig:corner_plotKelt20_D1} and \ref{fig:corner_plotKelt20_Ha}).\\
\begin{figure*}
 \centering
    \includegraphics[width=\textwidth]{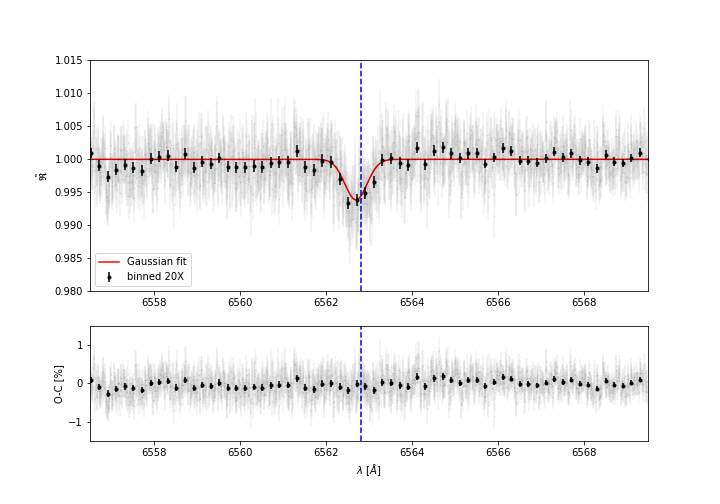}
    \caption{\label{fig:Kelt20_Halpha} Same as Fig. \ref{fig:Kelt20_transmission_spectrum} but for H$_\alpha$ line.}
\end{figure*}
\begin{figure*}
 \centering
 \includegraphics[width=\textwidth]{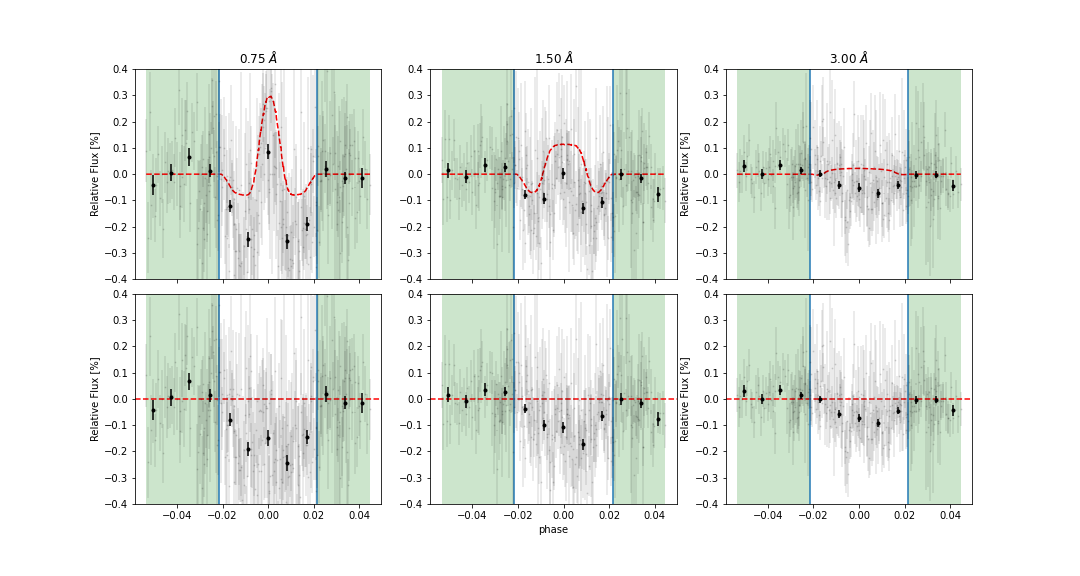}\\
 \includegraphics[width=\textwidth]{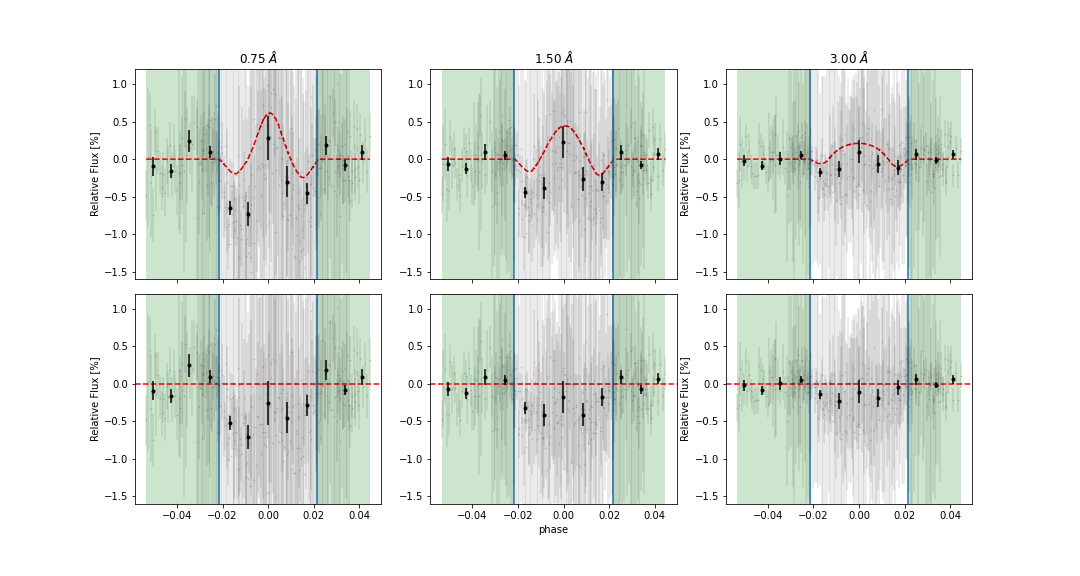}
     \caption{Same as Fig. \ref{fig:tlc_HD189733} but for KELT-20 b. \textit{Top}: \ion{Na}{i} doublet. \textit{(Bottom)}: H$_\alpha$ line.}
     \label{fig:tlc_Kelt20}
\end{figure*}
\indent Our results are compatible with the ones obtained by CB2019 most of the times. 
For both sodium lines and for all nights, the values of $K_p$, $r$ and $v_{wind}$ are always compatible within the error bars, while the contrasts and the FWHMs show some discrepancies. In particular, in the second and third night, CB2019 reports very high FWHMs and very low contrasts, unlike our results which instead show an almost constant trend. However, if we consider only the results relative to the combined nights, we are compatible with CB2019. Also in this case the $K_p$ value is compatible with that one theoretically predicted ($\sim$ 173 km$\, $s$^{-1}$). \\
\indent When comparing the best-fit parameters of H$_\alpha$ line, we find measurements in agreement with CB2019 except in the second night for the $v_{wind}$ value (we find an excessive blue-shift of the line) and the $r$ factor (our value is higher than CB2019's). The difference in the $r$ factor leads to incompatibility when combining all nights.  \\
\indent The corresponding TLCs for the \ion{Na}{i} doublet and for the H$_\alpha$ line are shown in Figure \ref{fig:tlc_Kelt20}; in the TLC corresponding at the 3 \AA \, passband, the CLV and RM effects are negligible, especially for the \ion{Na}{i} doublet.
Table \ref{tab:absorption_depths_kelt20} reports the ADs extracted from the TS and from the TLC, which are compatible with each other. 
As expected, for both the \ion{Na}{i} lines, the ADs are in agreement with CB2019 except in the second night where we found higher values with both methods. The ADs found when combining all three nights are compatible with CB2019 only if extracted from the TLC. Our best detection is at 0.75 \AA \, with a value of 0.18 $\pm$ 0.01 (corresponding to $\sim$ 18$\sigma$).
For the H$_\alpha$ line, whose reference passbands in the continuum have been taken at [6558.0 - 6561.0] \AA \, for the blue part and [6564.0 - 6567.0] \AA \, for the red part, we are perfectly compatible with CB2019, finding our best detection at 0.75 \AA \, with a value of 0.55 $\pm$ 0.05 (corresponding to $\sim$ 11$\sigma$).

\begin{table*}[ht]
\caption{\label{tab:absorption_depths_kelt20}Summary of the measured relative absorption depth in [\%] of atmospheric sodium on KELT-20 b extracted from the transmission spectrum (TS) and from the transmission light curve (TLC).}
 \centering
    \begin{tabular}{c c| c c| c c }
    \hline
    \hline
     & & \multicolumn{2}{c|}{TS} & \multicolumn{2}{c}{TLC}\\
     & & 0.75 \AA & 1.50 \AA & 0.75 \AA & 1.50 \AA \\
      \hline
     H$_\alpha$ & 2017-08-16 & 0.68 $\pm$ 0.11 & 0.47 $\pm$ 0.08 & 0.49 $\pm$ 0.58 & 0.36 $\pm$ 0.42 \\
      & 2018-07-12 & 0.48 $\pm$ 0.07 & 0.31 $\pm$ 0.05 & 0.42 $\pm$ 0.21 & 0.28 $\pm$ 0.15 \\
      & 2018-07-19 & 0.59 $\pm$ 0.09 & 0.28 $\pm$ 0.07 & 0.48 $\pm$ 0.33 & 0.23 $\pm$ 0.24 \\
      & all nights & 0.55 $\pm$ 0.05 & 0.34 $\pm$ 0.04 & 0.47 $\pm$ 0.25 & 0.30 $\pm$ 0.18\\
     \hline
     \ion{Na}{i} D$_2$ & 2017-08-16 & 0.06 $\pm$ 0.04 & +0.01 $\pm$ 0.03 & 0.04 $\pm$ 0.09 & 0.06 $\pm$ 0.07 \\
      & 2018-07-12 & 0.24 $\pm$ 0.03 & 0.16 $\pm$ 0.02 & 0.20 $\pm$ 0.05 & 0.14 $\pm$ 0.03\\
      & 2018-07-19 & 0.18 $\pm$ 0.03 & 0.09 $\pm$ 0.02 & 0.18 $\pm$ 0.05 & 0.11 $\pm$ 0.04 \\
      & all nights & 0.18 $\pm$ 0.02 & 0.11 $\pm$ 0.01 & 0.13 $\pm$ 0.04 & 0.06 $\pm$ 0.03\\
       \hline
     \ion{Na}{i} D$_1$ & 2017-08-16 & 0.18 $\pm$ 0.04 & 0.09 $\pm$ 0.03 & 0.12 $\pm$ 0.09 & 0.06 $\pm$ 0.07 \\
      & 2018-07-12 & 0.19 $\pm$ 0.03 & 0.13 $\pm$ 0.02 & 0.20 $\pm$ 0.05 & 0.12 $\pm$ 0.03\\
      & 2018-07-19 & 0.15 $\pm$ 0.03 & 0.10 $\pm$ 0.02 & 0.11 $\pm$ 0.05 & 0.08 $\pm$ 0.04 \\
      & all nights & 0.18 $\pm$ 0.02 & 0.11 $\pm$ 0.01 & 0.15 $\pm$ 0.04 & 0.08 $\pm$ 0.03\\
       \hline
      \ion{Na}{i} D$_{12}$ & 2017-08-16 & 0.12 $\pm$ 0.03 & 0.04 $\pm$ 0.02 & 0.08 $\pm$ 0.07 & 0.01 $\pm$ 0.05 \\
     & 2018-07-12 & 0.21 $\pm$ 0.02 & 0.14 $\pm$ 0.01 & 0.20 $\pm$ 0.03 & 0.13 $\pm$ 0.02 \\
     & 2018-07-19 & 0.17 $\pm$ 0.02 & 0.10 $\pm$ 0.02 & 0.14 $\pm$ 0.04 & 0.09 $\pm$ 0.03  \\
     & all nights & 0.18 $\pm$ 0.01 & 0.11 $\pm$ 0.01 & 0.14 $\pm$ 0.03 & 0.07 $\pm$ 0.02\\
    \hline
    \hline
    \end{tabular}
\end{table*}

\section{Summary and future perspectives}\label{sec:conclusions}
In this paper we present \texttt{SLOPpy} (Spectral Lines Of Planets with python), a standard, user-friendly tool which automatically extract and analyze the optical transmission spectrum of exoplanets.
The scientific aim of \texttt{SLOPpy} is the characterization of exoplanetary atmospheres in the visible through ground-based high-resolution transmission spectroscopy, one of the most robust techniques to constrain the composition of the atmosphere of a transiting planet. \\
\indent Extracting a transmission spectrum that is as reliable as possible is not an easy task. It requires a series of reduction steps, such as the removal of the sky emission or the correction of the differential refraction, that we have developed and implemented in \texttt{SLOPpy} specifically for this purpose. In addition of being detailed in this paper, all the details of the data reduction steps can be inspected and analyzed from the GitHub repository, where \texttt{SLOPpy} is available as an open-source code.\\
\indent One of the major difficulties in the extraction of the transmission spectrum from a ground-based facility is dealing with the telluric imprints from the Earth’s atmosphere, whose intensity changes continuously with time. Telluric correction is a very crucial step for the search of planetary signals, since that telluric features are very similar to the ones we look for on exoplanetary atmospheres. In \texttt{SLOPpy} three different approaches are implemented; however, we decided to apply telluric correction using the atmospheric transmission code \texttt{Molecfit} for all the targets analyzed in this work, because it is the most powerful method (see section \ref{sec:telluric_correction}).\\
\indent We also re-implemented the Center-to-Limb Variation (CLV) and the Rossiter-McLaughlin (RM) effect, which are the two main processes strongly affecting the transmission spectrum. For stronger absorption lines, such as the resonant \ion{Na}{i} doublet, their treatment become crucial in the detection of exoplanetary atmospheric species. \\ 
\indent The eventual absorption signals retrieved from the transmission spectrum can be interpreted as equivalent relative altitudes. The pipeline is already optimized for calculating them and to extract the relative absorption depths.\\
\indent To assess the validity of the pipeline on HARPS and HARPS-N data, we applied it to several datasets relative to four favorable targets for atmospheric characterization, namely HD 189733 b, WASP-76 b, WASP-127 b and KELT-20 b. In particular, we focused on the detection of the sodium doublet, which is one of the most prominent atmospheric signature thanks to its large absorption cross section. Comparing our results with those obtained by other research groups who independently analyzed the same data, we found that our results are compatible most of the time with other works within 1$\sigma$, and that with \texttt{SLOPpy} we get a similar or higher significance. \\
\indent The only target on which the pipeline obtains contrasting results with independently measured physical parameters, (e.g., the RV semi-amplitude of the planet) is HD 189733 b. Being the only target among those analyzed here to show this kind of incompatibility, we are confident that our overall methodology is correct, although we cannot exclude that an unidentified error in the input parameters or data processing are negatively affecting our results exclusively in this case, despite our best efforts. Rather, differences with literature results are likely due to the different telluric correction and/or stellar models used to correct the CLV and RM effects. This fact stress the necessity of a public and standard tool, like \texttt{SLOPpy}; the publication of a code is the first (and perhaps only) step for an open and independent comparison that allows to highlight possible errors in the analyses or in the models, despite an accurate and careful analysis. \\
\indent Since our code is written in separate computing and plotting modules, the user can check the effects of each individual step on the final transmission spectrum and to see how the results change if a particular correction is not applied. Thus, \texttt{SLOPpy} can be used as a tool to understand and analyze the scientific effects of each data reduction step on the observed data. In this work, for example, we show how the correction for the sky emission in WASP-127 b is crucial in the extraction of the master-out.
Besides, we developed this code trying to be as general and user-friendly as possible: indeed, all the parameters that can ultimately affect the final transmission spectrum are recorded and stored in the configuration file, which is compiled by the user according to his requirements.\\
\indent Thanks to the modularity of SLOPpy, improvements and developments can be easily implemented at any time. For example, at the moment of writing the pipeline supports HARPS and HARPS-N, but there are many other high-resolution facilities such as PEPSI on the LBT \citep{Strassmeier_2015}, CARMENES on the Calar Alto Observatory \citep{CARMENES}, and ESPRESSO on the VLT \citep{ESPRESSO} that are increasingly used in the study of exoplanetary atmospheres. While in this work we focused on the Na doublet and H$_{\alpha}$, \texttt{SLOPpy} can perfectly work over the entire visible range. In principle, the pipeline can be easily adapted to any spectral range covered by an instrument, as long as the wavelength solution during a night is sufficiently stable to apply the division by the master-out without problems, as for example CRIRES+ \citep{CRIRES+}.\\
\indent Other analysis techniques are often applied for identifying atomic and molecular species in the atmospheres of both transiting and nontransiting exoplanets, such as the Doppler tomography \citep{Watson_2019} and the cross correlation technique \citep{Pino_ccf}. Although our initial focus in SLOPpy development was on single lines and simple line systems, like Na doublet and Mg triplet, new approaches such as those mentioned above are under development and they will be made available through the public repository.\\
\indent In the long term, we plan to extend the same philosophy behind \texttt{SLOPpy} - flexibility, portability, reproducibility - to other tools dedicated to the comparison of the transmission spectrum with model atmospheres (e.g., \citealt{Guillot_2010, Kawashima-Ikoma}). In this way, beyond detecting atomic and molecular species from the transmission spectrum, \texttt{SLOPpy} and its successors could also characterize physical conditions in exoplanetary atmospheres.

\begin{acknowledgements} 
This work is based on observations made with the ESO Telescopes at the La Silla Paranal Observatory and with the Italian Telescopio Nazionale Galileo (TNG) operated on the island of La Palma by the Fundación Galileo Galilei of the INAF (Istituto Nazionale di Astrofisica) at the Spanish Observatorio del Roque de los Muchachos of the Instituto de Astrofísica de Canarias. Part of this work has been carried out within the frame of the PhD Programme at the University of Padova. 
We thank the anonymous referee for their thoughtful comments which helped to improve the quality of this manuscript.  DS acknowledges the funding support from Italian Space Agency (ASI) regulated by ‘Accordo ASI-INAF n. 2019-29-HH.0 del 26 novembre 2019'. We thank A. Wyttenbach, and J. V. Seidel for discussions and insights.
\end{acknowledgements}

\bibliographystyle{aa}
\bibliography{sample} 

\begin{appendix}

\section{Other techniques implemented in \texttt{SLOPpy} for telluric correction}\label{Appendix_telluric_correction}

\subsection{Empirical approach for spectral time series}\label{appendix_empirical}
\begin{figure}[h]
    \centering
    \includegraphics[width=0.5\textwidth, keepaspectratio]{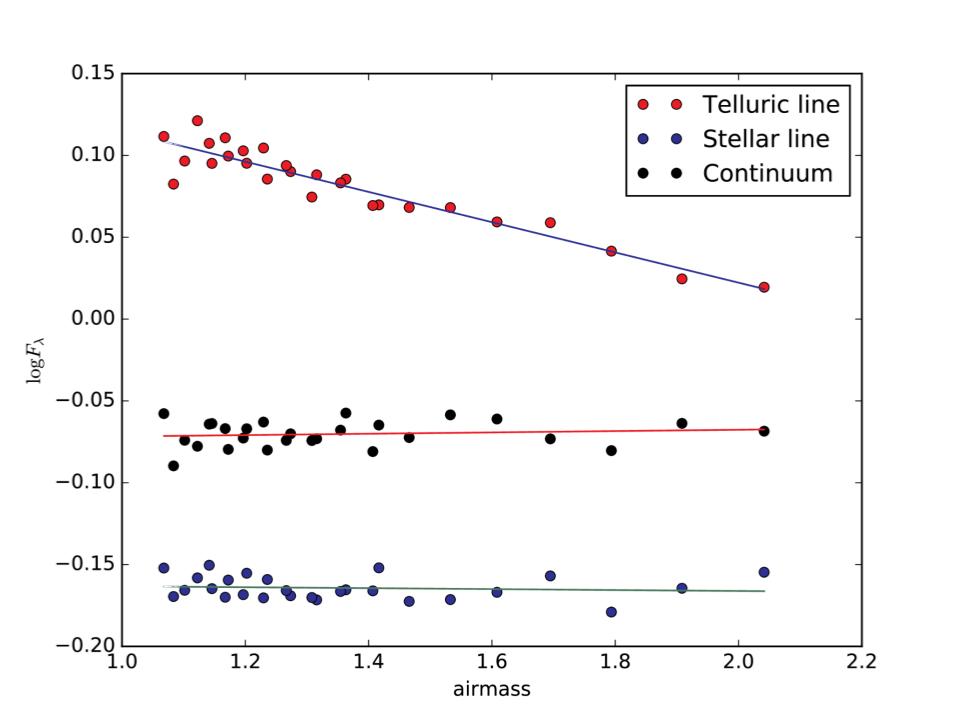}
    \caption{Logarithm of flux versus airmass for a telluric line (at 5883.91 Å in the observer reference frame), a stellar line (5883.66 Å) and the continuum (at 5903.02 Å). The straight lines are the best fits to each line.}
    \label{fig:intensity_telluric_airmass}
\end{figure}

This method was first described in \citet{Vidal-Madjar2010} and \citet{Astudillo-Defru}. It is entirely empirical, since the telluric spectrum is extracted from the data themselves, in particular from the out-of-transit observations. The technique exploits the fact that the intensity of telluric lines is a linear function of the airmass, in a logarithmic scale, while the intensity of stellar features and the continuum are essentially the same during a night (Fig. \ref{fig:intensity_telluric_airmass}); this is a consequence of the usual hypothesis of radiative transfer in a plane-parallel atmosphere. 

Denoting with $I(\lambda)$ the line intensity (i.e., the contrast between the continuum and the core of a telluric line) in the spectrum of the star before entering the Earth atmosphere and with $T(\lambda)$ the atmospheric transmission function \citep{Vidal-Madjar2010}, for a given airmass the observed spectrum $O_a(\lambda)$ is given by:
\begin{equation}\label{eq:observed_product}
    O_a(\lambda) = I(\lambda) \times T_a(\lambda).
\end{equation}
For airmass smaller than $\sim$ 5, we can assume a plane parallel atmosphere. In this case, the airmass is $\sim \sec\theta$, with $\theta$ the zenithal angle, and the atmospheric transmission function is then simply $T^{\sec\theta}$. 

From the radiative transfer equation, $T(\lambda)$ at the zenith ($\sec\theta$ = 1) can be expressed as a function of the number of particles per surface unit $N$ and  the opacity at a given wavelength $k_\lambda$ \citep{Wyttenbach_2015}:
\begin{equation} \label{exp}
T(\lambda)=T^1(\lambda)\equiv e^{Nk_\lambda} 
\end{equation}
So, the telluric spectrum at a given airmass is given by
\begin{equation}\label{eq:transmission_airmass}
T^a=\exp(Nk_\lambda a).
\end{equation}
By including equation \ref{eq:transmission_airmass} into equation \ref{eq:observed_product} we obtain:
\begin{equation}
I_\lambda=I_{0,\lambda}e^{Nk_\lambda a}=I_{0,\lambda}e^{\tau_{\lambda,0}\sec\theta},
\end{equation}
which can be rewritten as 
\begin{equation}\label{best-fit}
\ln I_\lambda=\ln I_{0,\lambda}+ \tau_{\lambda,0}\sec\theta.
\end{equation}
\begin{figure}
    \centering
    \includegraphics[width=0.5\textwidth, keepaspectratio]{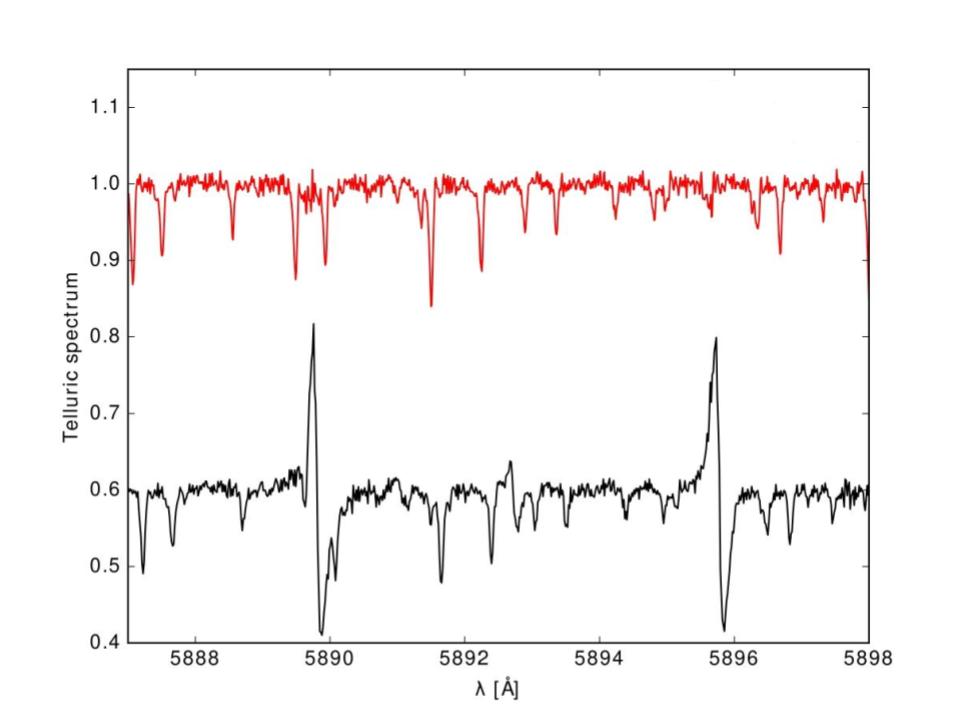}
    \caption{Telluric spectrum around the \ion{Na}{i} D doublet. The red one is obtained using spectra referred to the BRF, the black one using spectra referred to the ORF. The latter spectra is shifted vertically for clarity.} \label{fig:s1d_vs_e2ds}
\end{figure}
There are two unknowns in this equation, $I_{0,\lambda}$ and $\tau_{\lambda,0}$, neither of the two can be determined with a single observation of the received intensity $I_\lambda$. However, as time passes and as Earth rotates on its axis, the angle $\theta$ will change while $\tau_{\lambda,0}$ and $\ln I_{0,\lambda}$ will stay the same, thus allowing us to solve the system. Plotting $\ln {I_\lambda}$ as a function of $\sec\theta$, the slope of the best-fitting straight line is $\tau_{\lambda,0}=Nk_\lambda$. Extrapolating the best-fitting line to $\sec\theta=0$ provides the value of $I_{0,\lambda}$. Knowing $\tau_{\lambda,0}$, we can build the reference telluric spectrum $T(\lambda)$.

Finally, for the correction of the telluric contamination, each observed spectrum must be divided by the telluric reference spectrum, given by equation \ref{exp}, to the power $a_i-a_{ref}$  where $a_{ref}$ is the average airmass of in-transit spectra:
\begin{equation}\label{reference spectrum}
F_{ref}(\lambda)=\frac{O_a(\lambda)}{T(\lambda)^{a_i-a_{ref}}} = \frac{O_a(\lambda)}{e^{Nk_\lambda(a_i-a_{ref})}}.
\end{equation}
In this way, all spectra are rescaled as if they had been observed at same airmass, with the telluric absoprtion lines still present but now stationary in time. To completely remove them, a second telluric correction can be applied by linearly fitting the final transmission spectrum and $T(\lambda)$.    However, unlike other authors who used the same technique \citep{Wyttenbach_2015, Wyttenbach_2017}, 
sometimes with slight modifications \citep{Yan_2017}, the equation implemented in \texttt{SLOPpy} is not exactly the same as \ref{reference spectrum}. In \texttt{SLOPpy}, the telluric reference spectrum is raised to the exact airmass of the observed spectrum, considering the average airmass $a_{ref}$ equal to zero, in other words, \texttt{SLOPpy} completely removes the telluric absorption lines from the observed spectra without running the second correction (which is, however, implemented). 

\begin{figure}
    \centering
    \includegraphics[width=0.5\textwidth, keepaspectratio]{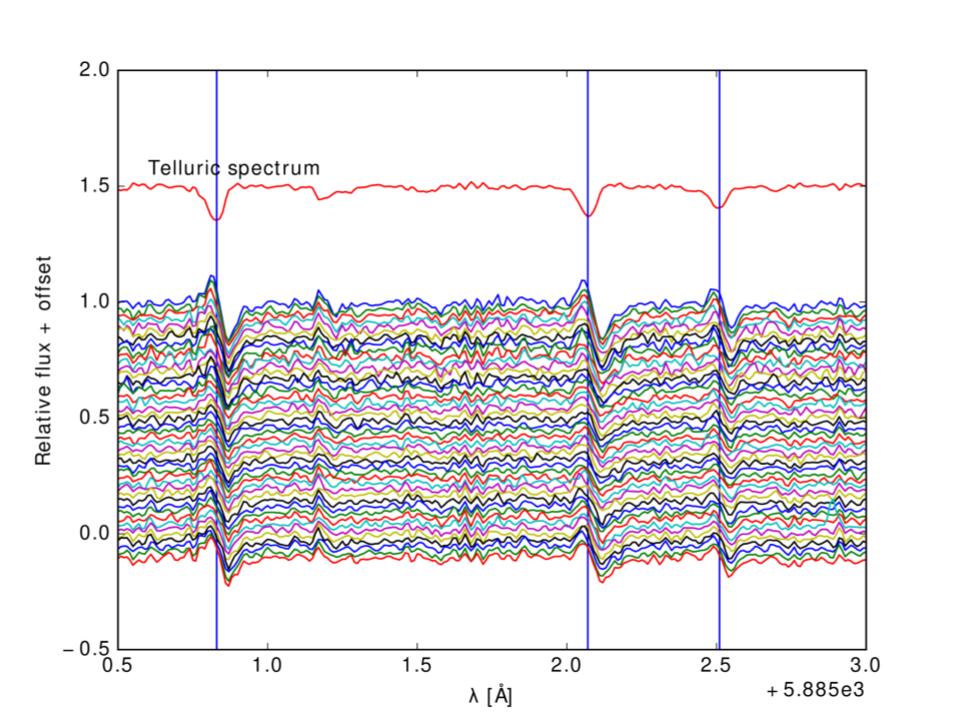}
    \caption{Stellar spectra in the BRF corrected with the telluric spectrum plotted in the top of the figure. The telluric spectrum is obtained using \texttt{s1d} out-of-transit spectra. The vertical blue lines identify telluric lines.}
    \label{fig:s1d_corrected}
\end{figure}

\begin{figure}[ht]
    \centering
    \includegraphics[width=0.5\textwidth]{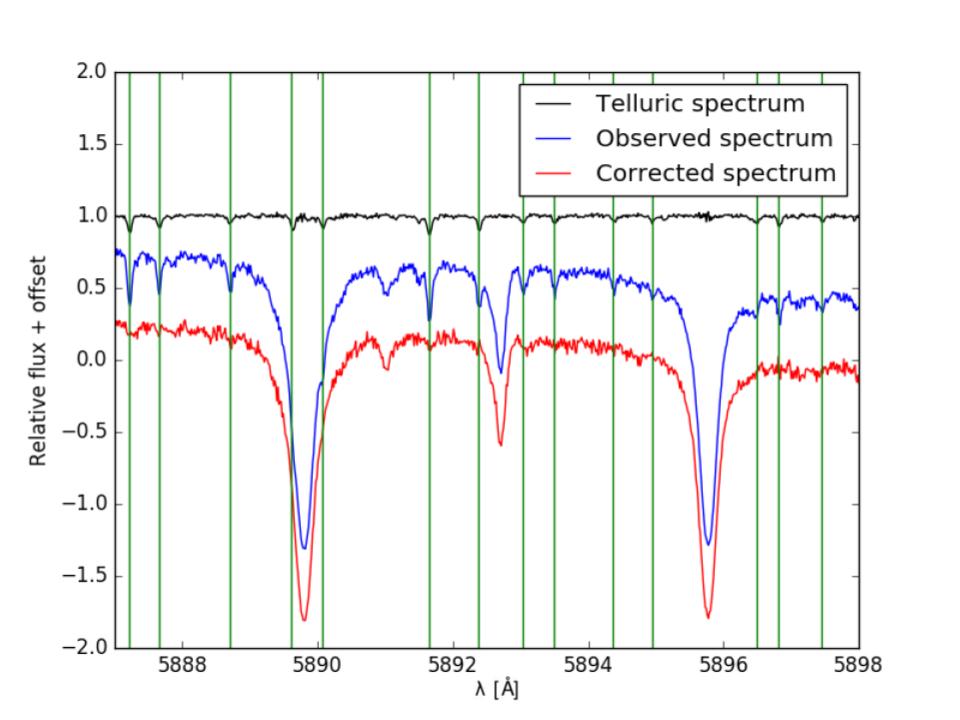}
    \caption{Example of telluric correction in the region of the \ion{Na}{i} doublet using the empirical approach. The green lines indicate the position of the telluric features. Spectra are in the ORF.}
    \label{fig:telluric_correction_lines}
\end{figure}

Furthermore, other authors applied equation \ref{best-fit} and extracted the telluric spectrum from spectra referred to the Barycentric Solar System reference frame (BRF), where the stellar lines are fixed at the same position, but they did not take into account the shift of the telluric lines due to the BERV, improperly assuming that the shift of the telluric lines during the night is negligible. Likewise, in the observer reference frame (ORF), telluric lines are fixed while stellar lines shift their positions. Figure \ref{fig:s1d_vs_e2ds} shows the difference between the telluric spectrum obtained in the BRF (in red) and the one obtained in the ORF (in black): the second presents some spikes just in correspondence of the stellar features, with intensity proportional to that of the stellar line at that position. 
The presence of the spikes is due to the variation of the BERV which is not taken into account in the ORF.\\
\indent We tested the solution to correct the out-of-transit spectra for the BERV before telluric correction, even if this introduces noise to the analysis. However, dividing the observations by the telluric spectrum, does not correct the telluric lines, which are still visible and red-shifted (see Fig.  \ref{fig:s1d_corrected}). Besides, at the same position, spikes appear in the corrected spectra, while they were not present in the telluric spectrum. This confirms that the spikes observed are generated by the shift of the telluric lines in the BRF and of the stellar lines in the ORF.\\
\indent To remove these spikes, which would inevitably affect the final transmission spectrum, \texttt{SLOPpy} simultaneously fits a linear trend as a function of the airmass and a linear trend as a function of the BERV, with the latter being a new term with respect to other authors, in the ORF:
\begin{equation}
\ln I_\lambda= C_{0,\lambda} \, \sec\theta + C_{1,\lambda} \, BERV + C_{2,\lambda},
\end{equation}
where $C_0$, $C_1$, and $C_2$ are the coefficients to be determined, while $\sec\theta$ and $BERV$ are the independent variables. In particular, $C_0$ should represent $\tau_{\lambda,0}$. In this way, for each point, both effects are considered. This means that the telluric spectrum is built using both airmass and BERV as independent variables instead of using just the airmass like in equation \ref{best-fit}. \\
\indent Figure \ref{fig:telluric_correction_lines} shows the telluric spectrum (in black) obtained using the last equation, an 
observed spectrum without any correction (in blue) and with the telluric correction (in red). 
Even if this technique presents several advantages (i.e., the fact to be model-independent or that no additional time to observe the reference star is required), the calculated telluric spectrum is often very noisy.

\subsection{Pregenerated telluric template}\label{appendix_template}
Another approach for the telluric correction implemented in \texttt{SLOPpy} involves the use of a template as telluric reference spectrum.\\
\indent When a model is used to correct the telluric contamination, one of the main problems is the variation in the transmission of the Earth’s atmosphere. The line depth variation is not equal for all the telluric lines, leading to poor results in the removal
of the telluric signatures when scaling the model, and possibly introducing small residuals into the final transmission spectrum.
However, one of the best points of this method is that no additional noise is introduced in the corrected spectra.\\
\indent The synthetic template is created using SkyCalc\footnote{\url{https://www.eso.org/observing/etc/bin/gen/form?INS.MODE=swspectr+INS.NAME=SKYCALC}}, an ESO Sky model Calculator to predict the atmospheric telluric absorption for given observational conditions, given as input by the user. 
This tool relies on HITRAN (HIgh-resolution TRANsmission molecular absorption database), a database that contains atomic lines data which may be used to simulate the spectrum of the Earth’s atmosphere, while the absorption spectrum is computed by using a line-by-line radiative transfer model (LBLRTM), with the weather conditions. 
Thanks to a command-line interface (CLI), the user can perform requests directly to SkyCalc which is hosted on the ESO web server instead of using the usual web form. This is useful if one wants to calculate the sky for many different observation conditions, or to integrate ESO SkyCalc in another astronomical tool, for instance.\\
\indent We verified that the synthetic template built using La Silla characteristics (where the HARPS spectrograph is placed) is nearly identical to a reference telluric spectrum (i.e., at airmass = 1) obtained from 184 spectra of the telluric standard HIP63901 retrieved with HARPS-N located in La Palma (Fig. \ref{fig:comparison_telluric}). After adjusting the FWHM of La Silla SkyCalc model to HARPS-N resolution, a rescaling is sufficient to fit the telluric lines. 
\begin{figure*}[ht]
    \includegraphics[width=\textwidth]{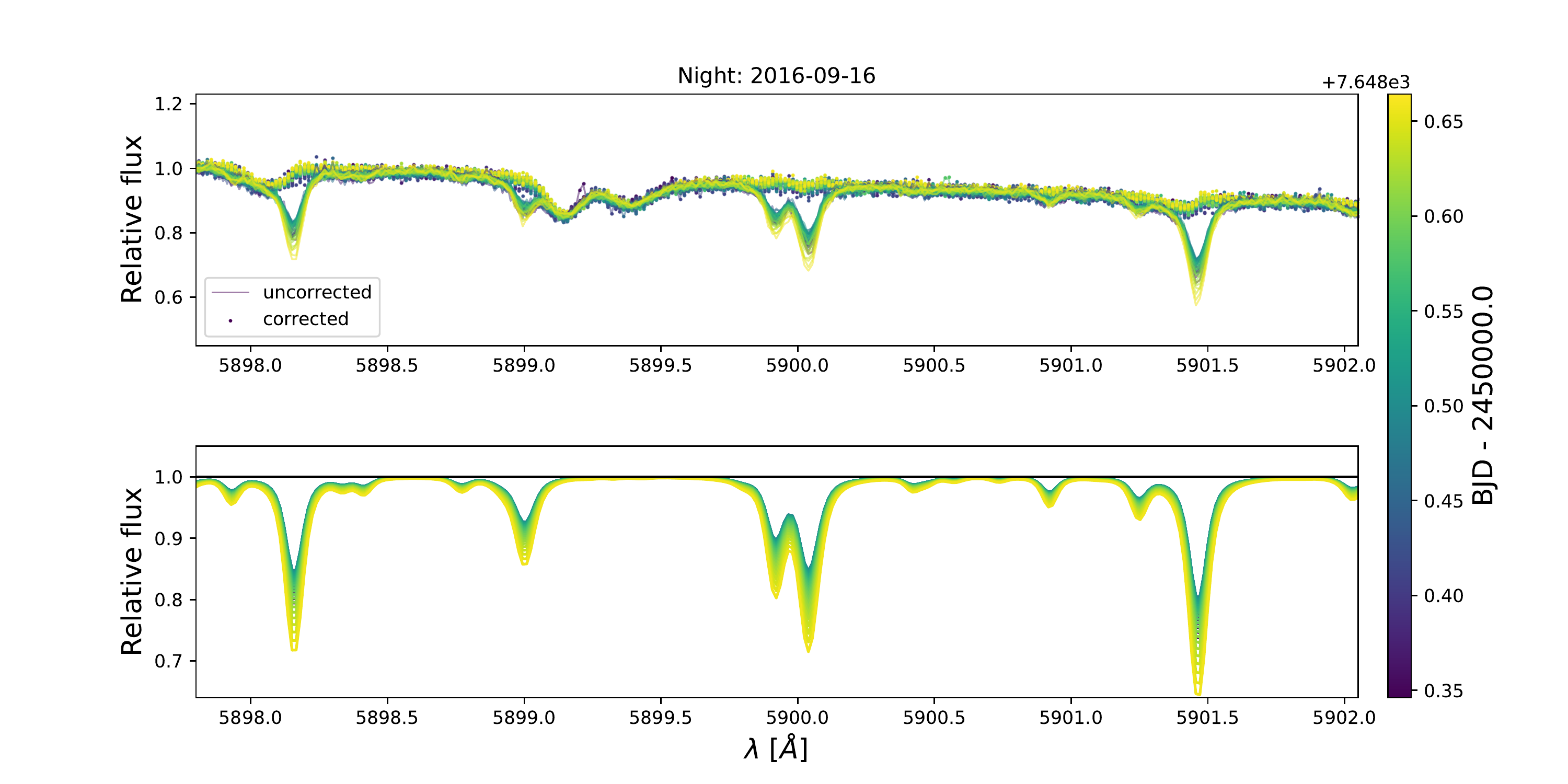}
    \caption{\label{fig:telluric_correction_template} Example of telluric correction using the template applied to a spectral time series  of the planet-host star HD 209458 b. \textit{Top panel:} comparison between uncorrected and corrected spectra. \textit{Bottom panel:} telluric template. All spectra are color-coded according to the time.}
\end{figure*}
\begin{figure}
    \centering
    \includegraphics[width=0.5\textwidth, keepaspectratio]{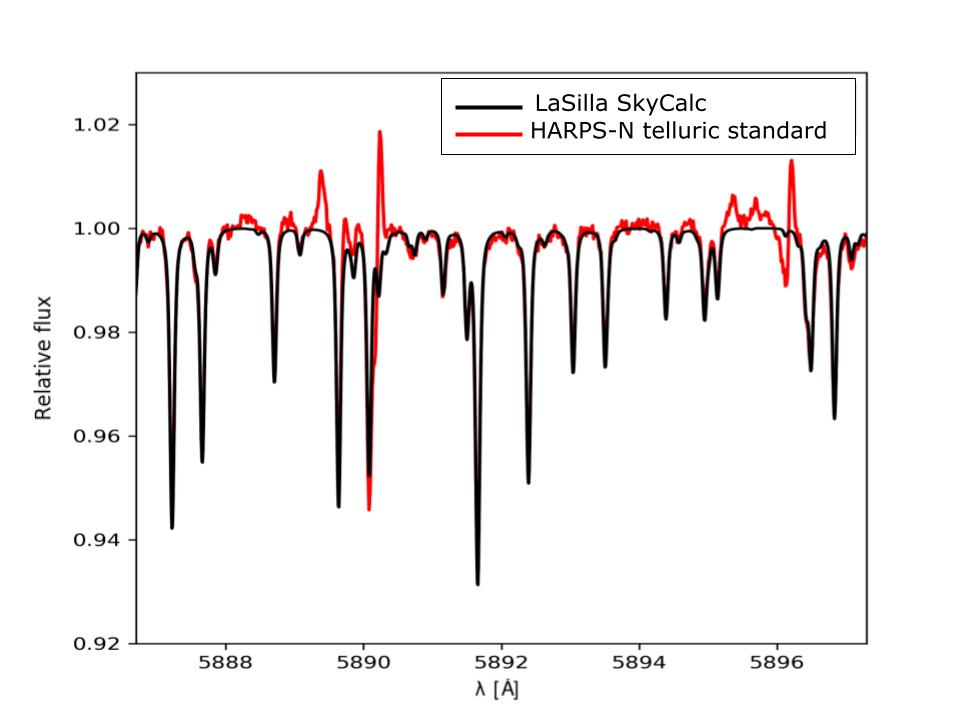}
    \caption{Comparison between the HARPS-N reference telluric spectrum obtained from 184 spectra of the telluric standard HIP63901, and the telluric template made with ESO SkyCalc using La Silla characteristics. The bumps in the red line corresponds to the center of the Na lines, where the removal of the stellar lines is more difficult.}
    \label{fig:comparison_telluric}
\end{figure}

Using the template results in a dramatic improvement in the telluric correction compared to the empirical approach. Furthermore, this method can be used even when the baseline of the observations is insufficient to compute a high-quality telluric spectrum for an efficient correction of telluric lines.
This approach has been tested for the Na doublet and the surrounding spectral regions, where the telluric lines are dominated by water vapor absorption (Fig. \ref{fig:telluric_correction_template}). Water vapor however is not the only constituent of the atmosphere which changes with time and altitude. If water vapor is the only constituent that is corrected appropriately with the template, as in our case, the features of the other constituents will not align properly in strength and width with the observer’s spectrum, producing an inaccurately corrected spectrum. Additionally, instruments with lower resolution than HARPS and HARPS-N may not be suitable for this approach,  since matching the line profiles could become too difficult for low resolution data since the template will show the lines individually resolved while atmospheric lines appear as bands in the spectrum. So while this approach is superior to the empirical one, it cannot be generalized to the whole spectral interval of HARPS and HARPS-N.

\section{Corner plots of the best-fit models}\label{Appendix_cornerplot}
\begin{figure*}
    \centering
    \includegraphics[width=\textwidth]{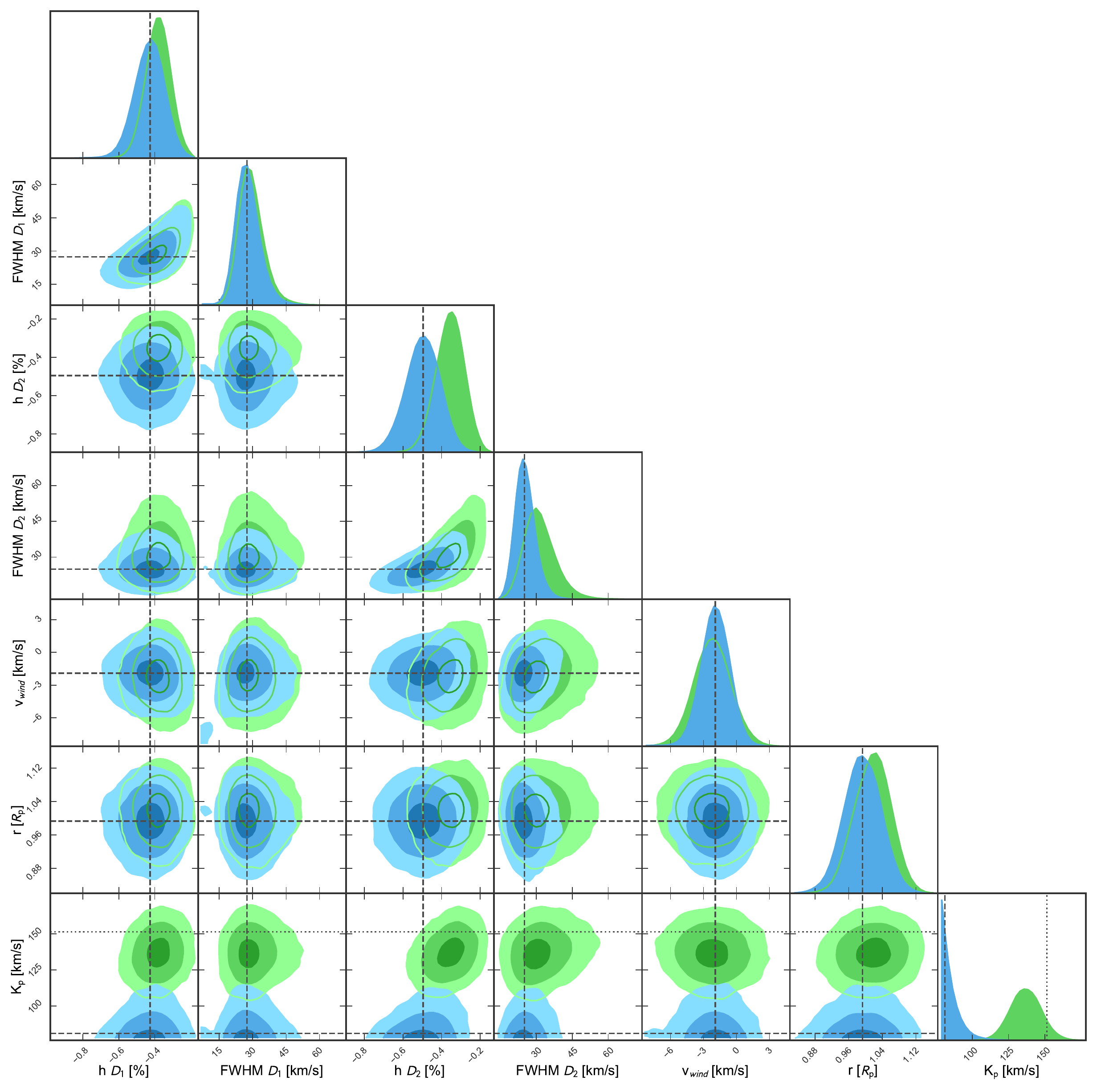}
    \caption{Corner plot of MCMC analysis of the \ion{Na}{i} D lines for HD 189733 b without a prior (\textit{blue}) and with a Gaussian prior on $K_p$ (\textit{green}). The dotted line in the posterior distribution of $K_p$ indicates the theoretical value ($\sim$ 150 km s$^{-1}$).}.
    \label{fig:corner_plotHD189}
\end{figure*}

\begin{figure*}
    \centering
    \includegraphics[width=\textwidth]{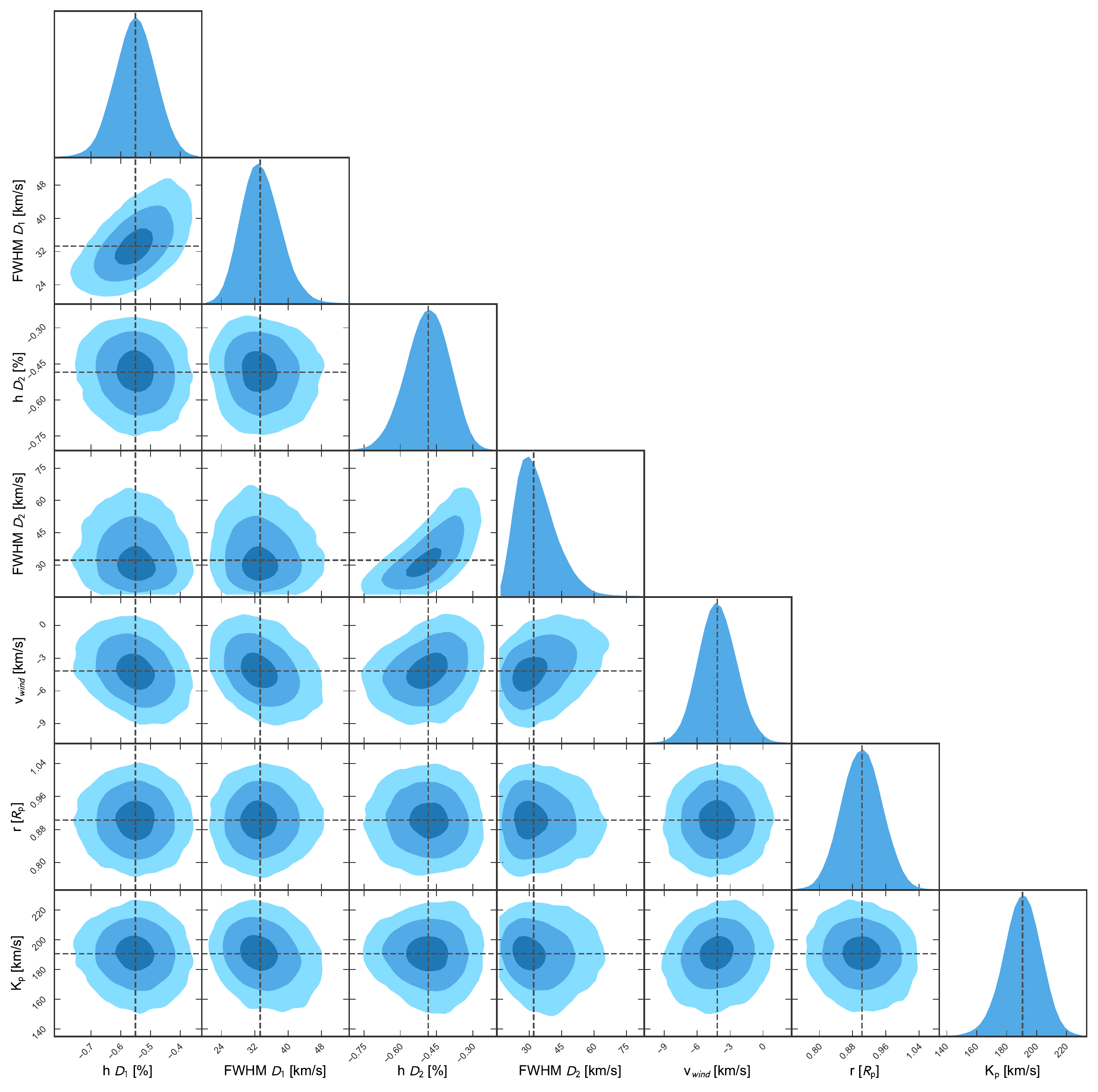}
    \caption{Corner plot of MCMC analysis of the \ion{Na}{i} D lines for WASP-76 b.}
    \label{fig:corner_plotWASP76}
\end{figure*}

\begin{figure*}
    \centering
    \includegraphics[width=\textwidth]{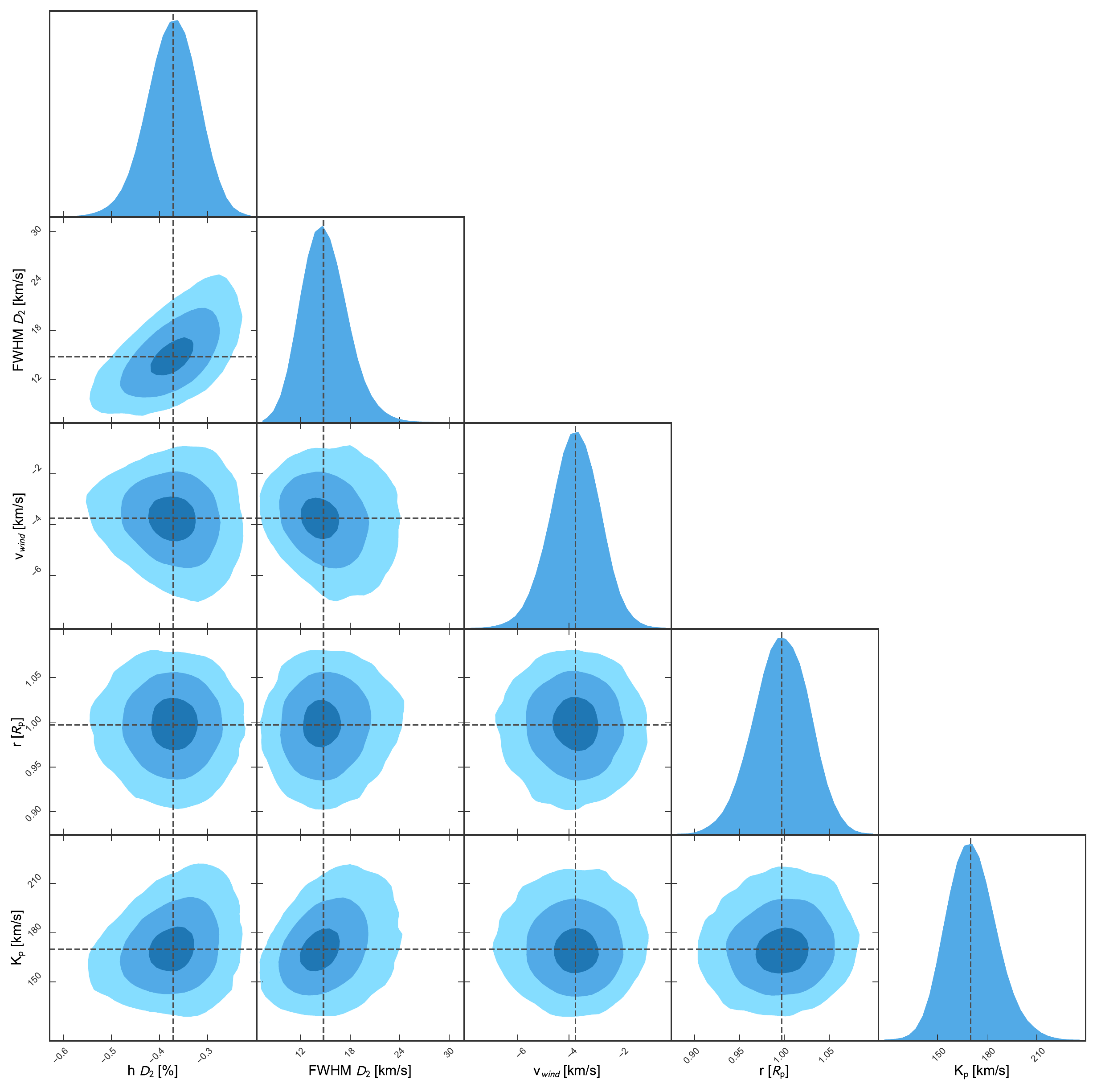}
    \caption{Corner plot of MCMC analysis of the Na D$_2$ line for KELT-20 b.}
    \label{fig:corner_plotKelt20_D2}
\end{figure*}

\begin{figure*}
    \centering
    \includegraphics[width=\textwidth]{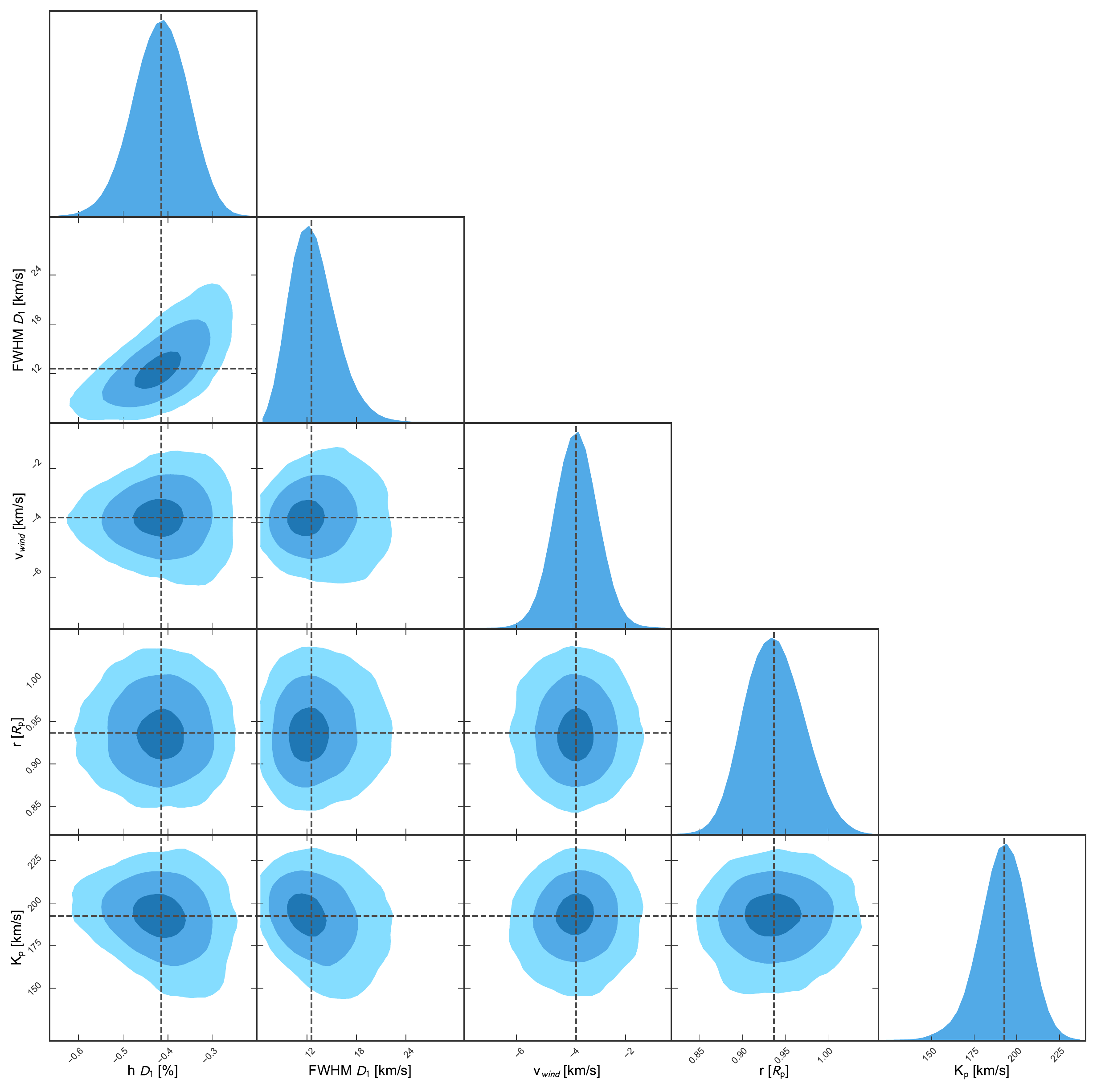}
    \caption{Corner plot of MCMC analysis of the D$_1$ line for KELT-20 b.}
    \label{fig:corner_plotKelt20_D1}
\end{figure*}

\begin{figure*}
    \centering
    \includegraphics[width=\textwidth]{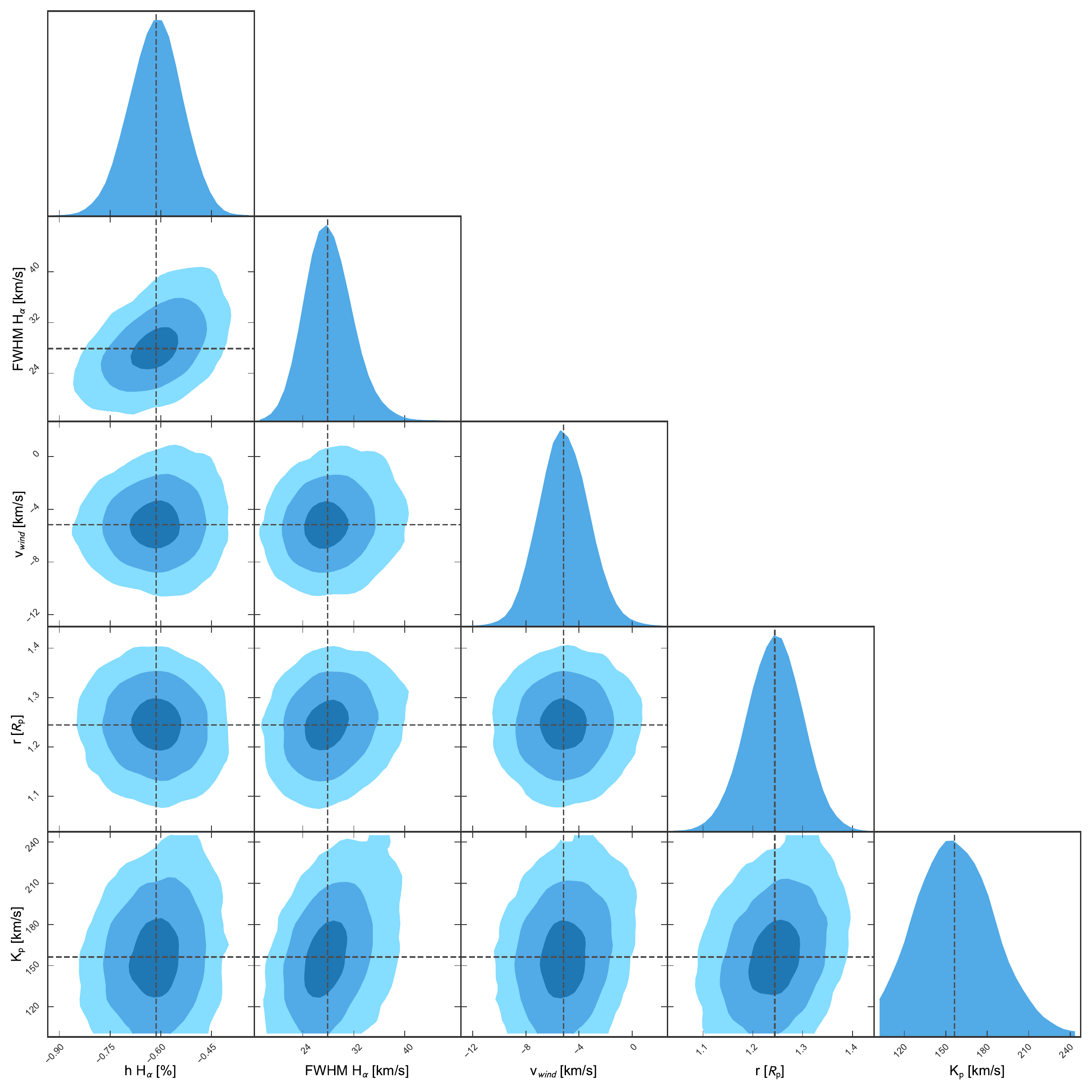}
    \caption{Corner plot of MCMC analysis of the H$_\alpha$ line for KELT-20 b.}
    \label{fig:corner_plotKelt20_Ha}
\end{figure*}

We present here the corner plots of the MCMC analysis obtained for HD 189733 (Fig. \ref{fig:corner_plotHD189}), WASP-76 (Fig. \ref{fig:corner_plotWASP76}) and KELT-20 b (Figs. \ref{fig:corner_plotKelt20_D2}, \ref{fig:corner_plotKelt20_D1} and \ref{fig:corner_plotKelt20_Ha}) combining all nights.

\end{appendix}

\end{document}